\documentclass[prd,showpacs,showkeys,preprintnumbers,floatfix,
nofootinbib,superscriptaddress]{revtex4}
\usepackage{float}
\usepackage{nicefrac}
\usepackage{slashed}
\usepackage{mathtools}
\usepackage{amsfonts} 
\usepackage{amssymb} 
\usepackage{amsmath} 
\usepackage{graphicx} 
\usepackage{subfigure} 
\usepackage{array} 
\usepackage{dcolumn} 
\usepackage{bm} 
\usepackage{latexsym} 
\usepackage{longtable} 
\usepackage{hyperref} 
\usepackage{verbatim}
\usepackage{epsfig}
\usepackage{color}
\newcommand{\Wdf}[0]{{\mathcal{W}}_{\rm df}}
\newcommand{\W}[0]{{\mathcal{W}}}
\newcommand{\Wtildf}[0]{{\mathcal{W}}_{L,{\rm df}}}

\newcommand{\nn}[0]{\nonumber}
\newcommand{\khs}[0]{\hat {\textbf{k}}^*}
\newcommand{\w}[0]{w}

\begin{document}

 \preprint{\vbox{
\hbox{JLAB-THY-15-2140} }}

\title{Relativistic, model-independent, multichannel\\ $2\to2$ transition amplitudes in a finite volume
}

\author{Ra\'ul A. Brice\~no}
\email[e-mail: ]{rbriceno@jlab.org}
\affiliation{
Thomas Jefferson National Accelerator Facility, 12000 Jefferson Avenue, Newport News, VA 23606, USA\\
}

\author{Maxwell T. Hansen}
\email[e-mail: ]{hansen@kph.uni-mainz.de}
\affiliation{
Institut f\"ur Kernphysik and Helmholz Institute Mainz, Johannes Gutenberg-Universit\"at Mainz,
55099 Mainz, Germany\\
}
\date{\today}
\begin{abstract}
We derive formalism for determining $\textbf{2} + \mathcal J \to \textbf{2}$ infinite-volume transition amplitudes from finite-volume matrix elements. Specifically, we present a relativistic, model-independent relation between finite-volume matrix elements of external currents and the physically observable infinite-volume matrix elements involving two-particle asymptotic states. The result presented holds for states composed of two scalar bosons. These can be identical or non-identical and, in the latter case, can be either degenerate or non-degenerate. We further accommodate any number of strongly-coupled two-scalar channels. This formalism will, for example, allow future lattice QCD calculations of the $\rho$-meson form factor, in which the unstable nature of the $\rho$ is rigorously accommodated.  
\end{abstract}
\pacs{13.40.Gp,14.40.-n,12.38.Gc,11.80.Jy}
\keywords{finite volume}
\maketitle

\section{Introduction \label{sec:intro}}

Theoretical predictions of hadron structure are entering a new era. The precise determination of form factors for stable hadronic states is already well underway~\cite{Shultz:2015pfa, Alexandrou:2011ga, Alexandrou:2011py, Green:2014xba} and resonant form factor studies are not far behind. Indeed, the first lattice QCD (LQCD) calculations of resonant $\mathcal J \rightarrow\textbf{2}$ and $\textbf{1}+\mathcal J \rightarrow\textbf{2}$ transition processes appeared earlier this year.
\footnote{Throughout this work, $\textbf{n}+\mathcal J\rightarrow\textbf{m}$ labels a process with $\textbf{n}$ incoming and $\textbf{m}$ outgoing stable hadrons in the presence of an external current $\mathcal{J}$.} 
These studies considered $\gamma^\star\to\pi\pi$~\cite{Feng:2014gba} and $\gamma^\star\pi\to\pi\pi$~\cite{Briceno:2015dca} transitions. In Ref.~\cite{Briceno:2015dca}, the Hadron Spectrum Collaboration determined the $\gamma^\star\pi\to\pi\pi$ amplitude for a range of energies and for various virtualities of the external photon. The resulting fit was analytically continued to the $\rho$-pole, thereby giving a first principles determination of the $\gamma^\star\pi\to\rho$ form factor. This result illustrates that resonance properties beyond masses and widths can be obtained from LQCD. Encouraged by the growing progress in this field, we present here the formalism needed to study generic $\textbf{2}+\mathcal J \rightarrow\textbf{2}$ transition processes in LQCD. This will make it possible to determine elastic form factors of resonances as well as various two-to-two transition amplitudes. Before describing the formalism derived in this work, we briefly motivate it in the context of LQCD studies of multi-particle observables.

In numerical LQCD the theory is placed in a finite, discretized Euclidean spacetime. For simple observables, such as single hadron masses and space-like form factors, truncation and discretization of spacetime, together with the restriction to Euclidean time, have little effect on the extracted observables. For matrix elements of two-or-more-hadron states, by contrast, these modifications have significant consequences.  The first issue is that, in a compactified spacetime, it is no longer possible to define asymptotic states. Thus the QCD eigenstates which arise in finite- and infinite-volume are fundamentally different objects. In addition, LQCD calculations can only provide numerical results for Euclidean correlators with nonzero statistical uncertainties. For such results, the analytic continuation required to access Minkowski-time transition amplitudes is an ill-posed problem (see for example Ref.~\cite{Bertero}).

It turns out that one can overcome these issues in certain cases, by deriving a model-independent relation between finite- and infinite-volume observables. For example, the finite-volume energy spectrum of two-  
\cite{Luscher:1986pf, Luscher:1990ux, Rummukainen:1995vs, He:2005ey, Kim:2005gf, Christ:2005gi, Leskovec:2012gb,  Briceno:2012yi, Hansen:2012tf, Briceno:2014oea} and three-particles \cite{Hansen:2014eka, Hansen:2015zga, Briceno:2012rv,  Polejaeva:2012ut} can be used to determine, or at least constrain, infinite-volume scattering amplitudes. In the two-particle sector, this formalism has made it possible to determine scattering amplitudes in channels with resonances from numerical LQCD~\cite{Wilson:2015dqa, Wilson:2014cna, Dudek:2014qha, Dudek:2012xn, Lang:2011mn, Lang:2015hza, Lang:2014yfa, Torres:2014vna, Feng:2010es, Pelissier:2012pi, Prelovsek:2013ela, Aoki:2011yj, Aoki:2007rd, Bolton:2015psa}. By parametrizing and analytically continuing the scattering amplitudes into the complex energy plane, some of these investigations also offer systematic determinations of resonance pole positions. 

The focus of the present work is an observation closely associated with the relation between finite-volume energy spectra and scattering observables, namely that finite-volume matrix elements can be used to extract infinite-volume matrix elements with two-particle asymptotic states~\cite{Lellouch:2000pv,Lin:2001ek, Kim:2005gf, Christ:2005gi, Hansen:2012tf, Agadjanov:2014kha, Meyer:2012wk, Bernard:2012bi, Briceno:2012yi, Feng:2014gba, Briceno:2015csa, Briceno:2014uqa}. The latter are referred to throughout this work as transition amplitudes. In earlier work we have derived the relation needed to map finite-volume matrix elements to arbitrary $\textbf{1}+\mathcal J \rightarrow\textbf{2}$ processes~\cite{Briceno:2014uqa, Briceno:2015csa}, thereby summarizing and generalizing previous studies on the topic. It was partly this formalism that made the calculation of the $\gamma^\star\pi\to\pi\pi$ amplitude possible~\cite{Briceno:2015dca}. In this article we demonstrate how this formalism can be extended to extract $\textbf{2}+\mathcal J \rightarrow\textbf{2}$ transition amplitudes. In the context of our field theoretic analysis, these transition amplitudes, which we collectively denote $\mathcal W$, are defined as the sum of all infinite-volume Feynman diagrams with four external hadron legs and one external current [see Fig.~\ref{fig:W_iepsilon} below].

\begin{figure*}[t]
\begin{center}
\subfigure[]{
\label{fig:2to2_long_range_first}
\includegraphics[scale=0.40]{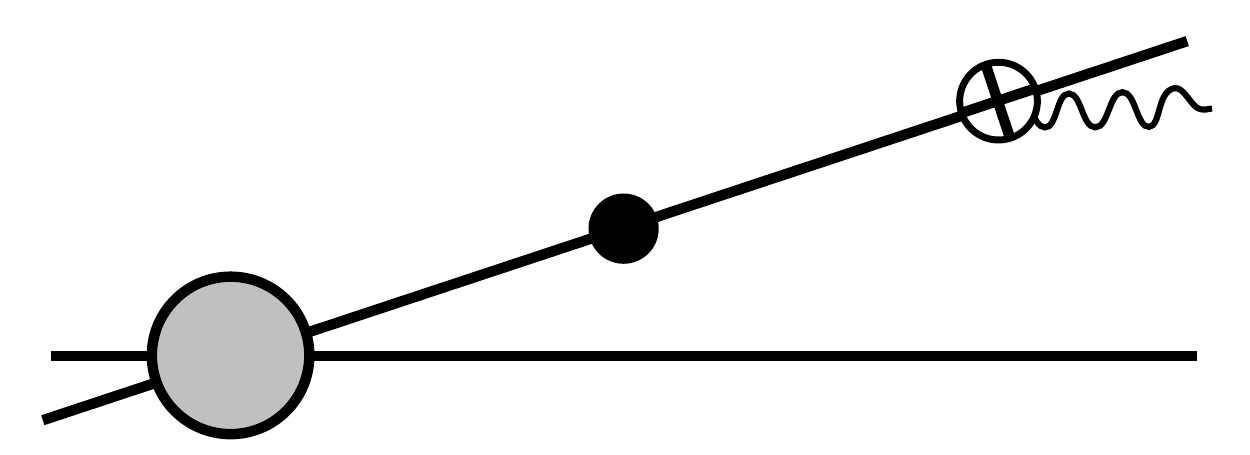}} 
\hspace{1cm}
\subfigure[]{
\label{fig:double_pole}
\includegraphics[scale=0.40]{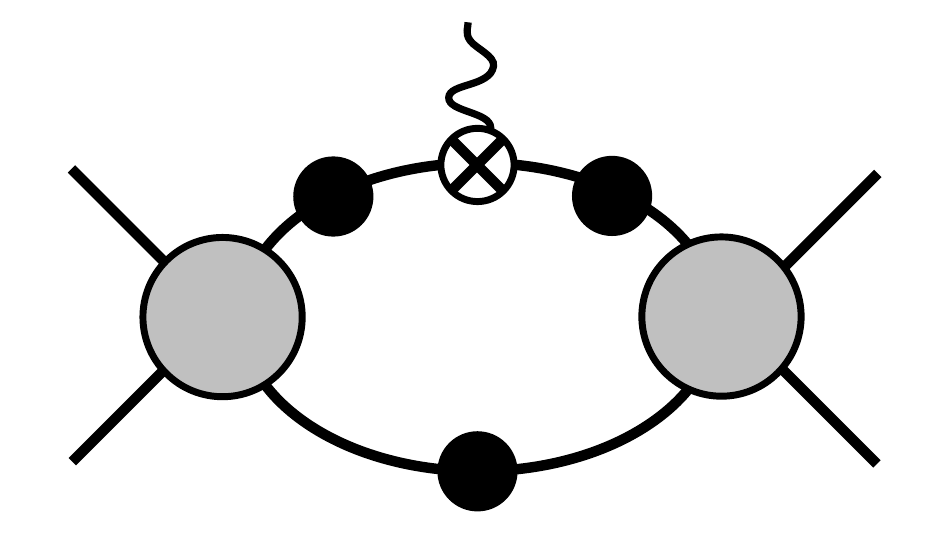}}\hspace{1cm}
\caption{Shown are types of subdiagrams that distinguish $\textbf{2} + \mathcal J \rightarrow \textbf{2}$ from the simpler $\textbf{1} + \mathcal J \rightarrow \textbf{2}$ processes. These include (a) divergent diagrams associated with intermediate particles going on-shell and (b) two-particle loops with an insertion of the external current.  The latter lead to a new finite-volume function, investigated for the first time in this work. This new object is defined in Eq.~(\ref{eq:Gmat}) and discussed in detail in Appendix \ref{app:Gnum}.}\label{fig:poles_infinite_volume}
\label{fig:2to2issues}
\end{center}
\end{figure*}

Although the study of $\textbf{2}+\mathcal J \rightarrow\textbf{2}$ systems bears similarities to that of $\textbf{1}+\mathcal J \rightarrow\textbf{2}$, the former is significantly more complicated for two reasons. The main sources of complication relative to the earlier analysis are summarized in Fig.~\ref{fig:2to2issues}. First, the infinite-volume $\textbf{2}+\mathcal J \rightarrow\textbf{2}$ amplitude, $\mathcal W$, possesses kinematic singularities that are absent in $\textbf{1}+\mathcal J \rightarrow\textbf{2}$ systems. These are due to diagrams in which a single hadron propagator connects a $\textbf 2 \rightarrow \textbf 2$ scattering amplitude, which we denote $\mathcal M$, with a $\textbf{1}+\mathcal J \rightarrow\textbf{1}$ transition amplitude, labeled $\w$ [see Fig.~\ref{fig:2to2_long_range_first}]. A divergence occurs if external kinematics are chosen to put the intermediate propagator on-shell. This divergence has nothing to do with bound-states but is instead due to the possibility of arbitrarily long lived intermediate states between physically observable sub-proceses.

The second complication in the finite-volume study of $\textbf{2}+\mathcal J \rightarrow\textbf{2}$ systems is that the summands of finite-volume loops include terms with two poles that share a common coordinate. These singularities arise from two-particle loops in which the current couples to one of the two-particles in the loop, possibly injecting energy and momentum [see Fig.~\ref{fig:double_pole}]. The new singularity structure leads to a new type of finite-volume function which is absent in studies of two-particle scattering and $\textbf{1}+\mathcal J \rightarrow\textbf{2}$ transitions. The issues of singularities in the infinite-volume transition amplitude, $\mathcal W$, and new pole structures in the finite-volume loops are in fact closely related. Understanding how to accommodate these new features is the primary focus of this work.

As the derivation presented in this article is lengthy, we think it helpful to summarize our main result here, 
\begin{equation}
\Big | \langle E_{n_f}, \textbf P_f, L \vert  \mathcal J(0)  \vert E_{n_i}, \textbf P_i, L \rangle \Big |^2_L 
=\frac{1}{L^6} {\rm{Tr}}\left[
 \mathcal R(E_{n_i}, \textbf P_i) 
\Wtildf(P_i,P_f,L)
\mathcal R(E_{n_f}, \textbf P_f)  
\Wtildf(P_f,P_i,L)
\right] \,,
  \label{eq:2to2_notdegen_intro}
\end{equation}
where $|E_{n}, \textbf P, L \rangle$ labels the finite-volume state with energy and momentum $E_{n}, \textbf P$ in a cubic, periodic volume with linear extent $L$. This relation is only valid if the center of mass (CM) frame energy, $E_n^{*2}\equiv E_n^2 - \textbf P^2$, is below the lowest multi-particle threshold. If this kinematic restriction is satisfied then the equality holds up to exponentially suppressed corrections of the form $e^{-mL}$, where $m$ is the physical mass of the lightest scalar in the theory. The trace here is over the direct product of angular-momentum and channel space, labeled by spherical harmonic indices $\ell, m$ and a channel index $a$. The matrix $\mathcal R(E_{n}, \textbf P)$ is the residue of a known function at the pole associated with the finite-volume state, and is defined in Ref.~\cite{Briceno:2014uqa} as well as in Eq.~(\ref{eq:Rdef}) of Sec.~\ref{sec:two-point} below. 

Suppose now that the finite-volume matrix element on the left-hand side of Eq.~(\ref{eq:2to2_notdegen_intro}) has been determined, in a numerical LQCD calculation or by some other method. If the on-shell scattering amplitude, $\mathcal M$, is also known, then one can determine the residue matrix $\mathcal R$, and with these inputs it is possible to constrain the remaining quantity, $\Wtildf$. This in turn can be expressed as a sum of two terms
\begin{equation}
 \Wtildf(P_f,P_i,L)  \equiv \Wdf(P_f,P_i) +  \mathcal  M(P_f) \ [G(L) \cdot \w](P_f,P_i) \ \mathcal M(P_i)  \,.
 \end{equation}
The first of these, $\Wdf(P_f,P_i)$, is the infinite-volume, divergence-free transition amplitude. This quantity, defined in Eq.~(\ref{eq:Wdfdef}) of Sec.~\ref{sec:1body} below, is given by subtracting the long-lived singularities [shown in Fig.~\ref{fig:2to2_long_range_first}] from the full transition-amplitude, $\mathcal W$. The second term, $\mathcal  M(P_f) \ [G(L) \cdot \w](P_f,P_i) \ \mathcal M(P_i)$, encodes the finite-volume effects of the double poles [shown in Fig.~\ref{fig:double_pole}]. We stress that the difference between $\Wtildf$ and the physically observable transition amplitude, $\mathcal W$, only depends on on-shell $\textbf 2 \rightarrow \textbf 2$ scattering amplitudes and $\textbf{1}+\mathcal J \rightarrow\textbf{1}$ transition amplitudes. 

\begin{figure*}[t]
\begin{center}
\includegraphics[scale=0.28]{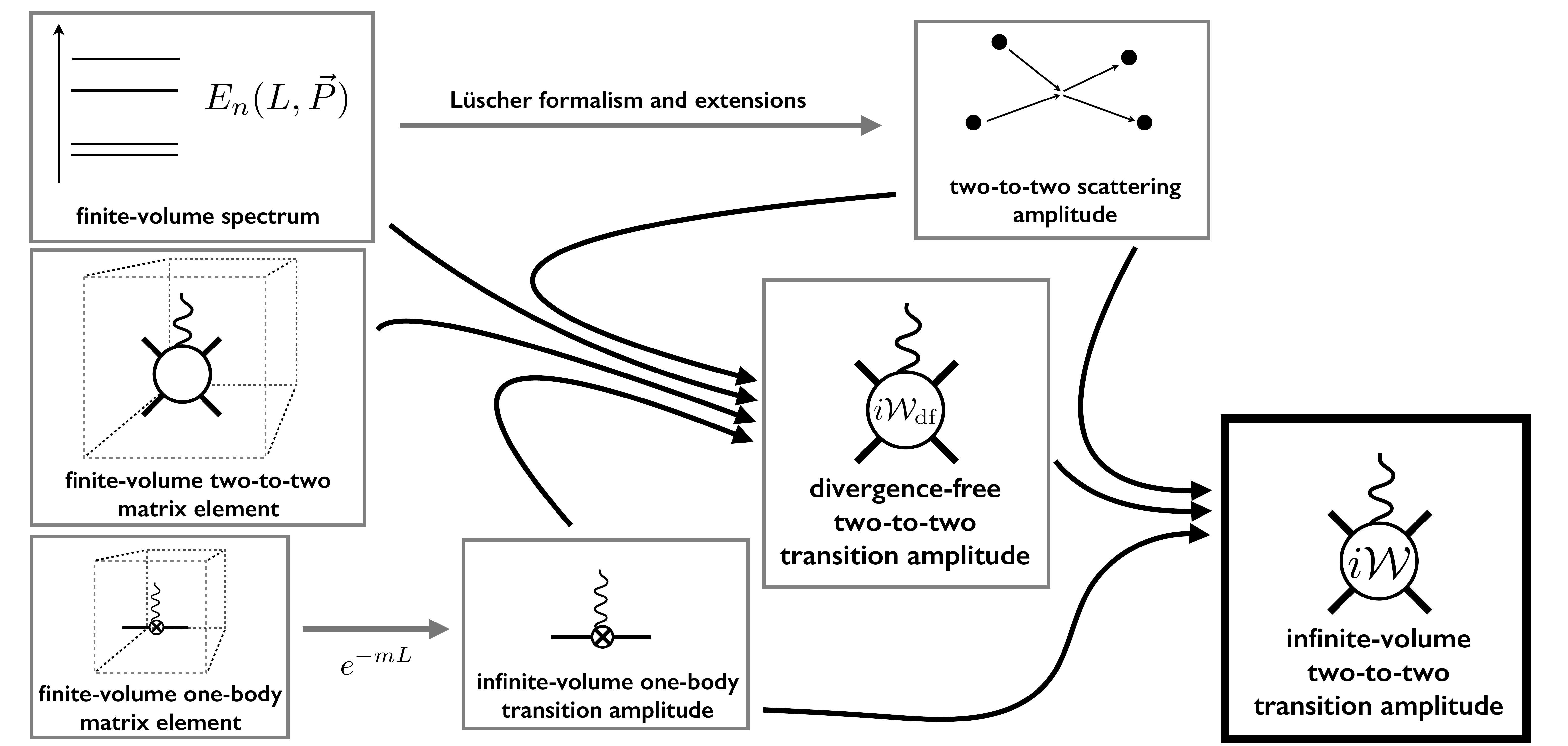}
\caption{Inputs needed to extract $\textbf{2}+\mathcal J \rightarrow\textbf{2}$ transition amplitudes using this formalism. In the first step one determines the infinite-volume, divergence-free transition amplitude $\Wdf$. Unlike the full transition amplitude, this quantity is a smooth function, which can be decomposed in harmonics and truncated at low energies. In a second step the divergence free quantity is combined with on-shell $\textbf 1 + \mathcal J \rightarrow \textbf 1$ amplitudes as well as $\mathcal M$, to determine the physical observable.}\label{fig:flowchart}
\end{center}
\end{figure*}

As is common in this type of formalism, the combined angular-momentum and channel space of the matrices in Eq.~(\ref{eq:2to2_notdegen_intro}) is formally infinite dimensional. Thus the result can only be made useful by truncating the infinite-volume observables to some finite-dimensional subspace. Such a truncation is well motivated at low energies, where the lowest partial waves are dominant, provided that the quantities in question are smooth functions of their directional degrees of freedom. This is true for $w$, $\mathcal M$ and $\Wdf$, and truncating these leads to simplified, useful expressions, as we demonstrate in Sec.~\ref{sec:simplim}. As we also discuss in that section, truncating $\mathcal W$ directly is not justified due to the singularities in that quantity. In Fig.~\ref{fig:flowchart} we summarize the information required to extract $\textbf{2}+\mathcal J \rightarrow\textbf{2}$ transition amplitudes using this formalism.

The relation between finite- and infinite-volume two-to-two matrix elements has already been studied in various contexts. In Ref.~\cite{Detmold:2014fpa}, Detmold and Flynn give a relation between finite-volume matrix elements of $n$-bosons and infinite-volume low-energy coefficients.  This work expands the finite-volume matrix elements in powers of $1/L$, keeping terms through $\mathcal O(1/L^5)$. In Refs.~\cite{Briceno:2012yi, Bernard:2012bi} the authors use two different effective field theories (EFTs) to find a relation between finite-volume matrix elements and infinite-volume observables in the lowest partial wave. This is done to all orders in the strong interaction but only keeping a finite order of the low-energy coefficients couplings the hadrons with the given external current. In the present article we present an all-orders, EFT-and-model-independent 
relation between finite- and infinite-volume quantities.  Furthermore, our result completely encodes the reduction of rotational symmetry, by accommodating partial wave mixing in accordance with the symmetry group of the system (octahedral group or little groups thereof). This study also goes beyond earlier derivations by accommodating any number of two-scalar channels, with identical or non-identical particles which, in the latter case, can be either degenerate or non-degenerate.

In addition to laying the foundation for the study of matrix elements of hadronic resonances, we envision that this result will have an impact in extracting other phenomenologically interesting quantities. One prominent example is related to the parity violating contribution to the two-nucleon scattering amplitude. It has been over half a century since Lee and Yang first suggested the possibility of parity non-conservation in the weak interaction~\cite{Lee:1956zz}, which was confirmed expermentially shortly thereafter by Wu et al.~\cite{Wu:1957my,tanner, Krane:1971zza, Krane:1971zz, Yuan:1991zz}. Modern day experimental \cite{Eversheim:1991tg,Balzer:1980dn,Balzer:1985au,Berdoz:2001nu,Kistryn:1987tq,Berdoz:2002sn,Cavaignac:1977uk,Gericke:2011zz,Knyazkov:1984zz, Snow:2011zza} and theoretical \cite{Phillips:2008hn, Griesshammer:2011md, Schindler:2009wd,Shin:2009hi, Savage:1999cm} studies have given attention to parity violating two-nucleon processes, where the strong interactions are most precisely understood. These include proton-neutron fusion, $p+n\rightarrow d+\gamma$, and elastic proton scattering, $p+p\rightarrow p+p$. 

There has been a great deal of theoretical progress in parametrizing low-energy parity-violating processes in terms of parity-conserving scattering parameters and the $N + \mathcal J_{\slashed P} \rightarrow N \pi$, $N \pi + \mathcal J_{\slashed P} \rightarrow N \pi$ and $N N + \mathcal J_{\slashed P} \rightarrow N N$ transition amplitudes, with $\mathcal J_{\slashed P}$ being the parity violating part of the weak hamiltonian.
\footnote{We point the reader to Refs.~\cite{Haxton:2013aca, Schindler:2013yua} for recent reviews on the topic.}
The first attempt to study such processes in LQCD was made by Wasem in Ref.~\cite{Wasem:2011zz}, where an exploratory calculation of $N\to N\pi$ was performed. This has inspired the CalLat Collaboration to begin efforts to determine all relevant matrix elements directly from LQCD. Recognizing that two-to-two scattering phase shifts and their derivatives are needed to relate finite- and infinite-volume matrix elements, CalLat has recently given the first determination of nucleon elastic scattering in higher partial waves, up to $\ell = 3 $~\cite{Berkowitz:2015eaa}. This study relied on the the two-nucleon finite-volume formalism derived in Refs.~\cite{Briceno:2013lba, Briceno:2013bda}.\footnote{NPLQCD has also recently performed a thorough study of S-wave nucleon elastic scattering in Ref.~\cite{Orginos:2015aya}. In it the authors expands on their previous efforts~\cite{Beane:2006mx, Beane:2013br}, by placing the first constraint of the tensor nuclear force via lattice QCD. }

 A final application of great interest would be the study of two-particle QCD states in fixed background fields. Recently the NPLQCD collaboration exploited the use of auxiliary fields to determine the $np \to d \gamma$ cross section~\cite{Beane:2014ora} and magnetic moments of light nuclei~\cite{Beane:2015yha}. This approach used the fact that, at unphysically heavy quark masses, the ground state of the channels considered are deeply bound and exponentially suppressed finite-size effects can be safely ignored. In order to  use the auxiliary field method for scattering states, and to account for the finite-size effects of shallow bound states (such as the deuteron), the formalism presented here and subsequent extensions will be needed.~\footnote{In Ref.~\cite{Detmold:2004qn}, Detmold and Savage used EFT methods to study two-nucleon states in the presence of an auxiliary field. Combining the work presented there with this general formalism could lead to an EFT-independent formalism for auxiliary fields in finite volume}

The remainder of this article is organized as follows: In the following section we describe the infinite-volume observables that enter this work. These include the $\textbf 2 \rightarrow \textbf 2$ scattering amplitude, $\mathcal M$, the $\textbf 1 + \mathcal J \rightarrow \textbf 1$ transition amplitude, $w$, and the $\textbf 2 + \mathcal J \rightarrow \textbf 2$ transition amplitude, $\mathcal W$. The latter is the key observable that we aim to extract with this formalism. In this section we also define the divergence free $\textbf 2 + \mathcal J \rightarrow \textbf 2$ transition amplitude, $\Wdf$, in which the singularities of Fig.~\ref{fig:2to2_long_range_first} are removed. In Sec.~\ref{sec:finite_volume_func} we derive two identities needed to analyze the finite-volume two- and three-point correlators studied in this work. In Sec.~\ref{sec:two-point} we use the first of these identities and review how to express the the finite-volume two-point correlator in terms of infinite-volume quantities and finite-volume kinematic functions. Then in Sec.~\ref{sec:3pt_func} we derive the analogous expression for the three-point correlator and reach our final result, Eq.~(\ref{eq:2to2_notdegen_intro}). In Sec.~\ref{sec:simplim} we describe various simplifying limits of our general result and also discuss subduction into irreducible representations of the finite-volume symmetry groups. We conclude in Sec.~\ref{sec:conclusion}.  In Appendices \ref{app:Fnum} and \ref{app:Gnum} we give important details about the finite-volume functions that enter our main result.

\begin{figure*}[t]
\begin{center}
\subfigure[]{
\label{fig:scat_ampa}
\includegraphics[scale=0.4]{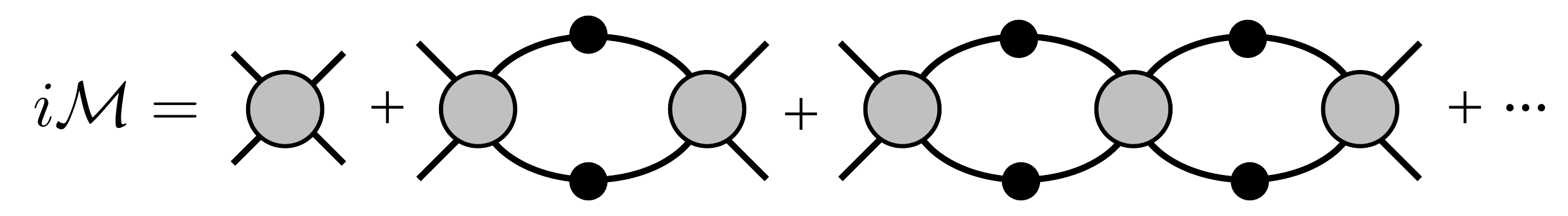}}\\
\subfigure[]{
\label{fig:kernel}
\includegraphics[scale=0.2]{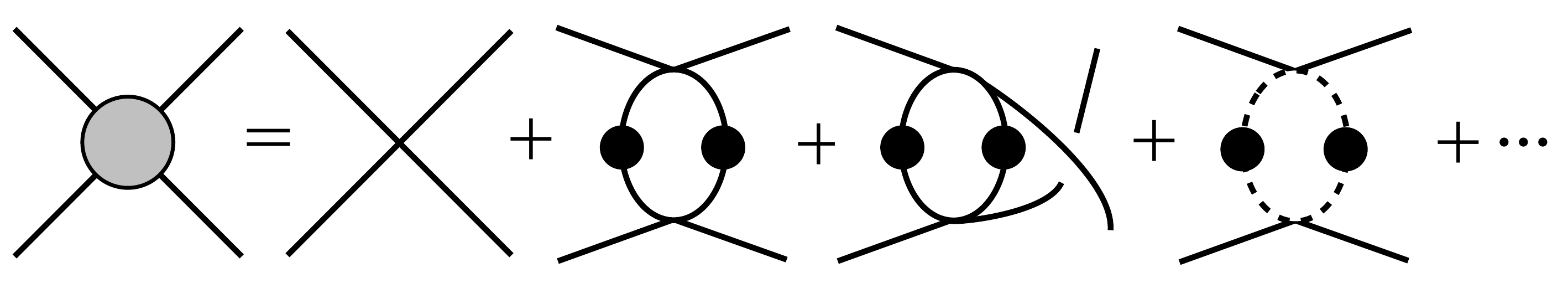}}
\subfigure[]{
\label{fig:1bodyprop}
\includegraphics[scale=0.25]{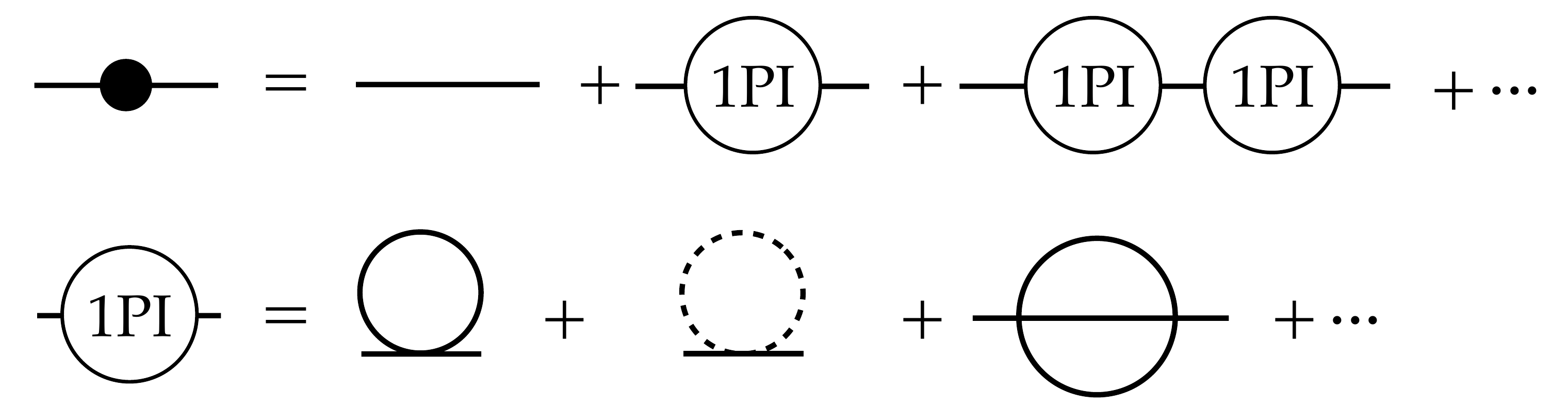}}

\caption{ (a) The scattering amplitude, $\mathcal M$, is defined as the sum over all on-shell, amputated four-point diagrams. This can be written in terms of the Bethe-Salpeter kernel (b) and the fully-dressed single body propagator (c). The Bethe-Salpeter kernel is given by the sum of all amputated four-point diagrams which are two-particle irreducible in the channel carrying the total energy and momentum. This quantity is useful in the present context because, for the kinematics we consider, the difference between its finite- and infinite-volume form is exponentially suppressed in the box size. The same is true for the fully dressed propagator.}\label{fig:scat_amp}
\end{center}
\end{figure*}

\begin{figure*}[t]
\begin{center}
\subfigure[]{
\label{fig:W_ie}
\includegraphics[scale=0.35]{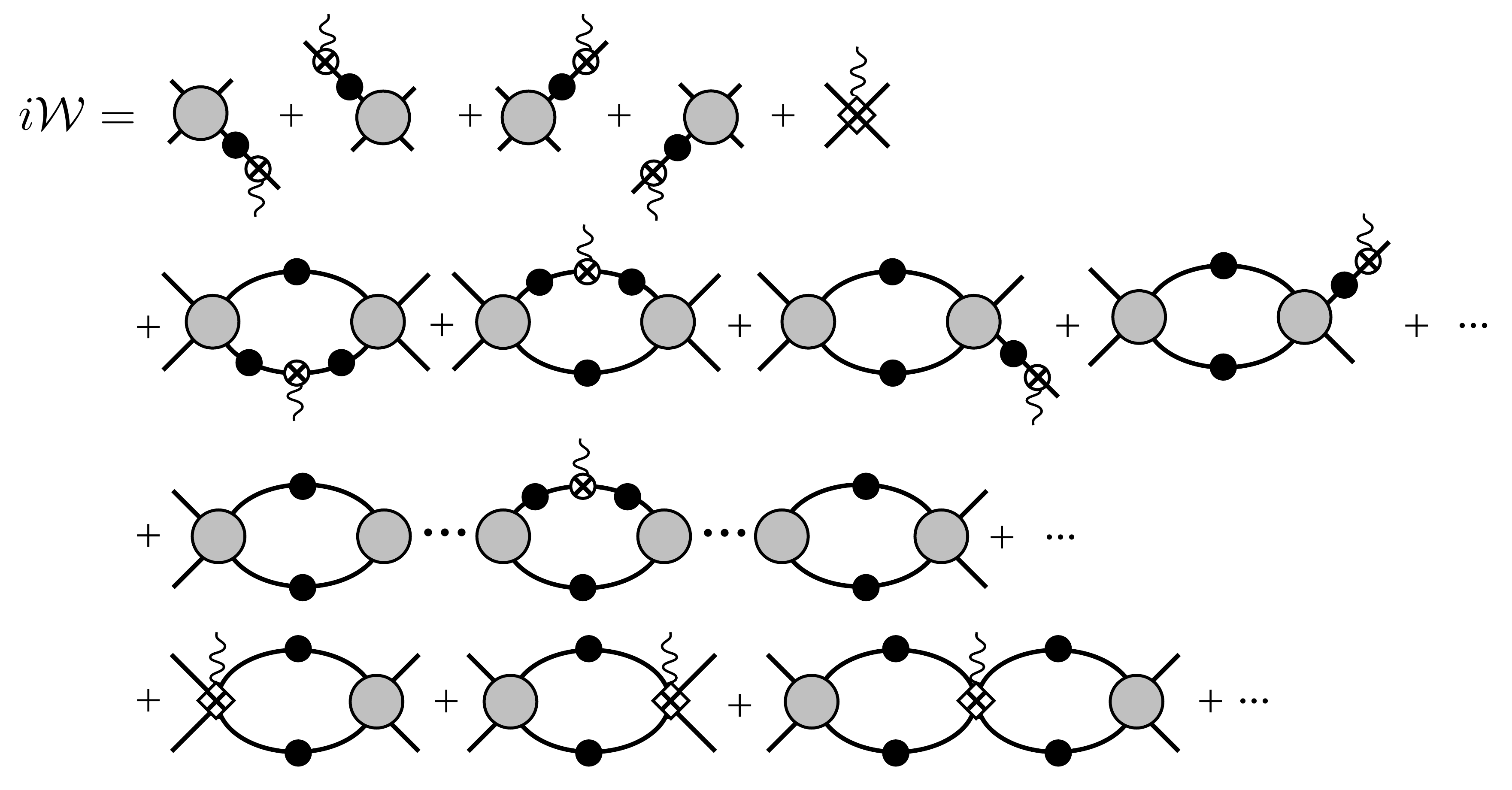}}\\
\subfigure[]{
\label{fig:1B-current}
\includegraphics[scale=0.30]{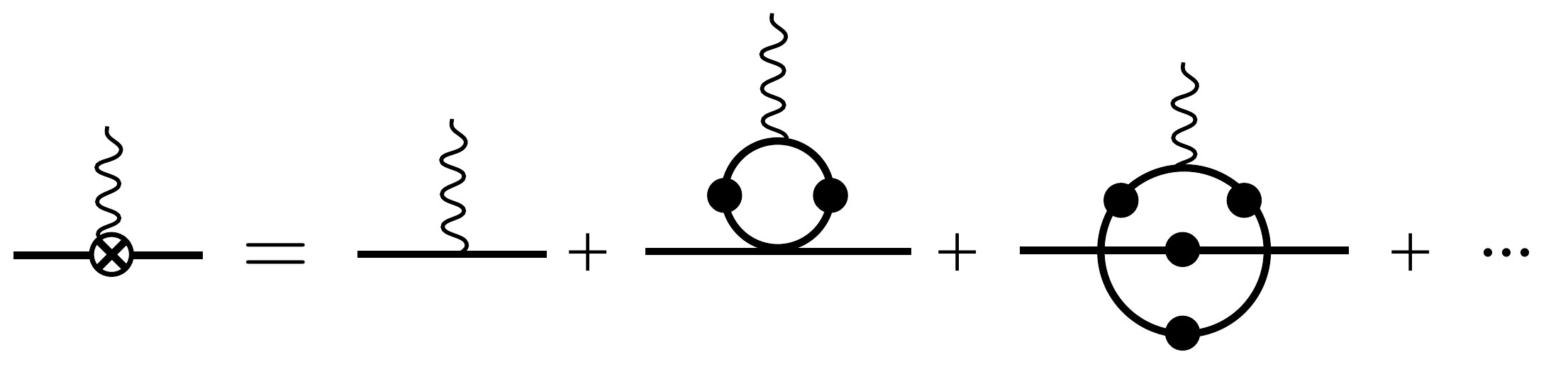}}
\subfigure[]{
\label{fig:2B-current}
\includegraphics[scale=0.30]{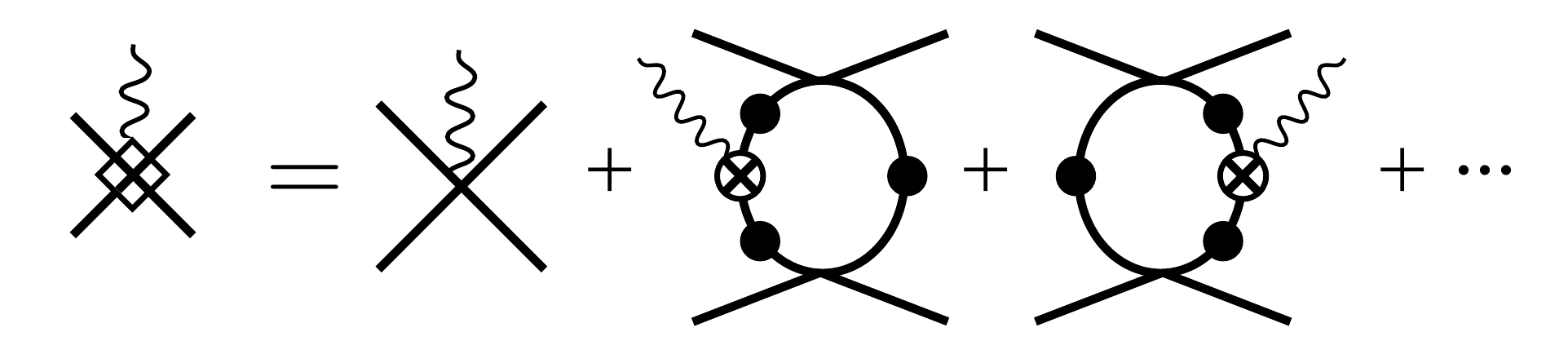}}
\caption{(a) The $\textbf{2}+\mathcal{J}\rightarrow \textbf{2}$ transition amplitude is defined as the sum of all $\textbf{2}+\mathcal{J}\rightarrow \textbf{2}$ amputated diagrams and can be written in terms of the (b) $\textbf{1}+\mathcal{J}\rightarrow \textbf{1}$ transition amplitude, (c) the weak kernel, and the QCD kernels and fully-dressed propagators defined in Fig.~\ref{fig:scat_amp}.}\label{fig:W_iepsilon}
\end{center}
\end{figure*}

\section{Infinite volume $\textbf{2} + \mathcal J \rightarrow \textbf{2}$ amplitudes \label{sec:trans_amps} }

In this work we present the relation between finite-volume matrix elements of two-particle states and infinite-volume $\textbf{2} + \mathcal J \rightarrow \textbf{2}$ transition amplitudes. We derive this relation using a generic, relativistic, scalar quantum field theory. Specifically we analyze the low-energy properties of finite-volume correlators in such a theory by summing a skeleton expansion to all orders in perturbation theory using the techniques developed by L\"uscher~\cite{Luscher:1986pf, Luscher:1990ux} and Kim, Sachrajda, and Sharpe~\cite{Kim:2005gf}. The analysis does not require defining a specific Lagrangian or power-counting scheme and is in this sense very general. We stress that, because we are interested in low-energy correlator properties, we work with fields that correspond to the low-energy degrees of freedom of the theory. For application to QCD, for example, meson and hadron fields, rather than quark fields, should be used. In the present article we only consider (pseudo)scalar particles, so that the applicability within QCD is limited to states composed of QCD-stable (pseudo)scalar mesons.

As we show in Secs.~\ref{sec:two-point} and \ref{sec:3pt_func} below, it turns out to be possible to group all finite-volume effects into known kinematic functions and to express the finite-volume correlator in terms of these functions together with infinite-volume on-shell observables. The finite-volume correlator can also be expressed in a spectral representation, by inserting a complete set of finite-volume states between fields. Equating the diagrammatic and spectral representations gives the relation between finite-volume matrix elements and transition amplitudes that we are after.

The infinite-volume quantities that emerge in our derivation are the on-shell $\textbf{2}\rightarrow \textbf{2}$ scattering amplitude, $\mathcal M$, the on-shell $\textbf{1} + \mathcal J \rightarrow \textbf{1}$ transition amplitude, $\w$, and the on-shell, {divergence-free} $\textbf{2} + \mathcal J \rightarrow \textbf{2}$ transition amplitude, $\Wdf$. We now explain each of these in some detail.

The scattering amplitude, $\mathcal M$, is a standard infinite-volume observable, which can be decomposed into definite angular-momentum contributions. For a system with $N$ open two-particle channels, each angular-momentum component can be expressed in terms of $N(N+1)/2$ scattering phase shifts and mixing angles. The scattering amplitude appears both in the quantization condition for the finite-volume energy spectrum~\cite{Luscher:1986pf, Luscher:1990ux, Rummukainen:1995vs, Kim:2005gf, Christ:2005gi, Briceno:2012yi, Hansen:2012tf, Briceno:2014oea} and in the relation between finite-volume matrix elements and infinite-volume transition amplitudes. This has already been demonstrated in studies of $\textbf{1} + \mathcal J \rightarrow \textbf{2}$~\cite{Lellouch:2000pv, Kim:2005gf, Christ:2005gi, Hansen:2012tf, Agadjanov:2014kha, Meyer:2012wk, Bernard:2012bi, Briceno:2012yi, Briceno:2014uqa, Briceno:2015csa}
and $\textbf{0} + \mathcal J \rightarrow \textbf{2}$~\cite{Feng:2014gba, Briceno:2015csa} transition processes. 

In the context of our field-theoretic analysis, $\mathcal M$ arises as the sum of all infinite-volume, amputated $\textbf{2}\rightarrow\textbf{2}$ Feynman diagrams, evaluated on-shell. This infinite series is organized in a skeleton expansion built from Bethe-Salpeter kernels connected by pairs of fully-dressed propagators [see Fig.~\ref{fig:scat_amp}]. The Bethe-Salpeter kernels are defined as the sum of all amputated four-point diagrams, which are two-particle irreducible in the $s$-channel ($s$-channel 2PI) [see Fig.~\ref{fig:scat_amp}(b)]. Here $s = -P^2$ refers to the Mandelstam variable. In other words the kernels are two-particle irreducible with respect to propagator pairs carrying the total energy-momentum. Alternatively, the kernels are defined by Fig.~\ref{fig:scat_amp}(a) directly. Given that the scattering amplitude on the left-hand side equals the sum of all four-point diagrams, one can infer which diagrammatic pieces must be included in the kernels. Note that it is only possible to accommodate all topologies by also using fully-dressed propagators [see Fig.~\ref{fig:scat_amp} (c)]. The motivation for this expansion is to explicitly display all intermediate states which can go on-shell, given the restriction that the total energy lies below the lowest three- or four-particle threshold. In the analysis of the finite-volume correlator, all power-law finite-volume effects are due to such on-shell intermediate states.

We now turn to the $\textbf{1} + \mathcal J \rightarrow \textbf{1}$ transition amplitude, which we denote $\w$. This is given by an infinite-volume matrix element of an external local current, $\mathcal J$, between one-particle states
\begin{equation}
\label{eq:littlewmedef}
w_{a2,b2}(P_f-k;P_i-k) \equiv \langle P_f-k, a2 \vert \mathcal J(0) \vert P_i - k, b2 \rangle \,,
\end{equation}
where $ \langle P_f-k, a2 \vert$ and $\vert P_i - k, b2 \rangle$ are infinite-volume single particle states with the first entry indicating the on-shell four-momentum and the second indicating particle flavor. These are assumed to have standard relativistic normalization
\begin{equation}
\langle P_f-k, a2  \vert P_i - k, a2 \rangle = 2 \omega_{a2f} (2 \pi)^3 \delta^3(\textbf P_f- \textbf P_i) \,,
\end{equation}
where $ \omega_{a2f} = \sqrt{(\textbf P_f - \textbf k)^2 + m_{a2}^2}$ is an example of notation used extensively below. The $\textbf{1} + \mathcal J \rightarrow \textbf{1}$ transition amplitude can also be defined as the sum of all diagrams with one incoming and one outgoing scalar, both amputated, together with one insertion of the current [see Fig.~\ref{fig:1B-current}]. In contrast to the $\textbf 2 \rightarrow \textbf 2$ scattering amplitude, this transition amplitude does not contain any on-shell intermediate states for the kinematics that we consider. For this reason the difference between the finite- and infinite-volume versions of the $\textbf{1} + \mathcal J \rightarrow \textbf{1}$ amplitude are exponentially suppressed.

\begin{figure*}[t]
\begin{center}
\subfigure[]{
\label{fig:3Body_long_range}
\includegraphics[scale=0.35]{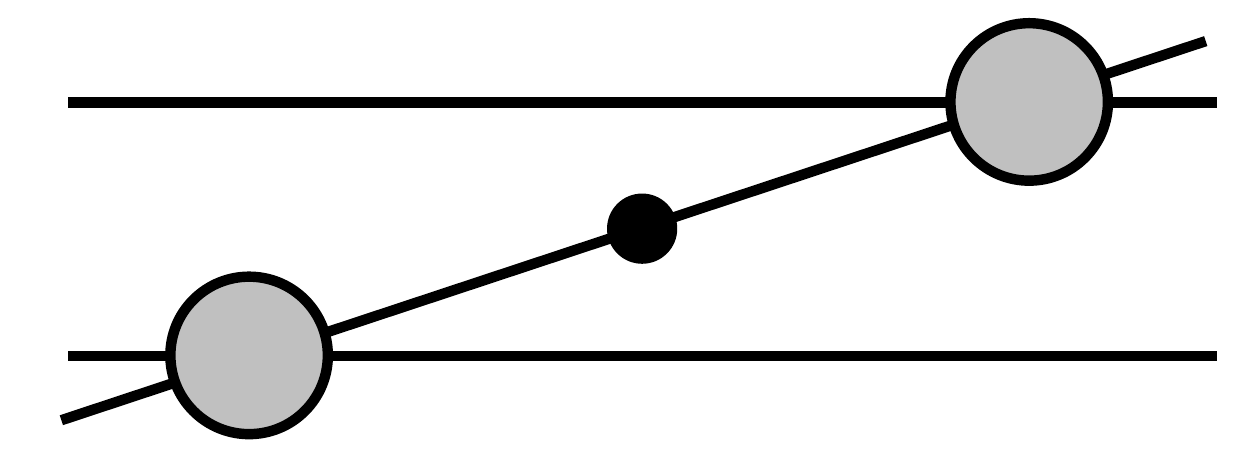}}\hspace{1cm}
\subfigure[]{
\label{fig:2to2_long_range}
\includegraphics[scale=0.40]{2to2_long_range}} 
\caption{Shown are divergent contributions to the (a) $\textbf{3}\rightarrow\textbf{3}$ scattering amplitude and (b) $\textbf{2} + \mathcal J \rightarrow \textbf{2}$ transition amplitudes. Both of these are associated with an intermediate hadron going on-shell, equivalently propagating for an arbitrarily long time. For the $\textbf{2} + \mathcal J \rightarrow \textbf{2}$ transition amplitudes, these divergences are only present if the $\textbf{1} + \mathcal J \rightarrow \textbf{1}$ subprocess is possible.  }\label{fig:long_range}
\end{center}
\end{figure*}
 
The remaining infinite-volume quantities that appear in our formalism are the $\textbf 2 + \mathcal J \rightarrow \textbf 2$ transition amplitude, $\mathcal W$, together with a subtracted, divergence free transition amplitude, $\Wdf$. The former quantity, $\mathcal W$, is a standard infinite-volume observable which may be expressed as a matrix element
\begin{equation}
\mathcal W_{ab}(P_f,p,P_i,k) \equiv \langle P_f, p, a, \mathrm{out} \vert \mathcal J(0) \vert P_i, k, b, \mathrm{in} \rangle \,.
\label{eq:bigWmedef}
\end{equation}
Here we have introduced $\vert P_i, k, b, \mathrm{in} \rangle$ as a two-particle {in}-state with $P_i$ denoting total four-momentum, $k$ the four-momentum of the particle with mass $m_{b1}$ and $b$ denoting particle flavor. Of course both $k$ and $P_i-k$ must be on-shell four-vectors in this asymptotic state. Similar definitions hold for the two-particle {out}-state. As with the single particle states, these are assumed to have standard relativistic normalization. $\mathcal W$ can also be expressed, in direct analogy to the scattering amplitude, as the sum of all infinite-volume, on-shell, amputated $\textbf{2}\rightarrow\textbf{2}$ Feynman diagrams with a single insertion of the external current included at all possible locations [see Fig.~\ref{fig:W_iepsilon}]. As compared to $\mathcal M$, the skeleton expansion for $\mathcal W$ includes two new functions in addition to the Bethe-Salpeter kernel. 

The first of these is the $\textbf{1} + \mathcal J \rightarrow\textbf{1}$ transition amplitude discussed above [see Fig.~\ref{fig:1B-current}]. When used in the skeleton expansion for $\mathcal W$ this quantity must be extended to off-shell four-momenta. The second new function in the expansion for $\mathcal W$ is an extension of the Bethe-Salpeter kernel, defined as the sum of all $\textbf 2 \rightarrow \textbf 2$, $s$-channel-2PI diagrams with an insertion of the external current [see Fig.~\ref{fig:2B-current}]. We will refer to the latter as the {weak Bethe-Salpeter kernel}. In EFTs it is common to replace these kernels with a finite number of low-energy coefficients that are expected to reproduce the dominant effects of the interactions. The EFT insertions are typically referred to as one- and two-body currents. In this work, we make no approximation on the functional form of these building blocks. Instead we take them to be general functions, assuming only that they are smooth and slowly varying.

Although the scattering amplitude only has poles when the energy of the particles coincides with a bound state, the transition amplitude has other kinematic singularities. This is reminiscent of the $\textbf{3}\rightarrow \textbf{3}$ scattering amplitude as discussed in earlier work by one of us~\cite{Hansen:2014eka, Hansen:2015zga}. For both the $\textbf{2} + \mathcal J \rightarrow \textbf{2}$ and $\textbf{3}\rightarrow \textbf{3}$ systems, the physical, infinite-volume scattering observable is known to diverge at certain kinematics due to arbitrarily long lived intermediate states. For three-to-three scattering the divergence arises from a diagram with two pairwise scatterings and a single internal propagator, see Fig.~\ref{fig:3Body_long_range}. If the external kinematics are chosen to put the intermediate propagator on shell then the amplitude diverges. Similarly, in the case of two-to-two scattering with an external current, the two-to-two amplitude diverges due to diagrams where the current is attached to an external leg. The divergence occurs when the external momenta are tuned such that the internal propagator, attached to the current, goes on-shell, see Fig.~\ref{fig:2to2_long_range}.

Also common between the $\textbf{2} + \mathcal J \rightarrow \textbf{2}$ and $\textbf{3}\rightarrow \textbf{3}$ systems is that, in each case, the observable of interest includes physically observable subprocesses. In the case of $\textbf{3}\rightarrow \textbf{3}$ scattering this is the $\textbf{2}\rightarrow \textbf{2}$ amplitude, and in the case of $\textbf{2} + \mathcal J \rightarrow \textbf{2}$ it is the $\textbf{1} + \mathcal J \rightarrow \textbf{1}$ subprocesses, as well as the $\textbf{2}\rightarrow \textbf{2}$ amplitude. These subprocesses completely dictate the form of the divergences exhibited in Fig.~\ref{fig:long_range}. Thus, by constraining them separately, one can determine a subtraction which renders the observable of interest finite. Indeed, it turns out that the finite-volume spectrum directly depends on these finite functions, in which the long range divergences have been subtracted off. In the case of three-to-three scattering the subtracted quantity introduced in Ref.~\cite{Hansen:2014eka} is denoted $\mathcal K_{\mathrm{df},3}$ and in the present work we denote the subtracted $\textbf{2} + \mathcal J \rightarrow \textbf{2}$ amplitude by $\Wdf$. We stress that, since the modifications contain only known subprocesses with on-shell kinematics, once the infinite-volume, divergence-free quantity is determined, one can add back in the long-distance piece to obtain the full, model-independent result. 

\begin{figure*}[t]
\begin{center} 
\includegraphics[scale=0.5]{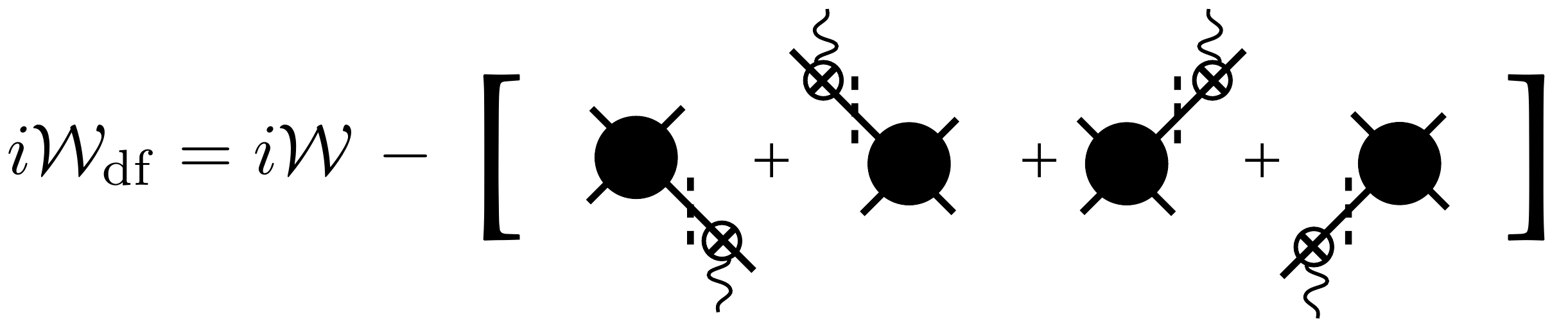}
\caption{Shown is the diagrammatic definition of the divergence free transition amplitude, $\Wdf$. This is written in terms of the full transition amplitude, $\W$ [defined in Fig.~\ref{fig:W_ie}], the $\textbf{1} + \mathcal J \rightarrow \textbf{1}$ amplitude [defined in Fig.~\ref{fig:1B-current}] and the scattering amplitude [defined in Fig.~\ref{fig:scat_ampa} and depicted here by the back circles]. The dashed cuts indicate that a simple pole is used in place of the propagator and that adjacent quantities are evaluated on-shell.}
\label{fig:divergence_free}
\end{center}
\end{figure*}

In Fig.~\ref{fig:divergence_free} we give the diagrammatic definition of $\Wdf$ and the explicit form is given in Eq.~(\ref{eq:Wdfdef}) of Sec.~\ref{sec:3pt_func} below. This turns out to be much more straightforward than the definition of $\mathcal K_{\mathrm{df},3}$. For $\mathcal W$, the only divergences that arise are those due to the tree-level graph of Fig.~\ref{fig:2to2_long_range}. Thus the subtraction needed to convert $\mathcal W$ to $\Wdf$ is a simple product of on-shell scattering amplitude $\mathcal M$, the $\textbf{1} + \mathcal J \rightarrow \textbf{1}$ transition amplitude, $\w$, and a simple pole. By contrast, the definition of $\mathcal K_{\mathrm{df},3}$ involves an integral equation, associated with the need to remove a more complicated singularity structure in the three-particle analysis.

In the following sections we analyze the finite-volume correlator to show how it can be written in terms of $\mathcal M$, $\w$, $\Wdf$ as well as two types of finite-volume functions. We postpone the detailed derivation of this to Sec.~\ref{sec:3pt_func}. To arrive at the final result, we must first understand how to evaluate the momentum sums that arise in the finite-volume correlators. This is done in Sec.~\ref{sec:finite_volume_func}. In Sec.~\ref{sec:FFunction} we review the necessary steps for evaluating the standard finite-volume two-particle loops already studied in Refs.~\cite{Kim:2005gf}. In Sec.~\ref{sec:1body_insertion} we evaluate the new type of loop which arises from the nonzero values of the $\textbf{1}+\mathcal{J}\rightarrow \textbf{1}$ amplitudes. We arrive at two identities, Eqs.~(\ref{eq:Fsum_final}) and (\ref{eq:GL_final}), which are then applied to reduce the finite-volume correlators.

\section{Loop functions in finite volume \label{sec:finite_volume_func}}
 
The main result of this work, Eq.~(\ref{eq:2to2_notdegen_intro}), follows directly from our analysis of two- and three-point correlation functions defined in a finite, cubic, spatial volume with periodic boundary conditions. In this section we derive the necessary tools to rewrite such correlation functions in a useful form. The finite-volume three-point function closely resembles the infinite-volume transition amplitude, Fig.~\ref{fig:W_iepsilon}. One can arrive at the finite-volume correlator from the transition amplitude by evaluating all loops in a finite volume (summing rather than integrating loop momenta) and attaching interpolating operators to the external legs. A diagrammatic representation of the three-point function is given in Fig.~\ref{fig:C2-to-2} below. Examining Fig.~\ref{fig:W_iepsilon} (or Fig.~\ref{fig:C2-to-2} below) makes clear that we must evaluate two classes of finite-volume loops, those with and without the $\textbf{1}+\mathcal{J}\rightarrow \textbf{1}$ subprocess. 

Defining $L$ to be the linear extent of the spatial volume, we recall that the periodic boundary conditions constrain the momenta of individual particles to be discretized, satisfying $\textbf{p}=2\pi\textbf{n}/L$, where $\textbf{n}\in \mathbb Z^3$. It is for this reason that spatial loop momenta are summed rather than integrated. The time-components of all momenta continue to be integrated since we take the coordinate time direction to have infinite extent.  In this section we are interested in evaluating the difference between finite-volume (summed) and infinite-volume (integrated) two-particle loops. We will see that the summands arising from such loops result in power law, $1/L^n$, corrections to 
\begin{equation}
 \left[ \frac{1}{L^3}\sum_{\mathbf{k}}\hspace{-.5cm}\int~\right]f(\textbf{k})\equiv\bigg [ \frac{1}{L^3} \sum_{\textbf k \in (2 \pi/L) \mathbb Z^3} - \int \frac{d\textbf k}{(2 \pi)^3} \bigg ] f(\textbf{k})\,.
\end{equation}
Generally speaking, if the function $f(\textbf k)$ is smooth (infinitely differentiable), one can show that this difference vanishes for large $L$ faster than any power of $L^{-n}$. As discussed extensively in the literature, this has an interesting physical consequence: power-law finite-volume corrections appear only in diagrams where the intermediate particles can go on-shell. The number of particles that can simultaneously go on-shell depends on the energy of the system as well as the masses of the asymptotic degrees of freedom. In this work, we restrict our attention to energies where only two-particle states can go on-shell. Consequently, $\mathcal{O}(L^{-n})$ corrections emerge only from two-particle intermediate states. In the context of QCD, the neglected exponentially suppressed corrections take the form $\mathcal{O}(e^{-m_\pi L})$, where $m_\pi$ is the pion mass. Thus the formalism derived here can only be applied to systems satisfying $m_\pi L \gg 1$.

\begin{figure*}[t]
\begin{center}
\label{fig:FVcorr}
\includegraphics[scale=0.4]{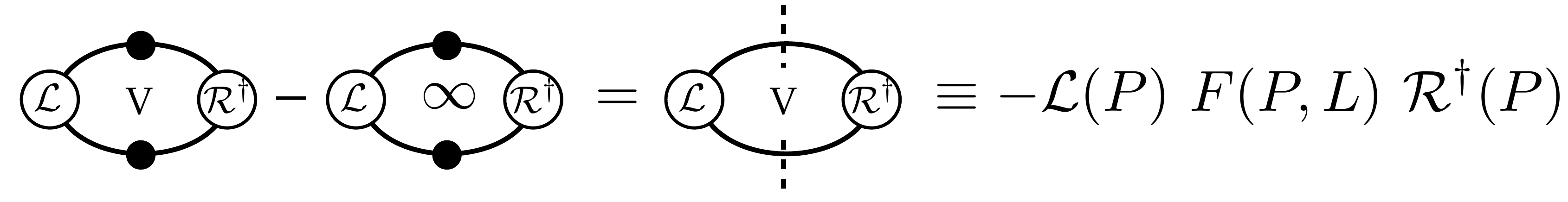}
\caption{ As discussed in the text, the difference between the finite- and infinite-volume two-particle loops 
can be written using the finite volume matrix $F(P,L)$, Eq.~(\ref{eq:Fscdef}), left- and right-multiplied by the on-shell endcaps $\mathcal L$ and $\mathcal R^\dagger$. 
}\label{fig:F_function}
\end{center}
\end{figure*}

As already mentioned above, in the analysis of finite-volume two- and three-point correlators there are two classes of subdiagrams that give rise to power-law corrections. The first correspond to standard two-particle $s$-channel loops [see Fig.~\ref{fig:F_function}]. This was first studied in Refs.~\cite{Luscher:1986pf, Luscher:1990ux, Rummukainen:1995vs, Kim:2005gf, Christ:2005gi} and we review the result in Sec.~\ref{sec:FFunction}. We stress that the finite-volume loops adjacent to the weak Bethe-Salpeter kernel [defined in Fig.~\ref{fig:2B-current}] are also accommodated using the more standard two-particle loops. 

The second class of subdiagrams is specific to three-point correlators for systems with $\textbf{1}+\mathcal{J}\rightarrow \textbf{1}$ subprocesses.
The presence of $\textbf{1}+\mathcal{J}\rightarrow \textbf{1}$ subprocesses in the intermediate loops and the resulting new class of power-law corrections is the central complication addressed in this work. These effects were first pointed out in Refs.~\cite{Briceno:2012yi, Bernard:2012bi}. Unlike in those references, in Sec.~\ref{sec:1body_insertion} we find a parametrization-independent expression for such finite-volume diagrams. Further, our formalism accommodates any number of two-scalar channels with identical or non-identical particles, which, in the latter case, can have either degenerate or non-degenerate masses.

\begin{figure*}[t]
\begin{center} 
\includegraphics[scale=0.4]{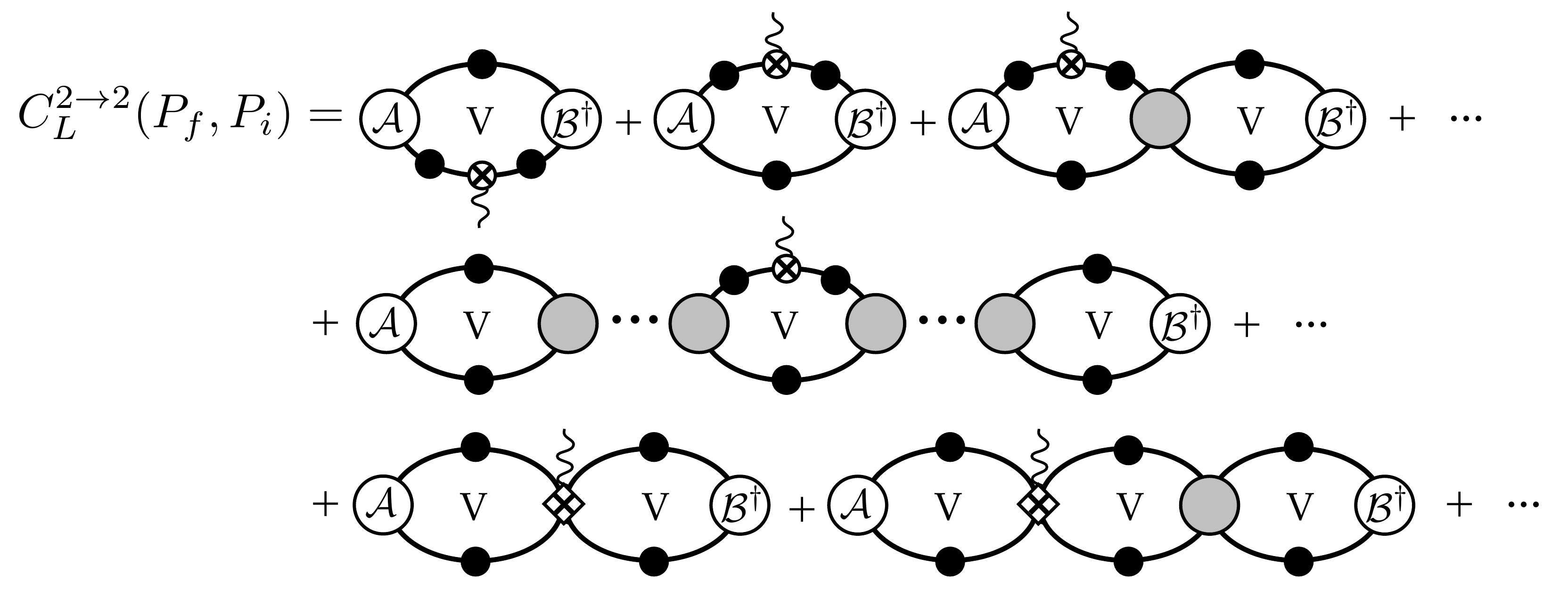}
\caption{Shown is the full two-to-two three-point function. The ``V" labels explicitly depict that the loops must be evaluated in a finite volume.  The one-particle propagators and Bethe-Salpeter kernel are defined in Fig.~\ref{fig:scat_amp}. The $\textbf{1}+\mathcal{J}\rightarrow \textbf{1}$ and weak kernels are defined in Fig.~\ref{fig:W_iepsilon}. The overlap factors with the source and sink, $\mathcal{B}^\dag$ and $\mathcal{A}$ respectively, will be defined in Sec.~\ref{sec:two-point}. }\label{fig:C2-to-2}
\end{center}
\end{figure*}
\subsection{Loop function without $\textbf{1}+\mathcal{J}\rightarrow \textbf{1}$ contributions \label{sec:FFunction}}

In this subsection we consider the standard $s$-channel two-particle loop with no $\textbf{1}+\mathcal{J}\rightarrow \textbf{1}$ subprocesses. With the exception of minor notational differences, this closely follows the derivation presented in Ref.~\cite{Kim:2005gf} and also discussed in our previous works~\cite{Briceno:2014uqa, Briceno:2015csa}. We are interested here in the difference between finite- and infinite-volume expressions, which we refer to throughout as the finite-volume residue.  We work with the Euclidean metric, $p^2=p_4^2+\textbf{p}^2$. With this convention the free scalar propagator is given by
\begin{equation}
\Delta_{i,{\rm free}}(p) \equiv \frac{1}{p^2 + m_i^2}.
\end{equation}
We label the fully dressed propagator as $\Delta(p)$, with the ``free'' subscript removed 
\begin{equation}
\label{eq:propdef}
\Delta_i(k) \equiv \int \! d^4 x  e^{-i k  x} \langle 0 \vert T \Phi_i(x) \Phi_i^\dag(0) \vert 0 \rangle \,,
\end{equation}
where $\Phi$ is the single particle interpolating field. We choose $\Phi$ with unit wave-function renormalization so that $\Delta$ and $\Delta_{\mathrm{free}}$ coincide at the pole. For the energies of interest, the difference between the finite- and infinite-volume propagators is exponentially suppressed, and we thus use the infinite-volume propagator throughout. To accommodate any number of two-particle channels, we introduce a channel label, $a$. Quantities that depends on the channel, will receive a subscript $a$. For single-particle quantities we must specify the particle in the given channel. We do so with the labels $a1$ and $a2$. For example, the $a1$ propagator will be defined as $\Delta_{a1}(k) $. 

We now proceed to analyze the general sum-integral difference 
\begin{align}
\label{eq:Fsum}
\mathcal F_L & = \sum_{a=1}^{N_c} \xi_a \left[\frac{1}{L^3}\sum_{\mathbf{k}}\hspace{-.5cm}\int~\right] \int \frac{d k_4}{2 \pi} \ \mathcal L_a(P, k) \Delta_{a1}(k) \Delta_{a2}(P-k)   \mathcal R_a^\dagger(P, k) \,,
\end{align}
where $\xi_a$ is the symmetry factor of the $a$th channel, equal to $1/2$ if the particles are identical and 1 otherwise. $\mathcal L_a(P, k)$ and $ \mathcal R_a^\dagger(P, k)$ are generic functions which we require to be smooth for total energy below the lowest lying three- or four-particle threshold. In the following section the Bethe-Salpeter kernel and weak kernels will appear in place of these functions. Since the endcap functions are smooth, we find that $\mathcal{O}(L^{-n})$ corrections arise only from the singularity of the single particle propagators. 

To identify these power-law contributions, we first perform the integral over $k_4$. We do this by closing the contour in the upper-half of the complex $k_4$ plane. The closed contour encircles a single particle pole at $k_4= i \omega_{a1}$, where $\omega_{a1}=\sqrt{m_{a1}^2+ \! {\color{white} (}{\textbf k}{\color{white})}\!\!^2}$, as well as an infinite tower of branch cuts associated with multi-particle states. However, as is demonstrated in Refs.~\cite{Luscher:1986pf, Luscher:1990ux}, the contributions from the latter are smooth functions of $\textbf k$ and thus result in exponentially suppressed corrections when one acts with the sum-integral difference. This leaves us with the sum-integral difference on the single-particle pole
\begin{align}
\mathcal F_L 
& =   \sum_{a=1}^{N_c} \xi_a  \left[\frac{1}{L^3}\sum_{\mathbf{k}}\hspace{-.5cm}\int~\right] \left. \mathcal L_a(P-k,  k)  
\frac{\Delta_{a2}(P-k) }{2 \omega_{a1}}    \mathcal R_a^\dagger(P, k)\right|_{k_4=i\omega_{a1}} \,.
\end{align}

\noindent 

\noindent Next we use the fact that $\Delta_{a2}(P-k)$ evaluated at $k_4 = i \omega_{a1}$ has a single-particle pole of the form\\ $-[2 \omega_{a2}(E - \omega_{a1} - \omega_{a2} + i \epsilon)]^{-1}$ where $\omega_{a2}= \sqrt{(\textbf P - \textbf k)^2 + m_{a2}^2}$ and where we have introduced the physical total energy in the moving frame $E = - i P_4$. Indeed the difference between $\Delta_{a2}$ and this single particle pole is a smooth function which results in an exponentially suppressed contribution to $\mathcal F_L$. We reach
\begin{equation}
\label{eq:Fonepartpole}
\mathcal F_L  =  -  \sum_{a=1}^{N_c} \xi_a  \left[\frac{1}{L^3}\sum_{\mathbf{k}}\hspace{-.5cm}\int~\right] \ \mathcal L_a(P - k,k)  
\frac{1}{2 \omega_{a1} 2 \omega_{a2}(E -   \omega_{a1} - \omega_{a2} + i \epsilon)}    \mathcal R_a^\dagger(P-k,k )  \bigg \vert_{k_4=i\omega_{a1}} \,.
\end{equation}

The final step in reducing $\mathcal F_L$ is to replace $\mathcal L_a(P - k,k)$ and $\mathcal R_a^\dagger(P-k,k )$ with projected forms, in which $P-k$ and $k$ are both on-shell four-vectors. This is justified because the difference between on- and off-shell values vanishes with the pole, resulting again in a smooth piece that can be neglected in the sum-integral difference. To define the on-shell projection we first introduce $\textbf k^*_{a}$ as the spatial part of the four-vector $(\omega_{a1}^*, \textbf k^*_a)$ which is reached by boosting $(\omega_{a1}, \textbf k)$ with boost velocity $- \textbf P/E$. In other words, $\textbf k^*_{a}$ is the momentum of particle one in the two-particle CM frame. We use this new coordinate to define new functions
\begin{equation}
\label{eq:coordchange}
\mathcal L_a(P,\textbf k^*_a) \equiv \mathcal L_a(P-k,k )  \bigg \vert_{k_4=i\omega_{a1}}\,, \ \ \ \ \ \mathcal R_a^\dagger(P,\textbf k^*_a) \equiv \mathcal R_a^\dagger(P-k,k )  \bigg \vert_{k_4=i\omega_{a1}} \,.
\end{equation}
The functions only differ in the frame used to define momentum coordinates.  We next note that $P-k$ is on-shell if and only if $\vert \textbf k^*_a \vert \equiv k^*_a = q^*_a$ where $q^*_a$ is defined via
\begin{equation}
E^* = \sqrt{q_a^{*2} + m_{a1}^2} + \sqrt{q_a^{*2} + m_{a2}^2} \,,
\end{equation}
where we have introduced $E^*$ for the center of mass (CM) frame energy, satisfying $E^{*2}= E^2 - \textbf P^2 = - P_4^2 - \textbf P^2 = - P^2$. Thus, the on-shell projection is effected by replacing $k^*_a \rightarrow q^*_a$ in $\mathcal R_a$ and $\mathcal L_a^\dagger$. The resulting functions depend only on $\khs_a$ and $E^{*}$ and it is convenient to decompose in spherical-harmonics, defining
\begin{align}
\label{eq:onshellcomp}
\mathcal L_a(P,  q^*_a \khs_a)=\sum_{\ell m }\sqrt{4 \pi} Y_{\ell m }(\hat {\textbf k}^*_a)\mathcal L_{a \ell m }(P),
\hspace{2cm}
\mathcal R_a^\dagger(P,  q^*_a \khs_a)
=\sum_{\ell m } \sqrt{4 \pi} Y^*_{\ell m }(\hat {\textbf k}^*_a)\mathcal R^\dagger_{a \ell m}(P) \,.
\end{align}

At this stage we encounter a subtlety with the on-shell projection. As we have already stressed, the difference between the functions $\mathcal L_a$ and $\mathcal R_a^\dagger$ appearing in Eq.~(\ref{eq:Fonepartpole}) and the on-shell projections of Eq.~(\ref{eq:onshellcomp}) vanishes for $E - \omega_{a1} - \omega_{a2} = 0$. As a result no power-law finite-volume effects appear from the one-particle pole in such an on-shell/off-shell difference. However the on-shell functions of Eq.~(\ref{eq:onshellcomp}) do have singularities near $\textbf k^*_a = 0$, due to the unit-vector varying rapidly in this region. These singularities, which are unphysical and were introduced by our projection, generate artificial power-law finite-volume effects if the on-shell functions are directly substituted into Eq.~(\ref{eq:Fonepartpole}). This motivates us to define a modified on-shell projection
\begin{equation}
\mathcal L_{a,\mathrm{on}}(P, \textbf k^*_a ) \equiv \sum_{\ell m }\sqrt{4 \pi} \bigg ( \frac{k^*_a}{q^*_a} \bigg )^{\ell} Y_{\ell m }(\hat {\textbf k}^*_a)\mathcal L_{a \ell m }(P),
\hspace{1cm}
\mathcal R_{a,\mathrm{on}}^\dagger(P, \textbf k^*_a )
\equiv \sum_{\ell m } \sqrt{4 \pi} \bigg ( \frac{k^*_a}{q^*_a} \bigg )^{\ell} Y^*_{\ell m }(\hat {\textbf k}^*_a)\mathcal R^\dagger_{a \ell m}(P) \,.
\label{eq:finalonfunc}
\end{equation}

We have presented a number of closely related definitions involving $\mathcal L_a$ and $\mathcal R_a^\dagger$ and so we think it is helpful to summarize these before giving our final form of $\mathcal F_L$. To avoid repetition, we describe all steps in terms of $\mathcal L_a$ only. Beginning with $\mathcal L_a(P-k,k)$, we first performed the $k_4$ integral and found that only the term with $k_4 = i \omega_{a1}$ gave power-law finite-volume effects. In this way one of the two four-vectors in $\mathcal L_a$ was put on-shell. We next defined a coordinate change to introduce $\mathcal L_a(P, \textbf k^*_a)$ in Eq.~(\ref{eq:coordchange}). This put us in position to define the on-shell partial wave contributions $\mathcal L_{a \ell m}(P)$ in Eq.~(\ref{eq:onshellcomp}). Finally we used these to define $\mathcal L_{a,\mathrm{on}}$ in Eq.~(\ref{eq:finalonfunc}). Only this final quantity has both desired properties of being everywhere smooth and only depending only on on-shell values of $\mathcal L_a$.

Finally we replace $\mathcal L_a(P-k,k) \mathcal R^\dagger_a(P-k,k) \longrightarrow \mathcal L_{a,\mathrm{on}}(P, \textbf k^*_a ) \mathcal R_{a,\mathrm{on}}^\dagger(P, \textbf k^*_a )$ in Eq.~(\ref{eq:Fonepartpole}), and deduce
\begin{align}
\label{eq:Fsum_final}
\mathcal F_L&= - \mathcal L_{a \ell m }(P)  F_{a \ell m ; a' \ell' m'} (P,L)\mathcal R^\dagger_{a '\ell' m'}(P)
\equiv 
- \mathcal L(P)  F (P,L)\mathcal R^\dagger(P) \,,
\end{align}
where the matrix elements of $F(P,L)$ are defined as
\begin{equation}
\label{eq:Fscdef}
F_{a \ell m ; a' \ell' m'}(P,L)  \equiv \delta_{aa'}\xi_a 
\left[\frac{1}{L^3}\sum_{\mathbf{k}}\hspace{-.5cm}\int~\right]
\frac{ 4 \pi  Y_{\ell m }(\hat {\textbf k}^*_a)Y^*_{\ell' m'}(\hat {\textbf k}^*_a)  }{2 \omega_{a1} 2 \omega_{a2}(E -  \omega_{a1} - \omega_{a2} + i \epsilon )} \left (\frac{k^{*}_a}{q^*_a} \right)^{\ell+\ell'} \,.
\end{equation}
In Appendix~\ref{app:Fnum} we give an alternative form of $F$ that is more convenient for numerical evaluation.


\subsection{Loop function with $\textbf{1}+\mathcal{J}\rightarrow \textbf{1}$ contributions \label{sec:1body_insertion}}

In this section we evaluate the finite-volume loop with a $\textbf{1}+\mathcal{J}\rightarrow \textbf{1}$ subprocess. Once again, we are interested in the difference between the finite- and infinite-volume expressions,
\begin{align}
\mathcal G_L & \equiv \sum_{a,b=1}^{N_c}  \left[\frac{1}{L^3}\sum_{\mathbf{k}}\hspace{-.5cm}\int~\right] \int \frac{d k_4}{2 \pi} \ \mathcal L_a(P_f, k) \Delta_{a1}(k) \left[\Delta_{a2}(P_f-k)  w_{a2,b2}(P_f-k;P_i-k) \Delta_{b2}(P_i-k) \right]\mathcal R_b^\dagger(P_i, k) + (1 \leftrightarrow 2),
\label{eq:GL0}
\end{align}
where $w_{a2,b2}(P_f-k;P_i-k)$ will play the role of the $\textbf{1}+\mathcal{J}\rightarrow \textbf{1}$ contributions in the finite-volume correlator analysis of the next section. We explain the $(1 \leftrightarrow 2)$ contribution in the paragraph after next. Note here that, since the external current can insert momentum, the incoming and outgoing two-particle states may have different momenta, which we label $P_i$ and $P_f$.  

Before starting the analysis of $\mathcal G_L$, we comment here on how the expression given above can be used to efficiently handle our general set-up with identical or non-identical scalars, possible non-degeneracy in the latter case, and also with any number of open two-scalar channels. Observe that we have included two channel indices, $a$ and $b$, to label separately the two-particle pairs appearing before and after the current. Of course the first particle, labeled $a1$, is not attached to the current and therefore cannot change. We will see below that it is convenient to nonetheless think in terms of two two-particle channels, and to identify $a1=b1$ so that labels can be exchanged to simplify expressions. Further, we require that the set of open channels used here is identical to that used for the simple loops in the previous subsection. This requires extending $\w$ by defining $w_{a2,b2}=0$ for all channels $a$ and $b$ which do not contain a common particle (or which contain particles that simply do not couple to the current).  Similarly we may need to include zeroes in the channel-space matrices for the Bethe-Salpeter kernel, to accommodate channels that only couple with the weak current. In short, always using the same (maximal) channel space and setting kernels to zero where necessary greatly simplifies the expressions that appear.

Along these same lines we note that not all possible cases can be accommodated using only $\textbf{1}+\mathcal{J}\rightarrow \textbf{1}$ transitions that couple to the particles labeled $a2$ and $b2$. For example, suppose that a given pair of channels $a$ and $b$ have exactly one particle in common, and therefore only admit a single such transition. Then we are free to label the non-identical particles $a2$ and $b2$. However, when these channels are coupled to a third channel, $c$, then transitions such as $w_{a1,c1}$, $w_{b2,c1}$ can arise. In addition, even in a two-channel system, if the particles are non-identical but the two channels are, then separate $w_{a1,b1}$ and $w_{a2,b2}$ transitions can arise. The most straightforward way to accommodate all possible scenarios is to include all four $\textbf{1}+\mathcal{J}\rightarrow \textbf{1}$ transitions $w_{a1,b1}$, $w_{a1,b2}$, $w_{a2,b1}$, and $w_{a2,b2}$ and define these to vanish as required. One subtlety with this approach is that redundant, identical contributions arise in channels with identical particles. These can be easily removed with symmetry factors, as we show below. In the following we first restrict attention to channels with a single $w_{a2,b2}$ coupling. We then show how the remaining terms can be easily included in our final result, Eq.~(\ref{eq:GL_final}) below.

As in the previous subsection, we first perform the $k_4$ integral and discard the smooth contributions to reach
\begin{align}
\mathcal G_L & =  \sum_{a,b=1}^{N_c}  \left[\frac{1}{L^3}\sum_{\mathbf{k}}\hspace{-.5cm}\int~\right] \frac{1}{2\omega_{a1}} \ \mathcal L_a(P_f, k)  \left[\Delta_{a2}(P_f-k)  w_{a2,b2}(P_f-k;P_i-k) \Delta_{b2}(P_i-k) \right]\mathcal R_{b}^\dagger(P_i, k) \bigg \vert_{k_4=i\omega_{a1}} .
\label{eq:GL1}
\end{align}
In order to reduce the remaining expression, we once again use the fact that the poles of the integrand give rise to all power-law scaling in the sum-integral difference. Unlike Eq.~(\ref{eq:Fsum}), this sum has two poles due to the two remaining propagators and for this reason it is more difficult to identify how all power-law contributions depend only on on-shell quantities.

To demonstrate this on-shell dependence nonetheless, we first define on-shell projections of $\w_{a2,b2}(P_f-k;P_i-k)$. This proceeds exactly as in the previous subsection, by first defining a new coordinate system for the $\textbf{1}+\mathcal{J}\rightarrow \textbf{1}$ amplitude. In contrast to above, however, here we have two frames to choose from. We thus define both $(\omega^*_{a1f}, \textbf k_{af}^*)$ and $(\omega^*_{b1i}, \textbf k_{bi}^*)$ by boosting $(\omega_{a1}, \textbf k)$ by $- \textbf P_f/E_f$ and $- \textbf P_i/E_i$ respectively. This allows us to introduce
\begin{equation}
\w_{a2,b2}(P_f,  \textbf k_{af}^*;P_i, \textbf k_{bi}^*) \equiv \w_{a2,b2}(P_f-k;P_i-k) \bigg \vert_{k_4 = i \omega_{a1}} \,.
\end{equation}

Here we have treated the $\textbf k$ dependence in $P_f-k$ differently from that in $P_i-k$ as this will be convenient in the following steps. Continuing as above, we now define on-shell spherical-harmonic components
\begin{align}
\w_{a2,b2 }(P_f,  q^*_{af} \khs_{af};P_i, q^*_{bi} \khs_{bi}) & \equiv 4 \pi \sum_{\ell',m',\ell,m}    Y^*_{\ell' m'}(\khs_{af}) \ \w_{a2,b2;\ell' m'; \ell m}(P_f,P_i) \   Y_{\ell m}(\khs_{bi})    \,, \\
\w_{a2,b2 }(P_f,  \textbf k_{af}^*;P_i, q^*_{bi} \khs_{bi}) & \equiv \sqrt{4 \pi} \sum_{\ell,m}   \w_{a2,b2; \mathrm{off}; \ell m}(P_f,\textbf k^*_{af};P_i) \   Y_{\ell m}(\khs_{bi})   \,, \label{eq:wshcomp2}\\
\w_{a2,b2 }(P_f, q^*_{af} \khs_{af};P_i, \textbf k_{bi}^*) & \equiv \sqrt{4 \pi} \sum_{\ell',m' }   Y^*_{\ell' m'}(\khs_{af}) \ \w_{a2,b2;\ell' m' ;\mathrm{off}}(P_f;P_i,\textbf k^*_{bi}) \label{eq:wshcomp3} \,.
\end{align}
Here we have introduced $q_{bi}^{*}$ and $q_{af}^{*}$, defined via
\begin{align}
\label{eq:qstardef}
E^*_i = \sqrt{q_{bi}^{*2} + m_{b1}^2} + \sqrt{q_{bi}^{*2} + m_{b2}^2} \,, \\
E^*_f = \sqrt{q_{af}^{*2} + m_{a1}^2} + \sqrt{q_{af}^{*2} + m_{a2}^2} \,.
\end{align}
In Eq.~(\ref{eq:wshcomp2}) the subscript ``off'' indicates that the final state is off-shell, whereas in Eq.~(\ref{eq:wshcomp3}) it refers to the initial state. All remaining coordinates are on-shell. We comment that these definitions are very similar to those of Eq.~(\ref{eq:onshellcomp}) above. The main difference is that we now have two sets of coordinates and have included the possibility that one set is off shell while the other is on shell and decomposed in harmonics. We are now ready to give the various on-shell projections which are also smooth near $\textbf k^*_{bi}, \textbf k^*_{af} = 0$
\begin{align}
\label{eq:wondef1}
\w_{a2,b2,\mathrm{on,on}} & = 4 \pi \sum_{\ell',m',\ell,m}    \bigg(\frac{k^*_{af}}{q^*_{af}} \bigg )^{\ell'} Y^*_{\ell' m'}(\khs_{af}) \ \w_{a2,b2;\ell' m'; \ell m}(P_f,P_i) \   Y_{\ell m}(\khs_{bi})   \bigg(\frac{k^*_{bi}}{q^*_{bi}} \bigg )^{\ell} \,, \\
\w_{a2,b2,\mathrm{off,on}} & = \sqrt{4 \pi} \sum_{\ell,m}   \w_{a2,b2; \mathrm{off}; \ell m}(P_f,\textbf k^*_{af};P_i) \   Y_{\ell m}(\khs_{bi})   \left(\frac{k^*_{bi}}{q^*_{bi}} \right )^{\ell} \,, \\
\w_{a2,b2,\mathrm{on,off}} & = \sqrt{4 \pi} \sum_{\ell',m' }    \bigg(\frac{k^*_{af}}{q^*_{af}} \bigg )^{\ell'} Y^*_{\ell' m'}(\khs_{af}) \ \w_{a2,b2;\ell' m' ;\mathrm{off}}(P_f;P_i,\textbf k^*_{bi}) \,.
\label{eq:wondef3}
\end{align}
Here we have included a pair of subscripts drawn from ``on'' and ``off'' on each quantity, indicating whether the incoming and outgoing coordinates are on- or off-shell.

Unlike in the previous subsection, we cannot replace $\w_{a2,b2}$ in Eq.~(\ref{eq:GL1}) with any of these quantities directly. The problem is the double pole structure. Here we explain in detail how to circumvent this challenge. We first rewrite the partially off-shell $\w$ as 
\begin{equation}
\w_{a2,b2}(P_f-k;P_i-k)\bigg \vert_{k_4=i\omega_{a1}} = \w_{a2,b2,\mathrm{on,on}} + [\delta \w]_{a2,b2,\mathrm{off,on}}  +  [\w\delta]_{a2,b2,\mathrm{on,off}}+ [\delta \w \delta]_{a2,b2,\mathrm{off,off}} \,,
\end{equation}
where
\begin{align}
[\delta \w]_{a2,b2,\mathrm{off,on}} & = \w_{a2,b2,\mathrm{off,on}}-\w_{a2,b2,\mathrm{on,on}}  \,, \\
[\w\delta]_{a2,b2,\mathrm{on,off}} & = \w_{a2,b2,\mathrm{on,off}}-\w_{a2,b2,\mathrm{on,on}}  \,, \\
\begin{split}
[\delta \w \delta]_{a2,b2,\mathrm{off,off}} & = \w_{a2,b2}(P_f-k;P_i-k)\bigg \vert_{k_4=i\omega_{a1}} + \w_{a2,b2,\mathrm{on,on}}   - \w_{a2,b2,\mathrm{off,on}} - \w_{a2,b2,\mathrm{on,off}} \,.
\end{split} 
\end{align}
Similarly we rewrite the endcap functions as
\begin{align}
\mathcal L_{a}(P_f,k)\bigg \vert_{k_4=i\omega_{a1}} & = \mathcal L_{a,\mathrm{on}} +  \mathcal L \delta_{a,\mathrm{off}} \,, \\
\mathcal R^\dagger_{b}(P_i,k) \bigg \vert_{k_4=i\omega_{a1}}& = \mathcal R^\dagger_{b,\mathrm{on}} + \delta \mathcal R^\dagger_{b,\mathrm{off}} \,.
\end{align}
where $\mathcal L_{a,\mathrm{on}}$ and $\mathcal R^\dagger_{b,\mathrm{on}}$ are defined in Eq.~(\ref{eq:finalonfunc}) above and where the definitions of $\delta \mathcal L_{a,\mathrm{off}}$ and $\delta \mathcal R^\dagger_{b,\mathrm{off}}$ can be trivially inspected from the preceding equations.

The utility of this notation is that any function with a $\delta$ on the left (right) side, vanishes precisely when the pole on the left (right) diverges. Thus we can rewrite $\mathcal G_L$ as
\begin{align}
\mathcal G_L & =  \sum_{a,b=1}^{N_c}  \left[\frac{1}{L^3}\sum_{\mathbf{k}}\hspace{-.5cm}\int~\right] \frac{1}{2\omega_{a1}} \ [\mathcal L + \mathcal L \delta ]_a \left [\mathcal D_{a2f} + \mathcal S_f \right ] \left[ \w + \delta \w   +  \w\delta  +  \delta \w \delta \right]_{a2,b2}  \left [\mathcal D_{b2i}+ \mathcal S_i \right ] [ \mathcal R^\dagger + \delta \mathcal R^\dagger ]_b \,,
\label{eq:GL1} \\[10pt]
\begin{split}
 & =  \sum_{a,b=1}^{N_c}  \left[\frac{1}{L^3}\sum_{\mathbf{k}}\hspace{-.5cm}\int~\right] \frac{1}{2\omega_{a1}} \bigg ( \mathcal L_a \mathcal D_{a2f}  w_{a2,b2} \mathcal D_{b2i} \mathcal R_b^\dagger \\[-5pt]
 & \hspace{120pt} 
 +    \left \{ [\mathcal L + \mathcal L \delta ]_a \left [\mathcal D_{a2f} + \mathcal S_f \right ] \left[ \w + \delta \w  \right]_{a2,b2}   -  \mathcal L_a  \mathcal D_{a2f} w_{a2,b2} \right    \}   \mathcal D_{b2i} \mathcal R_b^\dagger  \\ 
& \hspace{120pt} +    \mathcal L_a   \mathcal D_{a2f}      \left \{ \left[  \w   +  \w\delta \right ]_{a2,b2} \left [\mathcal D_{b2i} + \mathcal S_i \right ] [ \mathcal R^\dagger + \delta \mathcal R^\dagger  ]_b - \w_{a2,b2} \mathcal D_{b2i} \mathcal R_b^\dagger \right \} \bigg ) \,,
\label{eq:Gawful}
\end{split} 
\end{align}
where we have introduced
\begin{align}
\mathcal D_{a2f} & = - \frac{1}{2 \omega_{a2f} (E_f - \omega_{a1} - \omega_{a2f} + i \epsilon)} \,,\ \ \ \ \ \mathcal S_f \equiv \Delta_{a2}(P_f-k) \big \vert_{k_4 = i \omega_{a1}} - \mathcal D_{a2f}\,,\\
\mathcal D_{b2i} & = - \frac{1}{2 \omega_{b2i} (E_i - \omega_{b1} - \omega_{b2i} + i \epsilon)} \,,\ \ \ \ \ \ \ \ \mathcal S_i \equiv \Delta_{b2}(P_i-k) \big \vert_{k_4 = i \omega_{a1}} - \mathcal D_{b2i} \,,
\end{align}
with $\omega_{a2f} = \sqrt{(\textbf P_f - \textbf k)^2 + m_{a2}^2}$ and $\omega_{b2i}=\sqrt{(\textbf P_i - \textbf k)^2 + m_{b2}^2}$. Note that $\mathcal S_f$ and $\mathcal S_i$ are smooth, by construction, in the vicinity of the single-particle pole.

In Eq.~(\ref{eq:GL1}) we have simply substituted our definitions and in (\ref{eq:Gawful}) we have discarded smooth terms and arranged the remaining terms according to the number and type of poles. We have left the ``on" and ``off" labels implicit to reduce clutter, and note that $\mathcal L, \mathcal R^\dagger$ and $\w$ in the above expressions are completely projected on-shell. Similarly the incoming (right-side) coordinates of $\delta \w$ and the outgoing (left-side) coordinates of $\w \delta$ are on-shell. Thus, Eq.~(\ref{eq:Gawful}) makes explicit the fact that poles, together with sum-integral differences, project the neighboring functions on-shell.

We simplify further by rewriting the terms in curly braces in Eq.~(\ref{eq:Gawful}). At this stage we also return to the completely general case in which all possible $\textbf 1 + \mathcal J \rightarrow \textbf 1$ couplings are included. This means that we sum over $w_{a1b1}$, $w_{a1b2}$, $w_{a2b1}$ and $w_{a2b2}$, with the understanding that some of these will vanish in most cases. We define
\begin{align}
\begin{split}
[\mathcal L \Delta \w]_b \delta_{\mathrm{df}}& \equiv \sum_{a=1}^{N_c} \bigg [\mathcal L_a \Big ( \Delta_{a1}(P_f-k)  \w_{a1,b2,\mathrm{off,on}} + \Delta_{a2}(P_f-k)  \w_{a2,b2,\mathrm{off,on}}  \Big ) \\
& \hspace{190pt} -  \mathcal L_{a,\mathrm{on}}  \Big ( \mathcal D_{a1f} w_{a1,b2,\mathrm{on,on}}  + \mathcal D_{a2f} w_{a2,b2,\mathrm{on,on}}  \Big ) \\
&  \hspace{24pt} + \mathcal L_a \Big ( \Delta_{a1}(k - P_i + P_f) \w_{a1,b1,\mathrm{off,on}}+ \Delta_{a2}(k - P_i + P_f) \w_{a2,b1,\mathrm{off,on}} \Big ) \\
& \hspace{190pt}  -  \mathcal L_{a,\mathrm{on}} \Big ( \overline {\mathcal D}_{a1f} w_{a1,b1,\mathrm{on,on}} +  \overline {\mathcal D}_{a2f} w_{a2,b1,\mathrm{on,on}} \Big )  \bigg ]  \,, 
\end{split} \label{eq:deltaLdef} \\
\begin{split}
\delta_{\mathrm{df}}  [  \w \Delta \mathcal R^\dagger]_a & \equiv  \sum_{b=1}^{N_c}\bigg [   \Big ( \w_{a2,b1,\mathrm{on,off}}  \Delta_{b1}(P_i-k)  +  \w_{a2,b2,\mathrm{on,off}}  \Delta_{b2}(P_i-k) \Big ) \mathcal R^\dagger_b \\
& \hspace{190pt} - \Big ( \w_{a2,b1,\mathrm{on,on}} \mathcal D_{b1i} + \w_{a2,b2,\mathrm{on,on}} \mathcal D_{b2i} \Big ) \mathcal R_{b,\mathrm{on}}^\dagger \\
& \hspace{24pt} +   \Big ( \w_{a1,b1,\mathrm{on,off}}  \Delta_{b1}(k - P_f + P_i) +  \w_{a1,b2,\mathrm{on,off}}  \Delta_{b2}(k - P_f + P_i) \Big )  \mathcal R^\dagger_b  \\ 
& \hspace{190pt} - \Big ( \w_{a1,b1,\mathrm{on,on}} \overline{ \mathcal D}_{b1i} +  \w_{a1,b2,\mathrm{on,on}} \overline{\mathcal D}_{b2i}  \Big ) \mathcal R_{b,\mathrm{on}}^\dagger   \bigg ] \,, \label{eq:deltaRdef}
\end{split}
\end{align}
where we have introduced
\begin{align}
\overline {\mathcal D}_{a1f} & \equiv -\frac{1}{ 2\overline{\omega}_{a1f}(E_f - \overline{\omega}_{a1f} -  \overline{\omega}_{a2}  + i \epsilon ) } \,, \\
\overline {\mathcal D}_{b1i} & \equiv -\frac{1}{ 2 \overline \omega_{b1i}(E_i- \overline \omega_{b1i}  -  \overline \omega_{b2} + i \epsilon )}\,,  
\end{align}
with
\begin{alignat}{3}
\label{eq:firstroundomegas1} 
\overline \omega_{a2}  &\equiv \sqrt{(\textbf P_i - \textbf k)^2 + m_{a2}^2} \,,\ \ \ \ \  &  \overline \omega_{a1f}&  \equiv \sqrt{(\textbf P_f - \textbf P_i + \textbf k)^2 + m_{a1}^2} \,, \\
\overline \omega_{b2} & \equiv  \sqrt{(\textbf P_f - \textbf k)^2 + m_{b2}^2} \,, & \overline \omega_{b1i} & \equiv \sqrt{(\textbf P_i - \textbf P_f + \textbf k)^2 + m_{b1}^2} \,.
\label{eq:firstroundomegas2}
 \end{alignat}
 All other terms appearing in Eqs.~(\ref{eq:deltaLdef}) and (\ref{eq:deltaRdef}) can be obtained by switching the labels associated with the particle coupling to the external current with that of the spectator.  For example $\overline {\mathcal D}_{b2i}$ is defined as
 \begin{align}
\overline {\mathcal D}_{b2i} & \equiv -\frac{1}{ 2 \overline \omega_{b2i}(E_i- \overline \omega_{b2i}  -  \overline \omega_{b1} + i \epsilon )}\,,  
 \end{align}
 where 
 \begin{align}
\overline \omega_{b1} & \equiv  \sqrt{(\textbf P_f - \textbf k)^2 + m_{b1}^2} \,, 
\hspace{1cm} \overline \omega_{b2i}  \equiv \sqrt{(\textbf P_i - \textbf P_f + \textbf k)^2 + m_{b2}^2} \,.
 \end{align}
Note that these expressions are valid for all types of channels and further accommodate all possible couplings to $w$.

\begin{figure*}[t]
\begin{center}
 \includegraphics[scale=0.4]{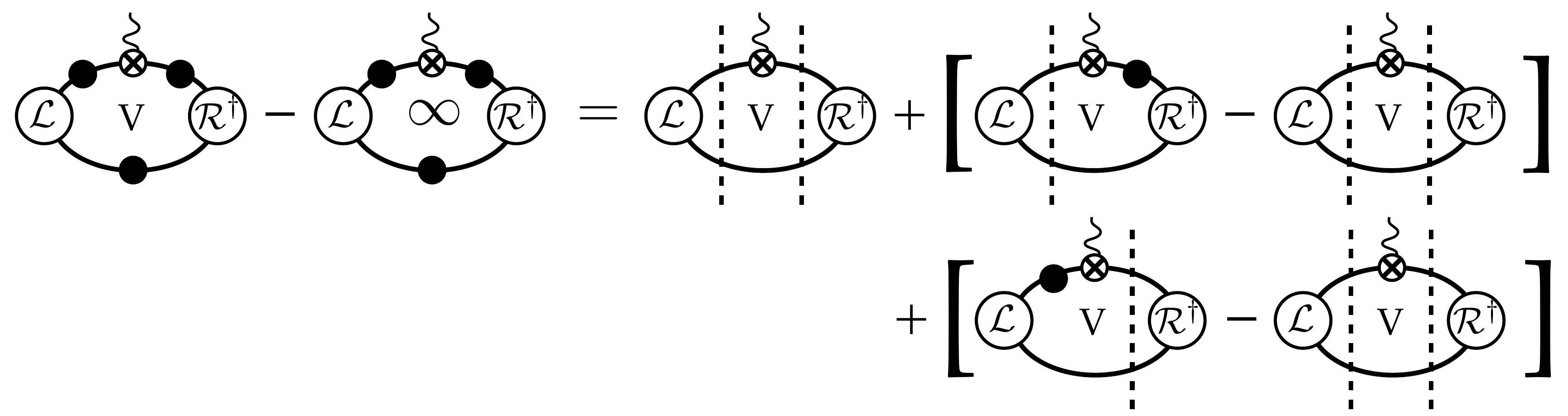}
\caption{Shown is the diagrammatic representation of Eq.~(\ref{eq:Gawfulreduced}), depicting a two-particle loop with an external current coupling to one of the intermediate particles. The first term on the right-hand side represents the finite-volume residue from the double pole, in which both the endcaps and the one-body current are projected on-shell. In the second and third terms, both in square brackets, only the momenta on one side of the current are on-shell. The careful analysis in the main text ensures that we have captured all power-law effects without overcounting.}\label{fig:1body_insertion}
\end{center}
\end{figure*}

Substituting these definitions, we reach
 \begin{equation}
\begin{split}
 \mathcal G_L & =  \sum_{a,b=1}^{N_c} \sum_{s,t=1,2} \xi_a \xi_b  \left[\frac{1}{L^3}\sum_{\mathbf{k}}\hspace{-.5cm}\int~\right] \ \mathcal L_{a,\mathrm{on}}  \frac{1}{2 \omega_{asf} (E_f - \omega_{a \slashed s} - \omega_{asf} + i \epsilon)}  \frac{w_{as,bt;
 \mathrm{on,on}}}{2\omega_{{a} \slashed{s}}}  \frac{1}{2 \omega_{bti} (E_i - \omega_{b\slashed t} - \omega_{bti} + i \epsilon)} \mathcal R_{b,\mathrm{on}}^\dagger \\  
&  + \sum_{b=1}^{N_c}  \xi_b \left[\frac{1}{L^3}\sum_{\mathbf{k}}\hspace{-.5cm}\int~\right] \frac{1}{2\omega_{a1}} { \ \left \{ [\mathcal L \Delta \w]_b \delta_{\mathrm{df}} \right    \}}  \left [- \frac{1}{2 \omega_{b2i} (E_i - \omega_{b1} - \omega_{b2i} + i \epsilon)}  \right ] \mathcal R_{b,\mathrm{on}}^\dagger  \\ 
&   + \sum_{a=1}^{N_c} \xi_a \left[\frac{1}{L^3}\sum_{\mathbf{k}}\hspace{-.5cm}\int~\right] \frac{1}{2\omega_{a1}} \ \mathcal L_{a,\mathrm{on}}  \left [- \frac{1}{2 \omega_{a2f} (E_f - \omega_{a1} - \omega_{a2f} + i \epsilon)}   \right ] {  \left \{ \delta_{\mathrm{df}}[  \w \Delta \mathcal R^\dagger]_a   \right \} }\,.
\label{eq:Gawfulreduced}
\end{split}
\end{equation}
Here we have also explicitly shown the form of the remaining poles. The symmetry factors $\xi_a$ and $\xi_b$ are included because, in the case of identical particles, the first term is overcounted. Finally, we have included particle indices $s,t$ which are summed over $1$ and $2$. The slashed notation indicates the particle not labeled by the index, for example for $s=1$ then $\slashed s=2$. This result is diagrammatically depicted in Fig.~\ref{fig:1body_insertion}. 

The quantities $ [\mathcal L \Delta \w]_b \delta_{\mathrm{df}}$ and $ \delta_{\mathrm{df}}[  \w \Delta \mathcal R^\dagger]_a $ in Eq.~(\ref{eq:Gawfulreduced}) are smooth functions which include off-shell coordinate dependence arising from the first two terms in Eqs.~(\ref{eq:deltaLdef}) and (\ref{eq:deltaRdef}). However since these factors only appear in terms with a single pole, we may proceed as in the previous subsection and replace them with on-shell projections. As explained previously, this is justified because the difference between on- and off-shell functions vanishes at the pole resulting in a smooth summand with a negligible sum-integral difference. We define
\begin{align}
[\mathcal L \Delta \w]_b\delta_{\mathrm{df}}(P_f,P_i,q^*_{bi} \hat{\textbf{k}}_{bi}^*) 
&\equiv\sqrt{4 \pi} \sum_{\ell,m}  \Big [[\mathcal L \Delta \w]_b \delta_{\mathrm{df}} \Big ]_{\ell m}(P_f,P_i) \   Y_{\ell m}(\khs_{bi})\,,\\
\delta_{\mathrm{df}}[  \w \Delta \mathcal R^\dagger]_a   (P_f,P_i,q^*_{af} \hat{\textbf{k}}_{af}^*) 
&\equiv\sqrt{4 \pi} \sum_{\ell,m}  \Big [\delta_{\mathrm{df}}[  \w \Delta \mathcal R^\dagger]_a \Big ]_{\ell m}(P_f,P_i) \   Y^*_{\ell m}(\khs_{af})\,.
\label{eq:delta_endcaps}
\end{align}
As above, due to singularities near $\textbf k^*_{bi}, \textbf k^*_{af} =0$, we cannot substitute this directly but instead take
\begin{align}
[\mathcal L \Delta \w]_b\delta_{\mathrm{df}}(P_f,P_i,\textbf{k}) 
&\longrightarrow \sqrt{4 \pi} \sum_{\ell,m}  \Big [[\mathcal L \Delta \w]_b \delta_{\mathrm{df}} \Big ]_{\ell m}(P_f,P_i) \   Y_{\ell m}(\khs_{bi}) \left(\frac{k^*_{bi}}{q^*_{bi}} \right )^{\ell},\\
\delta_{\mathrm{df}}[  \w \Delta \mathcal R^\dagger]_a   (P_f,P_i,\textbf{k}) 
&\longrightarrow \sqrt{4 \pi} \sum_{\ell,m}  \Big [\delta_{\mathrm{df}}[  \w \Delta \mathcal R^\dagger]_a \Big ]_{\ell m}(P_f,P_i) \   Y^*_{\ell m}(\khs_{af}) \left(\frac{k^*_{af}}{q^*_{af}} \right )^{\ell}.
\label{eq:delta_endcaps}
\end{align}

We reach our final form for $\mathcal G_L$ by substituting these projections for the endcaps as well as Eqs.~(\ref{eq:wondef1})-(\ref{eq:wondef3}) into Eq.~(\ref{eq:Gawfulreduced}) and grouping the spherical-harmonics into the finite-volume quantities. For the second and third terms this results in factors of $F$, defined in Eq.~(\ref{eq:Fscdef}) above. For the first term, a new quantity arises
 \begin{multline}
\label{eq:Gmat}
G^{st}_{a \ell_f m_{f},a' \ell_f' m_{f}'; b' \ell_i' m_{i}', b \ell_i m_{i}}(P_f,P_i,L)  \equiv \\
\hspace{-4cm}
\delta_{aa'} \delta_{bb'}
\left[\frac{1}{L^3}\sum_{\mathbf{k}}\hspace{-.5cm}\int~\right]
~\frac{1}{2\omega_{a\slashed s}}
\frac{ 4 \pi  Y_{\ell_fm_{ f}}(\hat {\textbf k}^*_{af})
Y^*_{\ell_f'm_{f}'}(\hat {\textbf k}^*_{af})
 }{2 \omega_{{asf}}(E_f -  \omega_{a\slashed s} - \omega_{asf} + i \epsilon )} 
  \bigg (\frac{k^{*}_{af}}{q^*_{af}} \bigg)^{\ell_f+\ell_f'}
\frac{ 4 \pi  Y_{\ell_i' m_{i}'}(\hat {\textbf k}^*_{bi})
Y^*_{\ell_i m_{i}}(\hat {\textbf k}^*_{bi})
 }{2 \omega_{bti}(E_i -  \omega_{b\slashed t} - \omega_{bti} + i \epsilon )} 
 \bigg (\frac{k^{*}_{bi}}{q^*_{bi}} \bigg)^{\ell_i+\ell_i'}
  \,.
\end{multline}
It is further convenient to introduce notation that contracts a tensor with four sets of channel and spherical-harmonic indices with a tensor that has two,
\begin{equation}
\label{eq:Gdotw}
 [G(L) \cdot \w]_{a \ell_fm_{f}; b\ell_im_{i}}(P_f,P_i) 
 \equiv 
\sum_{s,t=1,2}  \xi_a \xi_b  G^{st}_{a \ell_f m_{f}, a' \ell_f' m_{f}'; b' \ell_i' m_{i}', b \ell_i m_{i}}(P_f,P_i,L) 
~\w_{a'sb't; \ell_f' m_{f}';  \ell_i' m_{i}'}(P_f;P_i)  \,.
\end{equation} 
This leads to a compact result for $\mathcal G_L$ 
\begin{equation}
\mathcal G_L  =  
 \mathcal L(P_f) [G(L) \cdot \w](P_f,P_i) \mathcal R^\dagger(P_i)  
 -\mathcal L(P_f)
F(P_f,L)  \left \{ \delta_{\mathrm{df}}[w^{} \Delta  {\mathcal R}^\dag  ] (P_i)  \right  \}  
 - \left \{ [{\mathcal L}\Delta w^{}] \delta_{\mathrm{df}} (P_f) \right \}
F(P_i,L)  {\mathcal R}^\dagger(P_i)  .
\label{eq:GL_final}
\end{equation}
In Appendix \ref{app:Gnum} we describe how to reduce the function $G$ to a form which is more amenable for numerical evaluation. This analysis also shows that $G$ is a well defined function which is finite away from the free-particle poles.

\begin{figure*}[t]
\begin{center} 
\includegraphics[scale=0.4]{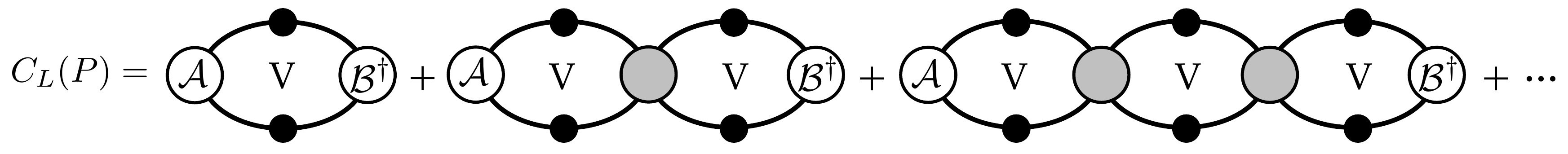}
\caption{Depicted is the diagrammatic representation of the two-point correlation function in a finite volume for energies where only two-particle states can go on-shell. Although not explicitly shown in the diagram, we accommodate any number of two-particle channels. $\mathcal B^\dag$ and $\mathcal A$ denote the creation and annihilation operators respectively. The kernels and propagators are defined in Fig.~\ref{fig:scat_amp}. }\label{fig:corr2}
\end{center}
\end{figure*}

\section{Two-body two-point function~\label{sec:two-point}}
In this section we review the derivation of the two-point correlation function. We closely follow our previous work, Ref.~\cite{Briceno:2014uqa}, which is a natural extension of Ref.~\cite{Kim:2005gf} for systems with arbitrary channels and generic masses. When defining a momentum space correlator we have the choice to project either the source or sink or both operators to the desired total momentum. We choose to project the sink and so define
\begin{equation}
C_L(P)  \equiv \int_L d^4 x \, e^{- i P x} \Big [ \langle 0 \vert T \mathcal A(x) \mathcal B^\dagger(0) \vert 0 \rangle \Big ]_L \,,
\label{eq:CLP}
\end{equation}
where $\mathcal A$ and $\mathcal B^\dag$ are two-body interpolating operators defined in position space. 
This is the definition of the correlator that is most easily represented diagrammatically. Another convenient definition is one where the source and sink are both projected to a definite spatial momentum and time,
\begin{align}
C_L(x_4-y_4, \textbf P) & \equiv \int_L \! d \textbf x \int_L \! d \textbf y \ e^{- i \textbf P \cdot (\textbf x - \textbf y)} \Big [ \langle 0 \vert T \mathcal A(x) \mathcal B^\dagger(y) \vert 0 \rangle \Big ]_L \,.
\label{eq:Cmonmon}
\end{align}
This definition is more closely related to that used in numerical lattice QCD calculations.

We begin by rewriting $C_L(x_4-y_4, \textbf P)$ by inserting a complete set of finite-volume states
\begin{align}
C_L(x_4-y_4, \textbf P) & \equiv \int_L \! d \textbf x \int_L \! d \textbf y \ e^{- i \textbf P \cdot (\textbf x - \textbf y)} \Big [ \langle 0 \vert T \mathcal A(x) \mathcal B^\dagger(y) \vert 0 \rangle \Big ]_L \,, \\
& = \int_L \! d \textbf x \int_L \! d \textbf y \ e^{- i \textbf P \cdot (\textbf x - \textbf y)} \sum_n \Big [ \langle 0 \vert \mathcal A(x_4, \textbf x)  \vert E_n, \textbf P, L \rangle \Big ]_L \Big [ \langle E_n, \textbf P, L \vert \mathcal B^\dagger(y_4, \textbf y) \vert 0 \rangle \Big ]_L\,, \\
& =  \int_L \! d \textbf x \int_L \! d \textbf y \ \sum_n e^{- E_{n,\textbf P,L}(x_4-y_4)}  \Big [  \langle 0 \vert  \mathcal A(0)  \vert E_n, \textbf P, L \rangle \Big ]_L \Big [ \langle E_n, \textbf P, L \vert \mathcal B^\dagger(0) \vert 0 \rangle \Big ]_L \,, \\
& =  L^6 \sum_n e^{- E_{n,\textbf P,L}(x_4-y_4)}   \Big [ \langle 0 \vert  \mathcal A(0)  \vert E_n, \textbf P, L \rangle \Big ]_L \Big [ \langle E_n, \textbf P, L \vert \mathcal B^\dagger(0) \vert 0 \rangle \Big ]_L \,.
\label{eq:completeset}
\end{align}
The $[]_L$ notation makes explicit that the states and operators have been defined in a finite volume. This spectral decomposition is used in analysis of lattice QCD calculations, to access the finite-volume spectrum and matrix elements.

To give meaning to these quantities in terms of infinite-volume observables, we proceed to evaluate $C_L(P)$ using finite-volume Feynman diagrams as depicted in Fig.~\ref{fig:corr2}. To reduce these we use Eq.~(\ref{eq:Fsum_final}) to separate finite- and infinite-volume quantities. Indeed for the two-point correlator it is possible to group all infinite-volume diagrams into two types of infinite-volume quantities. The first type consists of infinite-volume matrix elements
\begin{align}
\label{eq:Aldef}
A_{a \ell m}(P) & \equiv   \langle 0 \vert  \mathcal A(0) \vert \!-\!iP_4,  \textbf P,a,\ell ,m, \mathrm{in} \rangle \,, \\
B^*_{b \ell' m'}(P) &  \equiv \langle -iP_4, \textbf P, b,\ell' ,m', \mathrm{out} \vert  \mathcal B^\dagger(0) \vert 0 \rangle \,.
\end{align}
Here $\vert E, \textbf P,a ,\ell ,m,\mathrm{in} \rangle$ and $\langle E, \textbf P,b, \ell ,m,\mathrm{out} \vert$ are  in and out states that have been projected onto the $\ell$ partial wave. These are related to the states used in Eq.~(\ref{eq:bigWmedef}) above by
\begin{align}
\vert P_i,k,a, \mathrm{in} \rangle & \equiv \sqrt{4 \pi} \sum_{\ell, m} Y_{\ell,m}(\khs_{ai}) \vert \!-\!iP_{4,i},  \textbf P_i,a,\ell ,m, \mathrm{in} \rangle \,, \\
\langle P_f,k,b, \mathrm{out} \vert & \equiv \sqrt{4 \pi} \sum_{\ell, m} Y^*_{\ell,m}(\khs_{bf}) \langle -iP_{4,f},  \textbf P_f,b,\ell ,m, \mathrm{out} \vert \,.
\end{align} 
The second type of infinite-volume quantity which appears is the $\textbf{2}\rightarrow\textbf{2}$ scattering amplitude, which can also be decomposed into definite angular momentum states. In the single channel case each angular-momentum component of the scattering amplitude is directly related to the scattering phase shift, $\delta_\ell$, via
\begin{align}
\mathcal M_{\ell}(P)=\frac{8\pi E^*}{\xi q^*}\frac{1}{\cot\delta_{\ell}-i} \,.
\label{eq:scatamp}
\end{align}
For general coupled channels the relation is more complicated
\begin{equation}
\label{Smatrix}
i\mathcal{M}_\ell(P) \equiv \mathbb{P}^{-1} {\big [S_\ell(P)-\mathbb{I} \big ]} \mathbb{P}^{-1} \,.
\end{equation}
where for $N$ open two-particle channels $S_\ell$ is a unitary $N \times N$ matrix with $N(N+1)/2$ real degrees of freedom,  $\mathbb I$ is the $N \times N$ identity matrix, and
\begin{equation}
\mathbb{P}=\frac{1}{\sqrt{4\pi E^*}} \text{diag} \left (\sqrt{\xi_1q^{*}_1},\sqrt{\xi_2q^{*}_2},\ldots,\sqrt{\xi_Nq^{*}_N} \right ) \,.
\label{eq:Pmatrix}
\end{equation}
We view the $b \rightarrow a$ scattering amplitude, $\mathcal M_{ab}$, as a matrix in the same $a \ell m$ space on which $A^*$ and $B$ are defined
\begin{equation}
\mathcal M_{a \ell' m'; b \ell m}(P) \equiv \delta_{m'm} \delta_{\ell' \ell} \mathcal M_{a b, \ell}(P)  \,,
\end{equation}
with no sum on $\ell$ here.

With these matrices in hand we are ready to give the final result for the finite-volume correlator. We do not derive the expression for the momentum-space finite-volume correlator here, but simply state the result which is proven in Refs.~\cite{Kim:2005gf,Briceno:2014uqa}
 \begin{equation}
\label{eq:CLres}
C_L(P) = C_{\infty}(P) - A(P) \frac{1}{F^{-1}(P,L) + \mathcal M(P)} B^*(P) \,.
\end{equation}
The finite-volume correlator has poles whenever 
\begin{equation}
\label{eq:lumat}
\frac{1}{F^{-1}(P,L) + \mathcal M(P)} \,,
\end{equation}
has a divergent eigenvalue, or equivalently whenever
\begin{align}
\det[F^{-1}(P,L) + \mathcal M(P)]=0 \,.
\label{eq:QC}
\end{align}
This is the standard quantization condition for any number of two-boson channels in a finite volume~\cite{Luscher:1986pf, Luscher:1990ux, Rummukainen:1995vs, Kim:2005gf, Christ:2005gi}. This has also been generalized to systems with arbitrary spin in Ref.~\cite{Briceno:2014oea}, but here we restrict our attention to scalar particles. Having determined $C_L(P)$, we can obtain $C_L(x_4-y_4, \textbf P)$ by performing a Fourier transform in $P_4$ and multiplying by a factor of $L^3$~\cite{Briceno:2014uqa,Briceno:2015csa}
\begin{align}
C_L(x_4-y_4, \textbf P)& \equiv L^3 \int \frac{d P_4}{2 \pi} e^{i P_4(x_4-y_4)} C_L(P) \,, \\
&= L^3 \int \frac{d P_4}{2 \pi} e^{i P_4(x_4-y_4)} \left[ C_{\infty}(P) - A(P) \frac{1}{F^{-1}(P,L) + \mathcal M(P)} B^*(P) \right ] \,, \label{eq:steptwo} \\
&   = \sum_n e^{- E_{n,\textbf P,L}(x_4-y_4)}L^3  
A(E_{n}, \textbf P) \mathcal R(E_{n}, \textbf P)B^*(E_{n}, \textbf P) \,,
\label{eq:CL_diag}  
\end{align}
where $\mathcal R(E_{n}, \textbf P)$ is residue of the matrix in Eq.~(\ref{eq:lumat}) at the $n$th energy pole 
\begin{equation}
\label{eq:Rdef}
\mathcal R(E_{n}, \textbf P) \equiv  \lim_{P_4 \rightarrow i E_{n}} \left[ - (i P_4 + E_{n}) \frac{1}{F^{-1}(P,L) + \mathcal M(P)}\right] \,.
\end{equation}
This is a matrix in angular momentum and channel-space, which mixes different partial waves due to the breaking of continuous rotational symmetry in a cubic finite-volume.

Finally, by equating Eqs.~(\ref{eq:CL_diag}) and (\ref{eq:completeset}), we reproduce the relation between finite- and infinite-volume matrix elements 
\begin{align}
\label{eq:firstmaster}
\Big [ \langle 0 \vert    {\mathcal A}(0) \vert E_n, \textbf P, L \rangle \Big ]_L \Big [ \langle E_n, \textbf P, L  \vert  {\mathcal B}^\dagger(0) \vert 0\rangle \Big ]_L=
  \frac{1}{L^3}   A(E_{n}, \textbf P) \mathcal R(E_{n}, \textbf P)B^*(E_{n}, \textbf P) \,.
\end{align} 
In Ref.~\cite{Briceno:2015csa} we demonstrated how to use this relation to determine $\textbf{1}+\mathcal J \rightarrow \textbf{2}$ and $\textbf{0}+\mathcal J \rightarrow \textbf{2}$ transition amplitudes from finite-volume matrix elements of local currents. However the trick used to extract these quantities fails for $\textbf{2}+\mathcal J \rightarrow{2}$ transition amplitudes as explained in that reference. Thus in Sec.~\ref{sec:3pt_func} we directly consider three-point correlators and, using the techniques presented in Ref.~\cite{Briceno:2014uqa}, we derive the main result of this work.


\section{Two-body three-point function \label{sec:3pt_func}}

In this section we present an analysis of finite-volume three-point correlators. As in the case of two-point correlators discussed above, two closely related definitions of the correlation functions will be used. We begin with
\begin{equation}
C^{2\rightarrow 2}_{L}(P_i,P_f) =  \int_L \! d^4  x_f~ \! d^4  x_i  
 \ e^{- i   P_f   x_f   } 
  \ e^{+ i   P_i     x_i} 
    \Big [ \langle 0 \vert T 
  \mathcal A(x_f) 
  {\mathcal{J}}(0)\mathcal B^\dagger(x_i) \vert 0 \rangle \Big ]_L \,,
\end{equation}
where $\mathcal A$ and $\mathcal B^\dagger$ are the same interpolating operators defined in the previous section, and ${\mathcal{J}}$ is a local current. We contrast this with  
\begin{align}
C_L^{2\rightarrow 2}(x_{f,4}-y_{4},y_{4}-x_{i,4}, \textbf P_i, \textbf P_f) &\nn\\
&\hspace{-3cm}\equiv \int_L \! d \textbf x_f~ \! d \textbf x_i ~ \! d \textbf y
 \ e^{- i \textbf  P_f \cdot (\textbf x_f - \textbf y)} 
  \ e^{- i \textbf  P_i \cdot (\textbf y - \textbf x_i)} 
  \times  \Big [ \langle 0 \vert T 
  \mathcal A(x_f) 
  {\mathcal{J}}(y)\mathcal B^\dagger(x_i) \vert 0 \rangle \Big ]_L \,, \nn\\
  &\hspace{-3cm} = L^3 \int \frac{d P_{i,4}}{2 \pi} \int \frac{d P_{f,4}}{2 \pi} e^{iP_f(x_{f,4}-y_{4})}e^{iP_i(y_{4}-x_{i,4})} C^{2\rightarrow 2}_{L}(P_i,P_f).
  \label{eq:2to2_corrdef}
  \end{align}
  
As above, the second form of the correlator is most convenient for spectral decomposition 
\begin{align}
C_L^{2\rightarrow 2}(x_{f,4}-y_4,y_4-x_{i,4}, \textbf P_i, \textbf P_f) &\nn\\
&\hspace{-3cm}=  L^9 \sum_{n_i,n_f} 
  e^{- E_{n_f}(x_{f,4}-y_4)}
  e^{- E_{n_i}(y_4-x_{i,4})}\nn\\
 &\hspace{-2cm}
 \times
\Big [ \langle 0 \vert  \mathcal A(0)  \vert E_{n_f}, \textbf P_f, L \rangle \Big ]_L 
\Big [ \langle E_{n_f}, \textbf P_f, L \vert  \mathcal J(0)  \vert E_{n_i}, \textbf P_i, L \rangle \Big ]_L 
\Big [ \langle E_{n_i}, \textbf P_i, L \vert \mathcal B^\dagger(0) \vert 0 \rangle \Big ]_L.
\label{eq:3pt_completeset}
\end{align}
The matrix elements $\Big [ \langle 0 \vert  \mathcal A(0)  \vert E_{n_f}, \textbf P_f, L \rangle \Big ]_L$ and $\Big [ \langle E_{n_i}, \textbf P_i, L \vert \mathcal B^\dagger(0) \vert 0 \rangle \Big ]_L$ are the same as those appearing in Eq.~(\ref{eq:firstmaster}). In order to give a physical interpretation to the third matrix element, $\Big [ \langle E_{n_f}, \textbf P_f, L \vert  \mathcal J(0)  \vert E_{n_i}, \textbf P_i, L \rangle \Big ]_L$, we now evaluate the finite-volume three-point correlator diagrammatically. 


\begin{figure*}[t]
\begin{center}
\subfigure[]{
\label{fig:C2-to-2_no1B_a}
\includegraphics[scale=0.35]{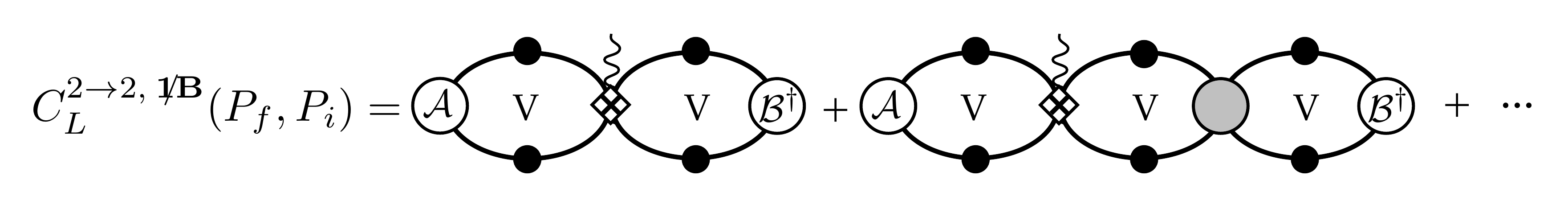}}\\ 
\subfigure[]{
\label{fig:W_no1B}
\includegraphics[scale=0.35]{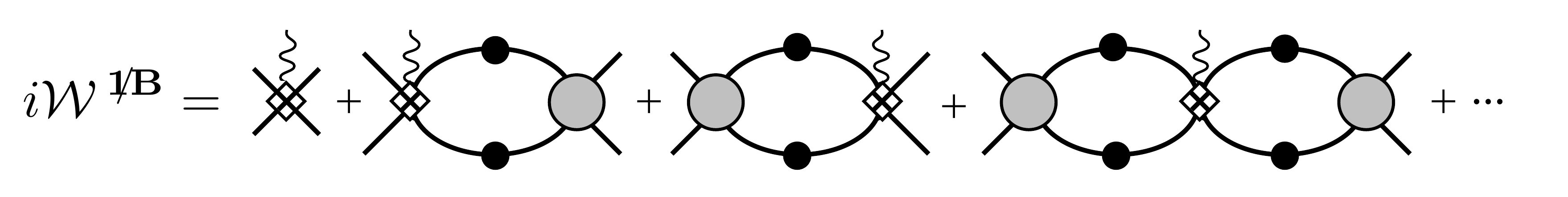}}
\caption{Shown is the finite-volume (a) three-point correlation function and (b) the infinite volume transition amplitude, both in the absence of $\textbf{1}+\mathcal{J}\rightarrow \textbf{1}$ subprocesses.}\label{fig:C2-to-2_no1B}
\end{center}
\end{figure*}
%

\subsection{Three-point functions and matrix elements: (a) For theories without $\textbf{1}+\mathcal{J}\rightarrow \textbf{1}$ contributions\label{sec:no1body}}

As a warm up, we first examine the three-point correlation function for transitions with no $\textbf{1}+\mathcal{J}\rightarrow \textbf{1}$ subprocesses. Although most processes involve such contributions, there are interesting examples where these are not allowed. One prominent case is parity violation in proton-proton scattering (see Ref.~\cite{Phillips:2008hn} and references within). Here we do not give details about how such systems arise, we simply envision a generic system where the weak interaction does not couple to single-particle states. In other words a system for which Eq.~(\ref{eq:littlewmedef}) vanishes
\begin{equation}
w_{a2,b2}(P_f-k;P_i-k) \equiv \langle P_f-k, a2 \vert \mathcal J(0) \vert P_i - k, b2 \rangle = 0 \,.
\end{equation}
In this subsection we show that, given this assumption, one can readily generalize the derivation of Ref.~\cite{Briceno:2014uqa} to find a relation between finite- and infinite-volume matrix elements. The result is given in Eqs.~(\ref{eq:2to2_no1B_degen}), (\ref{eq:2to2_no1B_notdegen}) and (\ref{eq:2to2_no1b_ratio}) below. 
 In the following subsection we include all possible interactions, in particular $\textbf{1} + \mathcal J \rightarrow \textbf{1}$ contributions, and show how this changes the relation. The results for this more complicated case, summarized in Eqs.~(\ref{eq:2to2_degen})-(\ref{eq:2to2_ratio}), are the main results of this paper. 
 
As discussed in Sec.~\ref{sec:trans_amps}, in the diagrammatic representation of the three-point function one must include all terms which have a single insertion of the weak current but any number of insertions of the strong-interaction vertices. As usual in this type of analysis, one can reduce the complexity of diagrams by identifying a skeleton expansion that explicitly displays all power-law finite-volume effects, but groups terms with exponentially suppressed volume dependence into kernels. For the three-point correlator defined in Eq.~(\ref{eq:2to2_corrdef}) and given the assumption of no $\textbf{1} + \mathcal J \rightarrow \textbf{1}$ contributions, only two types of kernels are needed. The first is the standard Bethe-Salpeter kernel, discussed in Sec.~\ref{sec:trans_amps}.  The second kernel, which includes the weak insertion, is referred to as the weak kernel. It is defined as the sum of all connected diagrams with four hadronic external legs and one current insertion, which are two-particle irreducible in the $s$-channel. In Fig.~\ref{fig:C2-to-2_no1B_a} we show how to express the full correlator in terms of these two building blocks.

We stress the similarities between this skeleton expansion and that of the two-point correlation function shown in Fig.~\ref{fig:corr2}, which was reviewed in the previous section. The only distinction is the presence of the weak kernel.
In fact, the finite-volume loops that appear here have the same structure as those studied previously. One may thus use Eq.~(\ref{eq:Fsum_final}) to determine the finite-volume correction to all of the diagrams appearing in Fig.~\ref{fig:C2-to-2_no1B_a}. In performing the separation between the finite- and infinite-volume terms, various important quantities emerge. First we recover the same objects that arise in the two-point correlator. These are the infinite-volume matrix elements $A$ and $B^*$, the infinite-volume $\textbf{2}\rightarrow\textbf{2}$ scattering amplitude, $\mathcal{M}(P)$, and the finite-volume function, $F$, defined in Eq.~(\ref{eq:Fscdef}). In addition we identify new infinite-volume quantities which contain the weak insertion. We will see below that, although ``weak endcap factors'' do arise (like $A$ and $B^*$ but with a weak current insertion) these play no role in our final result. Thus only one important new quantity appears, the fully-dressed infinite-volume $\textbf{2}+\mathcal{J}\rightarrow\textbf{2}$ transition amplitude, $\mathcal{W}^{~\slash\hspace{-.2cm} \textbf{1B}}(P_f,P_i)$ (see Fig.~\ref{fig:W_no1B}). Note that $\mathcal{W}^{~\slash\hspace{-.2cm} \textbf{1B}}(P_f,P_i)$ is a matrix in combined angular-momentum and channel space with matrix elements $\mathcal{W}^{~\slash\hspace{-.2cm} \textbf{1B}}_{a\ell_f m_{\ell_f};b \ell_im_{\ell_i}}(P_f,P_i)$. This matrix is not diagonal since the external current can couple different angular-momentum states and both the strong and weak interactions can couple the different channels. Finally, we have introduced the notation {~\slash\hspace{-.4cm} \textbf{1B}} to stress the absence of $\textbf{1} + \mathcal J \rightarrow \textbf{1}$ subprocesses.

Evaluating the correlation function to all orders in the strong interaction, one finds
\begin{align}
\label{eq:No1BodyCPP}
C^{2\rightarrow 2,~\slash\hspace{-.2cm} \textbf{1B}}_{L}(P_i,P_f) 
=
 A(P_f) 
 \frac{1}{F^{-1}(P_f,L) + \mathcal M(P_f)}
 \mathcal{W}^{~\slash\hspace{-.2cm} \textbf{1B}}(P_f,P_i)
 \frac{1}{F^{-1}(P_i,L) + \mathcal M(P_i)}
  B^*(P_i)+\cdots,
  \end{align}
where once again we have left implicit the summed angular-momentum and channel indices, and where the ellipses denotes contributions that do not contribute to the Fourier transform that we perform in the next step. These unimportant terms include the infinite-volume correlation function as well as terms where the weak current is attached to either $A$ or $B^*$. The expression for the right-hand side of Eq.~(\ref{eq:No1BodyCPP}) is straightforward to understand. For each two-particle state one obtains a factor of $[F^{-1}(P_j,L) + \mathcal M(P_j)]^{-1}$ and the two states are then coupled by the infinite-volume transition amplitude. To be able to compare this representation of the correlation function to Eq.~(\ref{eq:3pt_completeset}) we must perform two Fourier transforms, one each in $P_{i,4}$ and $P_{f,4}$. In each transform we pick up the residues of all poles defined by $\det[F^{-1}(P,L) + \mathcal M(P)]=0$. The neglected terms in which the weak current couples to either $A$ or $B^*$ will contain only one factor of $[F^{-1}(P,L) + \mathcal M(P)]^{-1}$. Thus although they contribute to one contour integral they do not contribute to the other and thus not to our final result. 

Using Eq.~(\ref{eq:2to2_corrdef}) we arrive at our final expression for the mixed-time-momentum correlator, in the absence of $\textbf{1} + \mathcal J \rightarrow \textbf{1}$ subprocesses
\begin{align}
C_L^{2\rightarrow 2,~\slash\hspace{-.2cm} \textbf{1B}}(x_{f,4}-y_4,y_4-x_{i,4}, \textbf P_i, \textbf P_f)
 & = L^3 \sum_{n_i,n_f} 
  e^{- E_{n_f}(x_{f,4}-y_4)}
  e^{- E_{n_i}(y_4-x_{i,4})}\nn\\
 &\times
A(E_{n_f}, \textbf P_f)~\mathcal R(E_{n_f}, \textbf P_f)
~
  \mathcal{W}^{~\slash\hspace{-.2cm} \textbf{1B}}(P_f,P_i)
~\mathcal R(E_{n_i}, \textbf P_i)
~
B^*(E_{n_i}, \textbf P_i).
\label{eq:2-to-2_no1b} 
\end{align}

We are now ready give an expression for $\Big [ \langle E_{n_f}, \textbf P_f, L \vert  \mathcal J(0)  \vert E_{n_i}, \textbf P_i, L \rangle \Big ]_L$. Equating Eqs.~(\ref{eq:3pt_completeset}) and (\ref{eq:2-to-2_no1b}) one finds
\begin{align}
\Big [ \langle E_{n_f}, \textbf P_f, L \vert  \mathcal J(0)  \vert E_{n_i}, \textbf P_i, L \rangle \Big ]_L 
= \frac{1}{L^6} \frac{A(E_{n_f}, \textbf P_f,)~\mathcal R(E_{n_f}, \textbf P_f)
~
  \mathcal{W}^{~\slash\hspace{-.2cm} \textbf{1B}}(P_f,P_i)
~\mathcal R(E_{n_i}, \textbf P_i)
~
B^*(E_{n_i}, \textbf P_i)}{\Big [ \langle 0 \vert  \mathcal A(0)  \vert E_{n_f}, \textbf P_f, L \rangle \Big ]_L \Big [ \langle E_{n_i}, \textbf P_i, L \vert \mathcal B^\dagger(0) \vert 0 \rangle \Big ]_L}.
\label{eq:2to2_no1b_master}
\end{align}
 Here we have used that the parametrically different time dependence allows one to match the coefficients term by term. We now stress an important point common to all analyses of this type. The momentum-space form of the correlator, Eq.~(\ref{eq:No1BodyCPP}), is only valid if $P_f$ and $P_i$ satisfy
\begin{equation}
- P^2 \equiv - P_4^2 - \textbf P^2 \equiv E^2 - \textbf P^2 \equiv E^{*2} < \Lambda^2 \,,
\end{equation}
where $\Lambda$ is the lowest lying three- or four-particle threshold not accounted for in our formalism.  For this reason, even though the expression contains an infinite tower of poles, the poles for which $- P^2 = E^{*2}> \Lambda^2$ suffer from neglected power-law corrections, due to on-shell multi-particle intermediate states. We can nevertheless formally perform the contour integral to reach Eq.~(\ref{eq:2-to-2_no1b}), but with the caveat that only the terms with $E_n$ satisfying the criterium above include all power-law finite-volume effects. Still we can unambiguously match these terms between Eqs.~(\ref{eq:3pt_completeset}) and (\ref{eq:2-to-2_no1b}). This leads to Eq.~(\ref{eq:2to2_no1b_master}) which is valid up to $e^{-m L}$ provided that $E_{n_i}^*, E_{n_f}^* < \Lambda$, where $m$ is the lightest particle mass in the spectrum.  
 
In order to simplify the right-hand side of this equation, we use an observation made in our previous work~\cite{Briceno:2014uqa}. The residue matrices, $\mathcal R$, have only one nonzero eigenvalue and can thus be written as an outer product 
\begin{align}
\mathcal R(E_{n_j }, \textbf P_j) 
&\equiv \lambda_j \mathbb{E}_{j} \mathbb{E}^\dag_{j},
\end{align}
where $\mathbb{E}_{j}$ is understood as a column vector in our combined angular-momentum and channel space.

We now apply this identity, first in the case where the initial- and final-channel spaces are the same and the incoming and outgoing states have the same energy and momentum. Then the denominator can be replaced using Eq.~(\ref{eq:firstmaster}), 
\begin{align}
\Big [ \langle E_{n}, \textbf P, L \vert  \mathcal J(0)  \vert E_{n}, \textbf P, L \rangle \Big ]_L 
&={ \frac{1}{L^3}}\frac{A(E_{n}, \textbf P)~\mathcal R(E_{n}, \textbf P)
~
  \mathcal{W}^{~\slash\hspace{-.2cm} \textbf{1B}}(P,P)
~\mathcal R(E_{n}, \textbf P)
~
B^*(E_{n}, \textbf P)}
{  A(E_{n}, \textbf P) \mathcal R(E_{n}, \textbf P)B^*(E_{n}, \textbf P)}\nn\\
&={ \frac{1}{L^3}} \lambda
{\mathbb E^\dag
~
  \mathcal{W}^{~\slash\hspace{-.2cm} \textbf{1B}}(P,P)\mathbb E}
\nn\\
&  ={ \frac{1}{L^3}}{\rm Tr}\left[
  \mathcal{W}^{~\slash\hspace{-.2cm} \textbf{1B}}(P,P)~\mathcal R(E_{n}, \textbf P)\right]. 
  \label{eq:2to2_no1B_degen}
\end{align}
If the initial- and final-channel spaces are distinct or if the current injects energy or momentum, we must multiply the denominator of Eq.~(\ref{eq:2to2_no1b_master}) with its complex conjugate to be able to use Eq.~(\ref{eq:firstmaster}). Following similar steps as above one finds 
\begin{align}
\Big | \langle E_{n_f}, \textbf P_f, L \vert  \mathcal J(0)  \vert E_{n_i}, \textbf P_i, L \rangle \Big |^2_L 
={ \frac{1}{L^6}}~{\rm{Tr}}\left[
 \mathcal R(E_{n_i}, \textbf P_i)
~  \mathcal{W}^{~\slash\hspace{-.2cm} \textbf{1B}}(P_i,P_f)
~\mathcal R(E_{n_f}, \textbf P_f)
 ~  \mathcal{W}^{~\slash\hspace{-.2cm} \textbf{1B}}(P_f,P_i)
\right].
  \label{eq:2to2_no1B_notdegen}
\end{align}
Of course these equations must be consistent when $E_{n_i}=E_{n_f}=E_{n}$,
\begin{align}
\Big | \langle E_{n}, \textbf P, L \vert  \mathcal J(0)  \vert E_{n}, \textbf P, L \rangle \Big |^2_L 
&={ \frac{1}{L^6}}~{\rm{Tr}}\left[
 \mathcal R(E_{n}, \textbf P)
~  \mathcal{W}^{~\slash\hspace{-.2cm} \textbf{1B}}(P,P)
~\mathcal R(E_{n}, \textbf P)
 ~  \mathcal{W}^{~\slash\hspace{-.2cm} \textbf{1B}}(P,P)
\right]\\
&={ \frac{1}{L^6}} \lambda^2~ \mathbb{E}^\dag
~  \mathcal{W}^{~\slash\hspace{-.2cm} \textbf{1B}}(P,P)
~\mathbb{E}\mathbb{E}^\dag
 ~  \mathcal{W}^{~\slash\hspace{-.2cm} \textbf{1B}}(P,P)\mathbb{E}
\nn\\
&=\left({ \frac{1}{L^3}}~{\rm{Tr}}\left[ \mathcal{W}^{~\slash\hspace{-.2cm} \textbf{1B}}(P,P)~\mathcal R(E_{n}, \textbf P)\right]\right)^2.
\end{align}
We have implicitly assumed equivalent channel spaces here by using the same $\mathbb E$ for the initial and final states.

Finally we comment that the absolute sign of matrix elements are \emph{not} physical observables, so the lack of sign information in Eq.~(\ref{eq:2to2_no1B_notdegen}) does not directly imply missing physical information. However, the relative sign between matrix elements is observable. To access this, we evaluate the matrix elements of two distinct currents $\mathcal J_{x}$ and $\mathcal J_{y}$ between the same initial and final states. This leads to two versions of Eq.~(\ref{eq:2to2_no1b_master}) with different transition amplitudes $\mathcal{W}_x$ and $\mathcal{W}_y$ on the right hand side. Taking the ratio of these two equalities we find [see also Ref.~\cite{Briceno:2015csa}]
\begin{align}
\frac{\Big [ \langle E_{n_f}, \textbf P_f, L \vert  \mathcal J_{x}(0)  \vert E_{n_i}, \textbf P_i, L \rangle \Big ]_L }
{\Big [ \langle E_{n_f}, \textbf P_f, L \vert  \mathcal J_{y}(0)  \vert E_{n_i}, \textbf P_i, L \rangle \Big ]_L }
&=
\frac{A(E_{n_f}, \textbf P_f)~\mathcal R(E_{n_f}, \textbf P_f)
~
  \mathcal{W}_x^{~\slash\hspace{-.2cm} \textbf{1B}}(P_f,P_i)
~\mathcal R(E_{n_i}, \textbf P_i)
~
B^*(E_{n_i}, \textbf P_i)}
{A(E_{n_f}, \textbf P_f)~\mathcal R(E_{n_f}, \textbf P_f)
~
  \mathcal{W}_y^{~\slash\hspace{-.2cm} \textbf{1B}}(P_f,P_i)
~\mathcal R(E_{n_i}, \textbf P_i)
~
B^*(E_{n_i}, \textbf P_i)}\nn\\
&=\frac{\chi_f^\dag~\mathcal R(E_{n_f}, \textbf P_f)
~
  \mathcal{W}_x^{~\slash\hspace{-.2cm} \textbf{1B}}(P_f,P_i)
~\mathcal R(E_{n_i}, \textbf P_i)~\chi_i^\dag}
{\chi_f^\dag~\mathcal R(E_{n_f}, \textbf P_f)
~
  \mathcal{W}_y^{~\slash\hspace{-.2cm} \textbf{1B}}(P_f,P_i)
~\mathcal R(E_{n_i}, \textbf P_i)
~\chi_i^\dag},
\label{eq:2to2_no1b_ratio}
\end{align}
where $\chi_i$ and $\chi_f$ are two generic vectors in our combined angular-momentum and channel space. These can be freely chosen at the user's convenience. 

We close this subsection by commenting that Eq.~(\ref{eq:2to2_no1B_notdegen}) closely resembles our $\textbf{1} + \mathcal J \rightarrow \textbf{2}$ result~\cite{Briceno:2014uqa}. One can in fact reproduce the $\textbf{1} + \mathcal J \rightarrow \textbf{2}$ result from Eq.~\cite{Briceno:2014uqa} by replacing $\mathcal R(E_{n_i}, \textbf P_i)$ with the appropriate one-particle propagator residue $1/2E_{n_i}$. In this limit, the residue becomes a one-dimensional matrix in angular momentum and channel space. Thus the trace above is converted to a product of a row-vector, a matrix, and a column vector, all defined in the combined angular momentum and channel space of the outgoing particle pair. In the next subsection we see that, in the presence of $\textbf{1} + \mathcal J \rightarrow \textbf{1}$ contributions, the expression for the two-body matrix element deviates substantially from that for the $\textbf{1} + \mathcal J \rightarrow \textbf{2}$ system.

\subsection{Three-point functions and matrix elements: (b) For general theories including $\textbf{1} + \mathcal J \rightarrow \textbf{1}$ contributions  \label{sec:1body}}

Having worked through the three-point function in the absence of a $\textbf{1} + \mathcal J \rightarrow \textbf{1}$ subprocesses, we now proceed to determine the more complicated and realistic scenario. As discussed extensively in Secs.~\ref{sec:trans_amps} and \ref{sec:1body_insertion}, this case is complicated by the appearance of singularities in the infinite-volume transition amplitude and by new finite-volume functions. The important distinction between the full three-point correlation function, Fig.~\ref{fig:C2-to-2}, and the simplified version without a $\textbf{1} + \mathcal J \rightarrow \textbf{1}$ amplitude, Fig.~\ref{fig:C2-to-2_no1B}, is the presence of finite-volume two-particle loops with the current coupling to one of the particles in the loop. This is depicted in Fig.~\ref{fig:1body_insertion} and the separation of finite-volume effects for these sections of diagrams is given by Eq.~(\ref{eq:GL_final}). The task of this section is to break all of the diagrams of Fig.~\ref{fig:C2-to-2} into finite- and infinite-volume parts and then to sum the terms into a useful expression. To achieve this we must use Eq.~(\ref{eq:GL_final}) for the two-particle loops with the weak insertion and must dress this expression on both sides by a series of finite-volume two-particle loops scattered by Bethe-Salpeter kernels. This same series also dresses the weak kernel as discussed in the previous section.  

In the analysis of the previous subsection, we argued that the only diagrams with poles in both $E_i$ and $E_f$ are those with at least one factor each of $F(P_i,L)$ and $F(P_f,L)$. In the present case, however, other types of poles arise due to the presence of the current and the corresponding finite-volume function, $G(L)$. For example, the sum of all terms with no insertions of $F(P_i,L)$ and $F(P_f,L)$ and exactly one insertion of $G(L)$ gives
\begin{equation}
C^{2\rightarrow 2}_{L}(P_i,P_f)  = A(P_f) \ [G  \cdot \w] \   B^*(P_i) +   \cdots \,.
  \end{equation}
Note that this term has poles in both $E_i$ and $E_f$ at the energies of two free particles in finite volume. If this term is Fourier transformed in isolation it will give Euclidean-time exponentials which decay according to these free-particle energies. As we see below, these poles cancel against poles in the terms not yet considered. 
 
 We now combine this with the set of all terms which have some number of insertions of either $F(P_i,L)$ or $F(P_f,L)$ but not both. These sum to give
 \begin{align}
 C^{2\rightarrow 2}_{L }(P_i,P_f)  & =  C^{2\rightarrow 2}_{L, \mathrm{FP}}(P_i,P_f)+   \cdots  \\[5pt]
 C^{2\rightarrow 2}_{L, \mathrm{FP}}(P_i,P_f)  & =  A(P_f) \ [G  \cdot \w] \   B^*(P_i)  - A(P_f) \ [G  \cdot \w]   \ \mathcal M(P_i) 
 \frac{1}{F^{-1}(P_i,L) + \mathcal M(P_i)}
    B^*(P_i) \\ & \hspace{150pt} - A(P_f)
     \frac{1}{F^{-1}(P_f,L) + \mathcal M(P_f)}
    \mathcal M(P_f) \  [G  \cdot \w]   \   B^*(P_i)    \,,
      \end{align}
where the subscript FP stands for free poles. Here the first term has free particle poles in both $E_i$ and $E_f$, the second has interacting and free poles in $E_i$ and free poles in $E_f$, respectively, and the third is as the second but with $E_i$ and $E_f$ exchanged. Thus the Fourier transform of all three terms gives unphysical time dependence. This will be cancelled by the final set of important terms, to which we now turn.

We now include those terms which have at least one insertion of both $F(P_i,L)$ and $F(P_f,L)$. Focusing first on those which have exactly one factor of each, we find that four types of terms can appear between the two $F$ factors
\begin{enumerate}
\item terms described by infinite-volume diagrams where the $\textbf 1 + \mathcal J \rightarrow \textbf 1$ transition amplitude is inserted between two Bethe-Salpeter kernels in an integrated two-particle loop,
\item terms described by infinite-volume diagrams which include the weak current via a weak Bethe-Salpeter kernel, inserted in some chain of strong-interaction Bethe-Salpeter kernels,
\item terms in which a factor of $G(L)$ separates the initial and final states,
\item terms described by infinite-volume diagrams where the $\textbf 1 + \mathcal J \rightarrow \textbf 1$ transition is directly adjacent to one of the $F$ insertions.
\end{enumerate}

Looking to Eq.~(\ref{eq:GL_final}) above, we see that this final class of terms necessarily contains an insertion of $\delta_{\mathrm{df}}$. Recall that this denotes a subtraction of the long distance poles that we have discussed throughout. This is shown explicitly in Eqs.~(\ref{eq:deltaLdef}) and (\ref{eq:deltaRdef}) above. Thinking of $\delta_{\mathrm{df}}$ as an operator which encodes the instruction to remove this on-shell divergence, it is convenient to extend the definition to act as the identity on any diagram that does not contain a current coupling to an external leg. Then the result for all terms with one factor each of $F(P_i,L)$ and $F(P_f,L)$ can be written
\begin{align}
C^{2\rightarrow 2}_{L}(P_i,P_f) & = A(P_f) [-F(P_f,L) ]
 \Big  ( \delta_{\mathrm{df}} \mathcal W(P_f,P_i)  \delta_{\mathrm{df}} +  \mathcal  M(P_f) \ [G  \cdot \w]  \ \mathcal M(P_i) 
 \Big ) [-F(P_i,L)]
  B^*(P_i)+\cdots \,, \\[5pt]
  & =   A(P_f) [-F(P_f,L) ] \Wtildf(P_f,P_i,L) [-F(P_i,L)]
  B^*(P_i)+\cdots \,,
  \label{eq:CLFfFi}
\end{align}
where
\begin{align}
\label{eq:WLdef}
\Wtildf(P_f,P_i,L) & \equiv \Wdf(P_f,P_i) +  \mathcal  M(P_f) \ [G(L) \cdot \w](P_f,P_i) \ \mathcal M(P_i) \,, \\[10pt]
\mathcal W_{\mathrm{df};ab;\ell' m'; \ell m}(P_f,P_i)  & \equiv  \big [\delta_{\mathrm{df}} \mathcal W_{ab}(P_f,P_i)  \delta_{\mathrm{df}}  \big ]_{\ell' m'; \ell m} \,.
\label{eq:Wdfdef}
\end{align}
We have left the indices implicit on all terms in Eq.~(\ref{eq:WLdef}).
 
 The definition of $\Wdf$ in terms of the $\delta_{\mathrm{df}}$ operator is very compact, so we now take some time to explain this quantity in detail by relating it to the standard $\textbf 2 + \mathcal J \rightarrow \textbf 2$ transition amplitude, $\mathcal W$. The first step is to contract with spherical harmonics
 \begin{equation}
 \label{eq:Wdfdef2}
  \mathcal W_{\mathrm{df};ab}(P_f,p,P_i,k) \equiv 4 \pi Y^*_{\ell' m'}(\hat {\textbf p}^*_{af}) \mathcal W_{\mathrm{df};ab;\ell' m'; \ell m}(P_f,P_i) Y_{\ell, m}(\hat {\textbf k}^*_{bi}) \,.
 \end{equation}   
Note that we have defined the quantity on the left-hand side with all vectors in the finite-volume frame. As is apparent from the expression on the right-hand side, all vectors are on-shell, meaning that the true degrees of freedom are only $E_f^*, E_i^*, \hat {\textbf p}^*_{af}$ and $\hat {\textbf k}^*_{bi}$. We next add back in the long distance poles to reach the standard transition amplitude
\begin{align}
\label{eq:Wdfdef3}
\mathcal W_{ab}(P_f,p,P_i,k) &  =\mathcal W_{\mathrm{df};ab}(P_f,p,P_i,k)\\
& \hspace{-90pt} - \xi_{a'} 4 \pi Y^*_{\ell' m'}(\hat {\textbf p}^*_{af}) \mathcal M_{aa';\ell' m';\ell_f' m_f'}(P_f)  \left [\frac{k^{*}_{a'f}}{q^*_{a'f}}\right ]^{\ell_f'}  
\frac{4 \pi  Y_{\ell_f' m_{f}'}(\hat {\textbf k}^*_{a'f})
Y^*_{\ell_f m_{f}}(\hat {\textbf k}^*_{a'f}) }{2 \omega_{a'sf} (E_f - \omega_{a'\slashed s} -  \omega_{a'sf}  + i \epsilon ) }
  \left [\frac{k^{*}_{a'f}}{q^*_{a'f}}\right ]^{\ell_f}   w_{a'sb2;\ell_f m_f; \ell m}(P_f,P_i)  Y_{\ell m}(\hat {\textbf k}^*_{bi}) \nn \\
& \hspace{-90pt} - \xi_{a'} 4 \pi  Y^*_{\ell' m'}(\hat {\textbf p}^*_{af}) \mathcal M_{aa';\ell' m';\ell_f' m_f'}(P_f)  \left [\frac{k^{*}_{a'f}}{q^*_{a'f}}\right ]^{\ell_f'} 
\frac{ 4 \pi  Y_{\ell_f' m_{f}'}(\hat {{\textbf k}}^*_{a'f})Y^*_{\ell_f m_{f}}(\hat {\overline{\textbf k}}^*_{a'f}) }
{ 2\overline{\omega}_{a'sf}(E_f - \overline{\omega}_{a'sf} -  \overline{\omega}_{a'\slashed s}  + i \epsilon ) } 
 \left [\frac{\overline k^{*}_{a'f}}{q^*_{a'f}}\right ]^{\ell_f}  w_{a'sb1;\ell_f m_f; \ell m}(P_f,P_i)  Y_{\ell m}(-\hat {\textbf k}^*_{bi}) \nn \\
& \hspace{-90pt} - \xi_{b'} 4 \pi  Y^*_{\ell' m'}(\hat {\textbf p}^*_{af}) w_{a2b't; \ell' m'; \ell_i' m_i'}(P_f,P_i) \left[\frac{p^*_{b'i}}{q^*_{b'i}} \right ]^{\ell_i'} 
\frac{ 4 \pi Y^*_{\ell_i' m_i'}(\hat {\textbf p}^*_{b'i}) Y_{\ell_i m_i}(\hat {\textbf p}^*_{b'i}) }
{2 \omega_{b'ti} (E_i - \omega_{b'\slashed t}-  \omega_{b'ti}  + i \epsilon )} 
 \left[\frac{p^*_{b'i}}{q^*_{b'i}} \right ]^{\ell_i} \mathcal M_{b'b; \ell_i m_i; \ell, m}(P_i) Y_{\ell m}(\hat {\textbf k}^*_{bi}) \nn \\
& \hspace{-90pt} - \xi_{b'} 4 \pi  Y^*_{\ell' m'}(- \hat {\textbf p}^*_{af}) w_{a1b't; \ell' m'; \ell_i' m_i'}(P_f,P_i) \left[\frac{\overline p^*_{b'i}}{q^*_{b'i}} \right ]^{\ell_i'} 
\frac{4 \pi Y^*_{\ell_i' m_i'}(\hat {\overline {\textbf p}}^*_{b'i}) Y_{\ell_i m_i}(\hat {\textbf p}^*_{b'i}) }
{ 2 \overline \omega_{b'ti}(E_i- \overline \omega_{b'ti}  -  \overline \omega_{b'\slashed t} + i \epsilon )}
  \left[\frac{p^*_{b'i}}{q^*_{b'i}} \right ]^{\ell_i} \mathcal M_{b'b; \ell_i m_i; \ell, m}(P_i) Y_{\ell m}(\hat {\textbf k}^*_{bi}) \nn   \,,
\end{align}
where
\begin{alignat}{3}
\omega_{a'2f} & \equiv \sqrt{(\textbf P_f - \textbf k)^2 + m_{a'2}^2} \,,  
& \omega_{a'1}  & \equiv \sqrt{ \textbf k^2 + m_{a'1}^2} \,, \\
\overline \omega_{a'2}  &\equiv \sqrt{(\textbf P_i - \textbf k)^2 + m_{a'2}^2} \,,\ \ \ \ \  &  \overline \omega_{a'1f}&  \equiv \sqrt{(\textbf P_f - \textbf P_i + \textbf k)^2 + m_{a'1}^2} \,, \\
\omega_{b'2i}  & \equiv \sqrt{(\textbf P_i-\textbf p)^2 + m_{b'2}^2} \,, & \omega_{b'1} &  \equiv \sqrt{ \textbf p^2 + m_{b'1}^2} \,, \\
\overline \omega_{b'2} & \equiv  \sqrt{(\textbf P_f - \textbf p)^2 + m_{b'2}^2} \,, & \overline \omega_{b'1i} & \equiv \sqrt{(\textbf P_i - \textbf P_f + \textbf p)^2 + m_{b'1}^2} \,.
 \end{alignat}
Note that the bars over omegas denote exchanging $\textbf k \rightarrow \textbf P_f- \textbf P_i + \textbf k$ or $\textbf p \rightarrow \textbf P_i-\textbf P_f + \textbf p$. This notation is required to denote the separate terms arising from the current attaching to each external leg. These definitions are closely related to those of Eqs.~(\ref{eq:firstroundomegas1}) and (\ref{eq:firstroundomegas2}) above, but here with $\textbf p$ in place of $\textbf k$ in certain cases. 

Here we have also introduced various starred momenta $k^*$, $p^*$ and $q^*$, with various subscripts and other decorations. Some of these quantities have been introduced above, but we review the entire set here. We first recall that $q^*_{a'f}$ is the magnitude of CM frame momentum for one of the particles with masses $m_{a'1}$ and $m_{a'2}$ and total four-momentum $P_f$ [see also Eq.~(\ref{eq:qstardef})]. This is distinct from $k^*_{a'f}$, which is the magnitude of the spatial part of $(\omega_{a'1f}^*, \textbf k^*_{a'f})$, given by boosting $(\omega_{a'1f},\textbf P_f- \textbf k)$ with boost velocity $- \textbf P_f/E_f$. The direction of $\textbf k^*_{a'f}$ also appears in the second and third lines of Eq.~(\ref{eq:Wdfdef3}), inside some of the spherical harmonics. We stress that both incoming mesons in channel $b$, with momenta $\textbf k$ and $\textbf P_i - \textbf k$, are on-shell. This means that if we boost these with $- \textbf P_i/E_i$ then the magnitude of each particle's spatial momenta is $k^*_{bi} = q^*_{bi}$. This is a constraint on $\textbf k$ that must be satisfied in Eq.~(\ref{eq:Wdfdef3}). However in the discussion of $k^*_{a'f}$ and $q^*_{a'f}$ we are using different masses (those of channel $a'$ instead of $b$) and a different boost ($-\textbf P_f/E_f$ instead of $- \textbf P_i/E_i$). For this reason, generally $k^*_{a'f} \neq q^*_{a'f}$. The two coincide only when the pole in the second line of Eq.~(\ref{eq:Wdfdef3}) diverges. We have also introduced $\overline k^*_{a'f}$ and $\hat{\overline{\textbf k}}^*_{a'f}$. As with the barred omegas, the bars here indicate that $\textbf k$ is to be exchanged with $\textbf P_f- \textbf P_i + \textbf k$. These new quantities are thus the magnitude and direction, respectively, of $(\omega_{a'2f}^*,\overline {\textbf k}^*_{a'f})$, given by boosting $(\omega_{a'2f}, \textbf P_f- \textbf P_i + \textbf k)$ with boost velocity $- \textbf P_f/E_f$. At this stage we have completely specified all momenta in the second and third lines of Eq.~(\ref{eq:Wdfdef3}). The definitions in the remaining lines are the same, but with $b'$ in place of $a'$, $\textbf p$ in place of $\textbf k$ and $i$ and $f$ everywhere switched.

In Eq.~(\ref{eq:Wdfdef3}), sums over the intermediate channels, $a'$ and $b'$, as well as the particles in the primed channel, $s$ and $t$, are understood. We recall that $w_{asbt}$ is defined for all channels but must vanish if the channels do not contain a common particle, or if the $\textbf 1 + \mathcal J \rightarrow \textbf 1$ transition does not couple the channels. Given this convention, the form of Eq.~(\ref{eq:Wdfdef3}) is valid for all types of channels, for identical and non-identical particles. In the case of identical particles, the two one-body currents $w_{a1b1}$ and $w_{a2b2}$ are identical functions, but both terms must be included since the external particles carry distinct momenta. However the sum over $s$ and $t$ still counts each of these contributions twice and for this reason the symmetry factors must be included to remove the redundancy. It is unfortunate that the definition takes such a complicated form, given that the basic idea [shown in Fig.~\ref{fig:divergence_free}] is straightforward. The main sources of complication are the two-different frames and the need to include ratios of $k^*/q^*$, to avoid spurious singularities near $\textbf k^* = 0$. 

The quantities defined in Eqs.~(\ref{eq:WLdef}) and (\ref{eq:Wdfdef}) are central to the main result of this paper. The first of these, $\Wtildf$, can be directly extracted from finite-volume matrix elements using Eq.~(\ref{eq:2to2_notdegen}) below. To convert this to the physical, infinite-volume, two-to-two transition amplitude, $\mathcal W$, two steps are needed. First one uses Eq.~(\ref{eq:Gdotw}) and (\ref{eq:WLdef}) to go from $\Wtildf$ to the divergence-free infinite-volume quantity $\Wdf$. This requires evaluating $G(L)$, as outlined in Appendix \ref{app:Gnum}, and combining this with on-shell values of $\mathcal M$ and $\w$. Finally to go from $\Wdf$ to the physical observable, $\mathcal W$, one must add back in the poles as dictated by Eq.~(\ref{eq:Wdfdef}). As with the evaluation of the $G(L)$-dependent term, this requires knowledge of on-shell $\mathcal M$ and $\w$. Together with Eq.~(\ref{eq:2to2_notdegen}) below, this prescription represents a model-independent, relativistic-field-theory approach for determining $\mathcal W$ from finite-volume observables.

To complete our calculation of $C^{2\rightarrow 2}_{L}(P_i,P_f)$ we must now include all terms which contain any number of factors of $F(P_i,L)$ and $F(P_f,L)$. Given Eq.~(\ref{eq:CLFfFi}), this modification is trivially implemented in analogy to the case of the previous subsection. Combining terms we reach our final result for the momentum-space, finite-volume correlator
\begin{align}
\label{eq:CLtwoterms}
C^{2\rightarrow 2}_{L}(P_i,P_f) 
& =
 C^{2\rightarrow 2}_{L,\mathrm{FP}}(P_i,P_f) + C^{2\rightarrow 2}_{L, \mathrm{IP}}(P_i,P_f)  + \cdots \,, \\
 C^{2\rightarrow 2}_{L, \mathrm{IP}}(P_i,P_f)  & =  A(P_f) 
 \frac{1}{F^{-1}(P_f,L) + \mathcal M(P_f)}
 \Wtildf (P_f,P_i)
 \frac{1}{F^{-1}(P_i,L) + \mathcal M(P_i)}
  B^*(P_i) \,,
  \end{align}
where the subscript IP stands for interacting poles. Here the ellipses denotes contributions that have no poles in either $E_i$ or $E_f$ (or both) and thus do not contribute to the Fourier transform that we perform in the next step.
  
We now argue that only the poles from $[F^{-1}(P_i,L) + \mathcal M(P_i)]^{-1}$ and $[F^{-1}(P_f,L) + \mathcal M(P_f)]^{-1}$ inside $C^{2\rightarrow 2}_{L, \mathrm{IP}}(P_i,P_f)$ contribute in the Fourier transform. This is because all free-particle poles cancel between the two terms in Eq.~(\ref{eq:CLtwoterms}). For example if both $E_i$ and $E_f$ are near free-particle poles then 
\begin{align}
\begin{split}
 C^{2\rightarrow 2}_{L, \mathrm{FP}}(P_i,P_f) &  \longrightarrow  A(P_f) \ [G  \cdot \w] \   B^*(P_i)  - A(P_f) \ [G  \cdot \w]   \ \mathcal M(P_i) 
 \frac{1}{ \mathcal M(P_i)}
    B^*(P_i)  \\ & \hspace{200pt}- A(P_f)
     \frac{1}{ \mathcal M(P_f)}
    \mathcal M(P_f) \  [G  \cdot \w]   \   B^*(P_i)    \,, 
    \end{split} \\
    &  \longrightarrow  - A(P_f) \ [G  \cdot \w] \   B^*(P_i)  \,.
\end{align}
and
\begin{equation}
C^{2\rightarrow 2}_{L, \mathrm{IP}}(P_i,P_f) \longrightarrow A(P_f) 
 \frac{1}{  \mathcal M(P_f)}
 \Wtildf (P_f,P_i)
 \frac{1}{  \mathcal M(P_i)}
  B^*(P_i) \rightarrow A(P_f) \ [G  \cdot \w] \   B^*(P_i) \,,
\end{equation}
resulting in perfect cancellation between the terms. Similar cancellations occur if one of either $E_i$ or $E_f$ is near a free pole and the other near an interacting pole. 

We deduce that the Fourier transform of Eq.~(\ref{eq:CLtwoterms}) is given by summing over the residues from the poles of $[F^{-1}(P_i,L) + \mathcal M(P_i)]^{-1}$ and $[F^{-1}(P_f,L) + \mathcal M(P_f)]^{-1}$ inside of $C^{2\rightarrow 2}_{L, \mathrm{IP}}(P_i,P_f)$ only. This is exactly the prescription used in the Fourier transform of the previous subsection where $\mathcal W^{~\slash\hspace{-.2cm} \textbf{1B}}$ has no poles and the full contribution with $E_i$ and $E_f$ poles has the form of $C^{2\rightarrow 2}_{L, \mathrm{IP}}(P_i,P_f)$. It follows that all of the Fourier transformed results from the previous subsection [Eq.~(\ref{eq:2-to-2_no1b}) on] can be used here with the simple modification $\mathcal W^{~\slash\hspace{-.2cm} \textbf{1B}}  \longrightarrow \Wtildf$. For example from Eq.~(\ref{eq:2to2_no1b_master}) we obtain the master equation for two-body matrix elements
\begin{align}
\Big [ \langle E_{n_f}, \textbf P_f, L \vert  \mathcal J(0)  \vert E_{n_i}, \textbf P_i, L \rangle \Big ]_L 
=\frac{1}{L^6}\frac{A(E_{n_f}, \textbf P_f)~\mathcal R(E_{n_f}, \textbf P_f)
~
\Wtildf(P_f,P_i,L) 
~\mathcal R(E_{n_i}, \textbf P_i)
~
B^*(E_{n_i}, \textbf P_i)}{\Big [ \langle 0 \vert  \mathcal A(0)  \vert E_{n_f}, \textbf P_f, L \rangle \Big ]_L \Big [ \langle E_{n_i}, \textbf P_i, L \vert \mathcal B^\dagger(0) \vert 0 \rangle \Big ]_L}.
\label{eq:2to2_master}
\end{align}
Following the steps taken in deriving Eqs.~(\ref{eq:2to2_degen}) and (\ref{eq:2to2_notdegen}) this can be used to derive the relation between the finite-volume matrix elements of an external current and $\Wtildf$. In the case of equivalent in and out channel spaces, with no energy or momentum inserted by the current, we find
\begin{align}
\Big [ \langle E_{n}, \textbf P, L \vert  \mathcal J(0)  \vert E_{n}, \textbf P, L \rangle \Big ]_L 
  =\frac{1}{L^3}{\rm Tr}\left[
\Wtildf(P,P,L)~\mathcal R(E_{n}, \textbf P)\right] \,.
  \label{eq:2to2_degen}
\end{align}
In the case of non-equivalent states we reach
 \begin{align}
\Big | \langle E_{n_f}, \textbf P_f, L \vert  \mathcal J(0)  \vert E_{n_i}, \textbf P_i, L \rangle \Big |^2_L 
=\frac{1}{L^6}~{\rm{Tr}}\left[
 \mathcal R(E_{n_i}, \textbf P_i)
~  
\Wtildf(P_i,P_f,L)
~\mathcal R(E_{n_f}, \textbf P_f)
 ~  
 \Wtildf(P_f,P_i,L)
\right].
  \label{eq:2to2_notdegen}
\end{align}
Finally we find the ratio of matrix elements of two currents satisfies,
\begin{align}
\frac{\Big [ \langle E_{n_f}, \textbf P_f, L \vert  \mathcal J_{x}(0)  \vert E_{n_i}, \textbf P_i, L \rangle \Big ]_L }
{\Big [ \langle E_{n_f}, \textbf P_f, L \vert  \mathcal J_{y}(0)  \vert E_{n_i}, \textbf P_i, L \rangle \Big ]_L }
&=\frac{\chi_f^\dag~\mathcal R(E_{n_f}, \textbf P_f)
~
\mathcal W_{L,\mathrm{df},x}(P_f,P_i,L)
~\mathcal R(E_{n_i}, \textbf P_i)~\chi_i^\dag}
{\chi_f^\dag~\mathcal R(E_{n_f}, \textbf P_f)
~
\mathcal W_{L,\mathrm{df},y}(P_f,P_i,L)
~\mathcal R(E_{n_i}, \textbf P_i)
~\chi_i^\dag},
\label{eq:2to2_ratio}
\end{align}
where, as above, $\chi_i$ and $\chi_f$ are general vectors in the space of $\mathcal{R}$. 

Unlike the result in the absence of $\textbf{1} + \mathcal J \rightarrow \textbf{1}$, Eq.~(\ref{eq:2to2_notdegen}) no longer resembles the $\textbf{1} + \mathcal J \rightarrow \textbf{2}$ result of Ref.~\cite{Briceno:2014uqa}. The nonzero value of $\w$ leads to the definition of a new object, $\Wtildf$, which includes the desired infinite-volume quantity $\Wdf$ as well as finite-volume effects. One can nonetheless recover the $\textbf{1} + \mathcal J \rightarrow \textbf{2}$ result from Eq.~(\ref{eq:2to2_notdegen}), by first setting $\w=0$ and then taking the steps discussed in the last paragraph of the previous subsection.

Finally, we reemphesize that the matrices appearing on the right-hand side of Eqs.~(\ref{eq:2to2_master})-(\ref{eq:2to2_ratio}) are formally infinite dimensional. To apply this result in the analysis of a LQCD calculation, it is necessary to truncate these to a finite subspace. This is justified at low-energies where the contributions from higher angular-momentum states are suppressed. More precisely $w$, $\mathcal M$, and $\Wdf$ are all smooth functions, which should induce a uniformly convergent partial wave expansion. As mentioned above, truncating an expansion of $\mathcal W$ would not be justified due to long distance singularities. We discuss this truncation and other simplifying limits in the next section.

\section{Simplifying Limits}

\label{sec:simplim}

In this section we consider various simplifying limits of the general result, derived in the last section. We begin by taking the energies considered to be very close to the lowest two-particle threshold.   In this case, the infinite-volume quantities $\w$, $\mathcal M$ and $\Wdf$ are all dominated by their S-wave values. We thus drop all higher partial waves in the matrices $w_{a1b1; \ell' m'; \ell  m }(P_f,P_i)$, $\mathcal M_{ab; \ell' m'; \ell m}(P)$ and $\mathcal W_{\mathrm{df};ab;\ell' m'; \ell m}(P_f,P_i)$. The second consequence of near-threshold energies is that only the lowest two-particle channel is open. In discussing this system it is convenient to introduce the shorthand
\begin{align}
w_{11}(P_f,P_i) & \equiv w_{a1b1; 00;00 }(P_f,P_i) \,, \\
\mathcal M (P) & \equiv \mathcal M_{ab; 00; 00}(P)\,, \\
\mathcal W_{\mathrm{df}}(P_f,P_i) & \equiv \mathcal W_{\mathrm{df};ab;00;00}(P_f,P_i) \,.
\end{align}
We comment here that, for a scalar form factor, symmetry and on-shell constraints guarantee that $w$ only depends on $(P_f-P_i)^2$ and thus not on $\textbf k$. In this case, the truncation of $w$ to the S-wave is exact.  
 Since all matrices have reduced to one dimensional, the trace may be dropped from Eq.~(\ref{eq:2to2_notdegen})
\begin{equation}
\label{eq:mainresSwave}
\Big | \langle E_{n_f}, \textbf P_f, L \vert  \mathcal J(0)  \vert E_{n_i}, \textbf P_i, L \rangle \Big |^2_L 
=\frac{1}{L^6}
 \mathcal R(E_{n_i}, \textbf P_i)
\Wtildf(P_i,P_f,L) \mathcal R(E_{n_f}, \textbf P_f) 
 \Wtildf(P_f,P_i,L) \,.
\end{equation}
 In addition, the residue matrix $\mathcal R$ simplifies significantly
\begin{align}
\mathcal R(E_{n}, \textbf P) & =   \left [ \frac{\partial}{\partial E} \left( F^{-1}(P,L) + \mathcal M(P) \right ) \right ]_{E=E_n}^{-1} = -  \left [ \mathcal M^2(P) \frac{\partial}{\partial E} \left( F(P,L) + \mathcal M^{-1}(P) \right ) \right ]_{E=E_n}^{-1} \,, \\
& = - \xi \frac{q^*}{8 \pi E^*}   \left [ \sin^2\! \delta\ e^{ 2i \delta}\  \frac{\partial}{\partial E} \left( \cot\phi^\textbf{d}_{}+ \cot\delta \right ) \right ]_{E=E_n}^{-1}    \,, \\
& = \xi \frac{q^*}{8 \pi E^*} e^{-2i \delta}  \left [  \frac{\partial}{\partial E} \left( \phi^\textbf{d}_{} + \delta \right ) \right ]_{E=E_n}^{-1}    \,,
 \end{align}
where $F = F_{a00;b00}$ is understood and where we have introduced the S-wave L\"uscher pseudophase
 \begin{equation}
 \cot \phi^{\textbf d} = \xi \frac{q^*}{8 \pi E^*} \mathrm{Re}F(P,L) \,.
 \end{equation}
Here we have also used the relation between scattering amplitude $\mathcal M$ and scattering phase shift $\delta$, given in Eq.~(\ref{eq:scatamp}) above. Substituting this result for $\mathcal R$ into Eq.~(\ref{eq:mainresSwave}) and rearranging gives
 \begin{multline}
 \label{eq:swaveWtoFV}
\Big [ e^{-i \delta_i}
\Wtildf(P_i,P_f,L) e^{- i \delta_f} \Big ] \Big [ e^{- i \delta_f}  \Wtildf(P_f,P_i,L) e^{- i \delta_i} \Big ]
\\[5pt] = \frac{8 \pi E^*_f}{q^*_f \xi} \frac{8 \pi E^*_i}{q^*_i \xi}  \left [  \frac{\partial}{\partial E_f} \left( \phi^\textbf{d} + \delta \right ) \right ]_{E_f=E_{f,n}} \left [  \frac{\partial}{\partial E_i} \left( \phi^\textbf{d} + \delta \right ) \right ]_{E_i=E_{i,n}} L^6 \Big | \langle E_{n_f}, \textbf P_f, L \vert  \mathcal J(0)  \vert E_{n_i}, \textbf P_i, L \rangle \Big |^2_L 
 \,.
\end{multline}
We thus see that a naive Lellouch-L\"uscher-like proportionality factor arises between the finite- and infinite-volume quantities. Since the right-hand side of this expression is manifestly pure real, this result also suggest a Watson-like theorem for $\Wtildf$, namely that its complex phases are the strong scattering phases associated with the incoming and outgoing two-particle states.

Finally the relations between $\Wtildf$, $\Wdf$ and $\mathcal W$ reduce to
 \begin{align}
\begin{split}
\mathcal W_{}(P_f, p,P_i, k) &  =\Wtildf - \xi \left[\frac{1}{L^3}\sum_{\mathbf{k'}}\hspace{-.5cm}\int~\right]  \frac{ \mathcal M(P_f)  w_{22}(P_f,P_i)  \mathcal M(P_i) }{2\omega_{1}' 2 \omega_{{2f}}' 2 \omega_{2i}' (E_f -  \omega_{1}' - \omega_{2f}' + i \epsilon )(E_i -  \omega_{1}' - \omega_{2i}' + i \epsilon )}  \\
&  - \xi \left[\frac{1}{L^3}\sum_{\mathbf{k'}}\hspace{-.5cm}\int~\right] \frac{  \mathcal M(P_f)    w_{11}(P_f,P_i)  \mathcal M(P_i) }{2\omega_{2}' 2 \omega_{{1f}}' 2 \omega_{1i}' (E_f -  \omega_{1f}' - \omega_{2}' + i \epsilon )(E_i -  \omega_{1i}' - \omega_{2}' + i \epsilon )}   \\
& -  \mathcal M_{ }(P_f)  \left [
\frac{w_{11}(P_f,P_i)}{2 \omega_{1f} (E_f -  \omega_{1f} - \omega_{2} + i \epsilon ) }
    +  
\frac{w_{22}(P_f,P_i)}
{ 2\overline{\omega}_{2f}(E_f -  \overline{\omega}_{1} - \overline{\omega}_{2f} + i \epsilon ) }  \right ] 
    \\
&   -   \left [    
\frac{w_{11}(P_f,P_i) }
{2 \omega_{1i} (E_i -  \omega_{1i} - \omega_{2} + i \epsilon )}  + 
\frac{w_{22}(P_f,P_i) }{ 2 \overline \omega_{2i}(E_i -  \overline \omega_{1} - \overline \omega_{2i} + i \epsilon )} \right ]
   \mathcal M_{}(P_i)   \,,
   \end{split}
   \label{eq:swaveWresult}
\end{align}
where $\xi$ is required to avoid double counting in the case of identical particles. The top two lines here give the expression for $\Wdf$ in terms of $\Wtildf$ and the reduced form of $\mathcal M [w \cdot G] \mathcal M$. In comparison to our general result, this gives a relatively simple prescription for accessing the physical observable, $\mathcal W$. We stress here that the result does not imply finite-volume poles in $\mathcal W$. The relation is only valid at the energies of the interacting spectrum, which generally differ from those of the free theory.

We emphasize also that the S-wave-only approximation has not been applied directly to $\mathcal W$ and that doing so would not make sense. The poles in Eq.~(\ref{eq:swaveWresult}) still depend on directional degrees of freedom, so that the full $\textbf 2 + \mathcal J \rightarrow \textbf 2$ transition amplitude receives contributions from all angular momenta. This is expected, since the long distance parts guarantee that all partial waves give important contributions, even arbitrarily close to the lowest threshold. By working with a truncation only on $w$, $\mathcal M$ and $\Wdf$ we have reached a solvable system, without requiring the ill-motivated truncation of $\mathcal W$ directly.


Next, it is instructive to take the non-interacting limit on our truncated result, Eqs.~(\ref{eq:swaveWtoFV}) and (\ref{eq:swaveWresult}). Here we first turn to the case where the $\textbf 1 + \mathcal J \rightarrow \textbf 1$ transition is absent, discussed in Subsec.~\ref{sec:no1body}. This special case can be reached from Eqs.~(\ref{eq:swaveWtoFV}) and (\ref{eq:swaveWresult}) by setting $w=0$. If we do so, and additionally take the strong interaction to vanish completely, then our result reduces to
\begin{align}
\mathcal W^{~\slash\hspace{-.2cm} \textbf{1B}}(P_{n_f}, p,P_{n_i}, k)^2   = \frac{8 \pi E^*_{n_f}}{q^*_{n_f} \xi} \frac{8 \pi E^*_{n_i}}{q^*_{n_i} \xi}  \left [  \frac{\partial}{\partial E_{f}}  \phi^\textbf{d}  \right ]_{E_f=E_{n_f}} \left [  \frac{\partial}{\partial E_i}  \phi^\textbf{d} \right ]_{E_i=E_{n_i}} L^6 \Big | \langle E_{n_f}, \textbf P_f, L \vert  \mathcal J(0)  \vert E_{n_i}, \textbf P_i, L \rangle \Big |^2_L \,.
\end{align}
We next substitute
\begin{equation}
\label{eq:freeF}
\frac{8 \pi E^*_n}{q^*_n \xi}  \left [  \frac{\partial}{\partial E}  \phi^\textbf{d}  \right ]_{E=E_{n}}  = \left[ \frac{\partial}{\partial E}  ( \mathrm{Re}\,F)^{-1}    \right ]_{E=E_{n}} = \frac{2 \omega_1 2 \omega_{2} L^3}{ \nu_{n}} \,,
\end{equation}
and also substitute the matrix element definition for $\mathcal W^{~\slash\hspace{-.2cm} \textbf{1B}}$, Eq.~(\ref{eq:bigWmedef}) above, to reach
\begin{equation}
\Big \vert \langle P_f, p, \mathrm{out} \vert \mathcal J(0) \vert P_i, k, \mathrm{in} \rangle \Big \vert = \sqrt{\frac{2 \omega_1 2 \omega_{2f} L^6}{ \nu_{n_f}}} \sqrt{ \frac{2 \omega_1 2 \omega_{2i} L^6}{ \nu_{n_i}} } \Big | \langle E_{n_f}, \textbf P_f, L \vert  \mathcal J(0)  \vert E_{n_i}, \textbf P_i, L \rangle \Big |_L \,,
\label{eq:freelimit}
\end{equation}
where $\nu_n$ counts the number of physically distinguishable finite-volume states with energy $E_n$. The value of $\nu_n$ depends on $E_n$ and $\textbf P$ and also on whether or not the particles are identical or non-identical, and degenerate or non-degenerate. Consider, for example, the case that $\textbf P=0$ and the energy coincides with $\sqrt{(2 \pi/L)^2 + m_1^2} + \sqrt{(2 \pi/L)^2 + m_2^2}$. Then $\nu=6$ for non-identical particles and $\nu = 3$ for identical particles. In the definition of $F$ this difference arises from the symmetry factor $\xi$. But the difference also reflects a physical property of the particles, namely the number of degenerate states. As a second example we consider $\textbf P = (2 \pi/L) \hat {\textbf{z}}$ and suppose the energy coincides with $\sqrt{2(2 \pi/L)^2 + m_1^2} + \sqrt{(2 \pi/L)^2 + m_2^2}$. Here three different scenarios arise, for non-identical non-degenerate particles $\nu=4$, for non-identical degenerate particles $\nu = 8$ and for identical particles $\nu=4$. In all cases this value emerges from direct evaluation of Eq.~(\ref{eq:freeF}), and is equal to the number of physically distinguishable finite-volume states.

 We now show how Eq.~(\ref{eq:freelimit}) can be confirmed by directly calculating the matrix elements on both sides in the free theory. In particular, we argue the prefactor on the right-hand side arises solely from the different normalization between finite- and infinite-volume states. Here two differences in the normalization must be accommodated. First, the finite-volume states that encode information about S-wave scattering, are constructed as symmetric combinations of the $\nu$ degenerate states with different individual particle momenta
\begin{equation}
\vert E_{n}, \textbf P, L \rangle \equiv \frac{1}{\sqrt{ \nu_n}} \sum_{\textbf k} \vert E_{n}, \textbf P-\textbf k, \textbf k , L \rangle \,.
\label{eq:spmom}
\end{equation}
The finite-volume states on the right-hand side here have definite individual particle momenta, $\textbf P - \textbf k$ and $\textbf k$, and the states on both sides have unit normalization. Substituting Eq.~(\ref{eq:spmom}) into Eq.~(\ref{eq:freelimit}) we find
\begin{equation}
\Big \vert \langle P_f, p, \mathrm{out} \vert \mathcal J(0) \vert P_i, k, \mathrm{in} \rangle \Big \vert = \sqrt{2 \omega_1 2 \omega_{2f} L^6} \sqrt{  2 \omega_1 2 \omega_{2i} L^6 } \Big | \langle E_{n_f}, \textbf P_f-\textbf p, \textbf p, L \vert  \mathcal J(0)  \vert E_{n_i}, \textbf P_i - \textbf k,\textbf k, L \rangle \Big |_L \,.
\end{equation}
Note that, since we have restricted attention to the S-wave dominated amplitude, we have $\nu_{n_i} \nu_{n_f}$ identical terms which, when combined with the normalization factor of Eq.~(\ref{eq:spmom}), perfectly cancel the $\nu$ factors in Eq.~(\ref{eq:freelimit}). The remaining factor arises because the finite-volume states have unit normalization whereas the infinite-volume states satisfy
\begin{multline}
\langle E' , \textbf P', \textbf k',a \vert E , \textbf P, \textbf k ,a \rangle \\=  2 \omega_{a1} 2 \omega_{a2} (2\pi)^6
\left[ \delta^3(\textbf k - \textbf k') \delta^3(\textbf P - \textbf k - \textbf P' + \textbf k') + \delta(a) \delta^3(\textbf k - \textbf P' + \textbf k') \delta^3(\textbf P - \textbf k - \textbf k') \right] \,,
\end{multline}
 where $\delta(a)=1$ if the particles are identical and $0$ otherwise.

We now return to the case where $\textbf 1 + \mathcal J \rightarrow \textbf 1$ are included, and examine how this effects the non-interacting limit. We begin by defining
 \begin{align}
 \label{eq:Wconndef}
 \mathcal W_{\mathrm{conn}}(P_{n_f}, p,P_{n_i}, k) & \equiv  \lim_{\mathcal M \rightarrow 0, E_i = E_{n_i}, E_f = E_{n_f}} \mathcal W_{\mathrm{df}}(P_f, p,P_i, k) \,, \\
  \begin{split}
 & \hspace{0pt} = \lim_{\mathcal M \rightarrow 0, E_i = E_{n_i}, E_f = E_{n_f}} \bigg \{ \mathcal W_{}(P_f, p,P_i, k)\\
 & \hspace{20pt} +  \mathcal M_{ }(P_f)  \left [
\frac{w_{11}(P_f,P_i)}{2 \overline\omega_{1f} (E_f -  \overline\omega_{1f} - \overline\omega_{2} + i \epsilon ) }
    +  
\frac{w_{22}(P_f,P_i)}
{ 2{\omega}_{2f}(E_f -  {\omega}_{1} - {\omega}_{2f} + i \epsilon ) }  \right ] 
    \\
&   \hspace{20pt} +   \left [    
\frac{w_{11}(P_f,P_i) }
{2 \overline\omega_{1i} (E_i - \overline \omega_{1i} -\overline \omega_{2} + i \epsilon )}  + 
\frac{w_{22}(P_f,P_i) }{ 2  \omega_{2i}(E_i -   \omega_{1} -  \omega_{2i} + i \epsilon )} \right ]
   \mathcal M_{}(P_i)  \bigg \} \,, 
   \end{split}\\[10pt]
   \begin{split}
 \mathcal W_{\mathrm{disc}}(P_{n_f}, p,P_{n_i}, k) & \equiv \lim_{\mathcal M \rightarrow 0, E_i = E_{n_i}, E_f = E_{n_f}} \xi  \left[\frac{1}{L^3}\sum_{\mathbf{k'}}\hspace{-.5cm}\int~\right]   \\  & \hspace{60pt} \times  \bigg ( \frac{ \mathcal M(P_f)  w_{22}(P_f,P_i)  \mathcal M(P_i) }{2\omega_{1}' 2 \omega_{{2f}}' 2 \omega_{2i}' (E_f -  \omega_{1}' - \omega_{2f}' + i \epsilon )(E_i -  \omega_{1}' - \omega_{2i}' + i \epsilon )}  \\
&  \hspace{70pt} + \frac{  \mathcal M(P_f)    w_{11}(P_f,P_i)  \mathcal M(P_i) }{2\omega_{2}' 2 \omega_{{1f}}' 2 \omega_{1i}' (E_f -  \omega_{1f}' - \omega_{2}' + i \epsilon )(E_i -  \omega_{1i}' - \omega_{2}' + i \epsilon )} \bigg )  \,.
\end{split}\label{eq:Wdiscdef}
 \end{align}
 Then the generalization of Eq.~(\ref{eq:freelimit}) can be written
 \begin{equation}
 \label{eq:freelimitdisc}
\bigg( \mathcal W_{\mathrm{conn}}(P_{n_f}, p,P_{n_i}, k) +  \mathcal W_{\mathrm{disc}}(P_{n_f}, p,P_{n_i}, k)  \bigg )^2 = \frac{2 \omega_1 2 \omega_{2f} L^6}{ \nu_{n_f}} \frac{2 \omega_1 2 \omega_{2i} L^6}{ \nu_{n_i}} \Big | \langle E_{n_f}, \textbf P_f, L \vert  \mathcal J(0)  \vert E_{n_i}, \textbf P_i, L \rangle \Big |^2_L \,.
 \end{equation}
 To show that this is the correct result in the non-interacting limit, we must argue that the various contractions of the finite-volume matrix element on the right-hand side precisely generate the terms on the left. These contractions can be divided into two parts, those which are connected, given by $\mathcal W_{\mathrm{conn}}$, and those which are disconnected, given by $\mathcal W_{\mathrm{disc}}$.

The connected contributions should generate the non-interacting version of the fully connected transition amplitude, $\mathcal W$, described in Sec.~\ref{sec:trans_amps} and summarized in Fig.~\ref{fig:W_iepsilon}. Note that, in the non-interacting limit, there is no distinction between $\mathcal W$ and $\Wdf$, since all terms in their difference contain factors of $\mathcal M$. However, a subtlety arises in Eq.~(\ref{eq:Wconndef}), because we are taking the limit with energies fixed at one of the values in the finite-volume spectrum. In this limit the difference between $\mathcal W$ and $\Wdf$ does not vanish, since the vanishing of the scattering amplitude is compensated by the divergence of the intermediate poles. Since we know that the non-interacting version of $\mathcal W$ should contain no contributions from this terms, we deduce that the correct definition is reached by the limit applied not to $\mathcal W$ but rather to the divergence-free version, as indicated. We conclude that $\mathcal W_{\mathrm{conn}}$ is precisely the full set of connected diagrams, with one insertion of $\mathcal J(0)$, in the non-interacting limit. In fact, the only diagram (class of diagrams) that persists in this limit is the contact interaction within the weak Bethe-Salpeter kernel (the first term in Fig.~\ref{fig:2B-current} inserted into the last term in the first line of Fig.~\ref{fig:W_ie}). Turning to the disconnected parts, we begin by evaluating $\mathcal W_{\mathrm{disc}}$. To do so we note that in the limit of vanishing interactions the energy shift vanishes as
\begin{equation}
\frac{L^3 2 \omega_1 2 \omega_2 (E - \omega_1 - \omega_2)}{\nu } = \mathcal M(P)   + \mathcal O[{\mathcal M(P)}^2]  \,.
\end{equation}
Substituting this into the definition of $\mathcal W_{\mathrm{disc}}$, Eq.~(\ref{eq:Wdiscdef}), gives
\begin{equation}
\label{eq:Wdiscred}
\mathcal W_{\mathrm{disc}}(P_{n_f}, p,P_{n_i}, k) = \frac{\xi L^3}{ \nu_{n_f} \nu_{n_i}}  \left[2 \omega_1 \nu^{(2)}_{n_{fi}} w_{22}(P_f,P_i) + 2 \omega_2  \nu^{(1)}_{n_{fi}} w_{11}(P_f,P_i) \right ]  \,,
\end{equation}
where $\nu^{(2)}_{n_{fi}}$ ($\nu^{(1)}_{n_{fi}}$) is the number of finite-volume momenta, $\textbf k$, for which both $E_i - \omega_1 - \omega_{2i}$ and $E_f - \omega_1 - \omega_{2f}$ ($E_i - \omega_{1i} - \omega_{2}$ and $E_f - \omega_{1f} - \omega_{2}$) vanish. This is indeed exactly the form of the disconnected, $\textbf 1 + \mathcal J \rightarrow \textbf 1$, contribution to the finite-volume matrix element. For example, assuming the particles are non-identical and focusing on the $w_{22}$ term, we reach
 \begin{equation}
    w_{22}(P_f,P_i)^2 = \frac{\nu_{n_f} \nu_{n_i}}{ (\nu_{n_{fi}}^{(2)})^2}   2 \omega_{2f}    2 \omega_{2i} L^6   \Big | \langle E_{n_f}, \textbf P_f, L \vert  \mathcal J(0)  \vert E_{n_i}, \textbf P_i, L \rangle \Big |^2_{L,\ 1\, \mathrm{disc}} \,.
 \end{equation}
 To see that the normalization has again been correctly accommodated we substitute Eq.~(\ref{eq:spmom}) to reexpress the right-hand side in terms of definite momentum states. We receive contributions from $\nu^{(2)}_{n_{fi}}$ different terms. Together with the normalization factors this then gives
 \begin{equation}
 \label{eq:littlewmatrixelement}
  w_{22}(P_f,P_i)^2 =   2 \omega_{2f}    2 \omega_{2i} L^6   \Big | \langle \textbf P_f - \textbf k, L, 2 \vert  \mathcal J(0)  \vert   \textbf P_i - \textbf k, L, 2 \rangle \Big |^2 \,.
 \end{equation}
Here the states on the right-hand side are single-particle finite-volume states. We conclude that the non-interacting limit of our general result gives the correct prediction, also in the case that the $\textbf 1 + \mathcal J \rightarrow \textbf 1$ transition is included. If the particles are identical then Eq.~(\ref{eq:Wdiscred}) becomes
 \begin{equation}
\mathcal W_{\mathrm{disc}}(P_{n_f}, p,P_{n_i}, k) = \frac{  \nu^{(2)}_{n_{fi}} }{ \nu_{n_f} \nu_{n_i} }   2 \omega_1 L^3 w_{22}(P_f,P_i)  \,,
\end{equation}
and substituting into Eq.~(\ref{eq:freelimitdisc}) again gives Eq.~(\ref{eq:littlewmatrixelement}).

Our final simplification of this section concerns subduction of the final result into irreps of the relevant symmetry group. If the total three-momentum of the system vanishes than this is the octahedral group, denoted ${\rm LG}(\textbf 0)$. Otherwise the symmetry breaks to a little group, denoted ${\rm LG}(\textbf P)$. In either case the residue matrices, $\mathcal{R}$, can be block diagonalized using the subduction coefficients obtained in Refs.~\cite{Dudek:2010wm, Thomas:2011rh, Dudek:2012gj}. These are denoted $\mathcal{S}^{[J,P,|\lambda|]}_{\Lambda\mu}$ where $J,P,\lambda$ are angular momentum, parity and helicity of the infinite-volume states, and $\Lambda,\mu$ are the irrep and row of interest for the finite-volume states. In this work we have written all angular momentum quantities in terms of $\ell,m_\ell$. Since the intrinsic spin of the individual particles discussed in this work is zero, $\ell=J$. The $|\ell,m_\ell\rangle$-basis is related to the $|\ell,\lambda\rangle$-basis via a unitary transformation
\begin{align}
|\ell,\lambda\rangle=\sum_{m_\ell} \mathcal{D}^{(\ell)}_{m_\ell\lambda}(\hat{R})|\ell,m_\ell\rangle \,,
\label{eq:WignerD}
\end{align}
where $\mathcal{D}^{(\ell)}_{m_\ell\lambda}$ are the standard Wigner-$\mathcal{D}$ matrices and $\hat{R}$ is an active rotation from the $\hat{z}$-axis to the direction of the total momentum of the two-particle system. Once $\mathcal R$ has been rotated to the helicity basis then $\mathcal S$ can be used to block diagonalize
\begin{align}
\mathcal{S}\mathcal{R}\mathcal{S}^\dag=\mathcal{R}_{\Lambda_1\mu_1}
\oplus\mathcal{R}_{\Lambda_1\mu_2}
\oplus
\cdots
\oplus\mathcal{R}_{\Lambda_n\mu_{n}},
\end{align} 
where we assume that the angular-momentum space has been truncated, such that $\mathcal R$ overlaps $n$ different irreps.

Finally note that one may formally attach projectors $P_{\Lambda \mu}$ to the current, $\mathcal J$, in order to subduce the full relation, Eq.~(\ref{eq:2to2_notdegen}), to a particular irrep
\begin{multline}
\Big | \langle E_{n_f}, \textbf P_f, L,\Lambda_f,\mu_{f} \vert  \mathcal J(0)  \vert E_{n_i}, \textbf P_i, L ,\Lambda_i,\mu_{i} \rangle\Big |^2_L 
=\\
\frac{1}{L^6}{\rm{Tr}}\left[
 \mathcal R_{\Lambda_i\mu_i}(E_{n_i}, \textbf P_i)
\Wtildf^{\Lambda_i\mu_i,\Lambda_f\mu_f}(P_i,P_f,L)
\mathcal R_{\Lambda_f\mu_f}(E_{n_f}, \textbf P_f)
 \Wtildf^{\Lambda_f\mu_f,\Lambda_i\mu_i}(P_f,P_i,L)
\right] \,,
  \label{eq:2to2_notdegen_explicit}
\end{multline}
where
\begin{align}
\Wtildf^{\Lambda_i\mu_i,\Lambda_f\mu_f}(P_i,P_f,L)
&\equiv
P_{\Lambda_i\mu_i} \Big [ \mathcal S\,
\Wtildf(P_i,P_f,L)\, \mathcal S^\dagger \Big ]
P_{\Lambda_f\mu_f}\,,
\end{align}
and similar with $i$ and $f$ exchanged. This expression demonstrates which elements of the transition amplitude contribute to a finite-volume matrix element with finite-volume states in a given irrep.

 \section{Final remarks and conclusion \label{sec:conclusion}}

In this work we have presented the first model-independent relation between two-body matrix elements and infinite volume $\textbf{2} + \mathcal J \rightarrow \textbf{2}$ transition amplitudes. The main result, Eq.~(\ref{eq:2to2_notdegen}), shows a multiplicative relation between finite- and infinite-volume observables. We find that a great deal of new technology is required here relative to the derivation for $\textbf{1} + \mathcal J \rightarrow \textbf{2}$ processes in Ref.~\cite{Briceno:2015csa, Briceno:2014uqa}. This is manifested, in part, by a new type of finite-volume function, which first appeared in Sec.~\ref{sec:1body_insertion}. Our final result, which holds for energies below the lowest open three- or four-particle threshold, accommodates any number of open two-particle channels. By including all angular momentum states we can also quantify the effects of reduced rotational symmetry, encoded in the mixing of different partial waves via our finite-volume functions. 

In order to implement this result in analyzing two-body matrix elements obtained from LQCD, one needs to first determine the $\textbf 2 \rightarrow \textbf 2$ scattering amplitude, $\mathcal{M}$ and the $\textbf{1} + \mathcal J \rightarrow \textbf{1}$ transition amplitude, $\w$. The former is accessible from the two-body spectrum using the L\"uscher formalism (or extensions thereof) and the latter can be obtained directly from one-body three-point functions. Given these, one can use an appropriate truncation of Eq.~(\ref{eq:2to2_notdegen}) to arrive at the infinite-volume divergence-free transition amplitude, $\Wdf$. This can be used to determine $\W$ since the divergence-free and full transition amplitudes only differ by terms which depend on on on-shell $\mathcal{M}$ and $\w$. This multistep procedure is summarized in Fig.~\ref{fig:flowchart}.

The presence of the divergence-free transition amplitude in our final result is conceptually related to the divergence-free quantity arising in the analysis of three-body systems by one of us in Refs.~\cite{Hansen:2014eka, Hansen:2015zga}. We suspect this is a very general observation. These otherwise unrelated systems both include experimentally observable subprocesses, giving rise to diagrams which contain two such processes separated by a long-lived intermediate state. In the present case one finds diagrams where two particles scatter and then propagate for an invterval before one couples to the external current (see Fig.~\ref{fig:2to2_long_range}). Similarly, the three-body sector includes diagrams where two (or more) pairwise scatterings are separated by potentially on-shell propagators. Similar divergences will be present for any tree-level process where intermediate particles go on-shell (as well as higher-order diagrams in certain cases).

\noindent
\subsection*{Acknowledgments}
RB acknowledges support from the U.S. Department of Energy contract DE-AC05-06OR23177, under which Jefferson Science Associates, LLC, manages and operates the Jefferson Lab. The authors would like to thank David Wilson, Christian Shultz and Andr\'e Walker-Loud for useful discussions. MTH would also like to thank Dalibor Djukanovic, Parikshit Junnarkar, and Harvey Meyer for useful discussion.
\noindent

 %

\appendix

 \section{Numerically evaluating $F$}
 
 \label{app:Fnum}

In this appendix we describe how to reduce and numerically evaluate the kinematic function $F$ that we introduced in Subsection \ref{sec:FFunction}.  We first note that it is more convenient to rewrite this in an alternative form that explicitly separates the real and imaginary parts of the $i \epsilon$ prescription. This is achieved with the identity
\begin{align}
\label{eq:PVdef}
\frac{1}{2 \omega_{a2}(E -  \omega_{a1} - \omega_{a2} + i \epsilon )}
= \frac{\omega_{a1}^*}{E^* (q_a^{*2}-k_a^{*2}+i\epsilon)} + \mathcal S_2
=\frac{\omega_{a1}^*}{E^*}\left[\mathcal{P}\frac{1}{(q_a^{*2}-k_a^{*2})}-i\pi\delta(q_a^{*2}-k_a^{*2})\right] + \mathcal S_2,
\end{align}
where $\mathcal S_2$ is a smooth function that will be annihilated by the sum-integral difference. Here $\mathcal P$ denotes the principal-value pole prescription. We further reduce the expression by combining the two spherical harmonics into one
\begin{equation}
4 \pi  Y_{\ell m}(\hat {\textbf k}^*_a)Y^*_{\ell' m'}(\hat {\textbf k}^*_a) =
{4 \pi}\sum_{\ell''m''}
  Y_{\ell'' m''}(\hat {\textbf k}^*_a)
\int d\Omega_{\textbf p}~Y^*_{ \ell m}(\hat {\textbf p}^*_a) Y^*_{\ell'' m''}(\hat {\textbf p}^*_a) Y_{\ell' m'}(\hat {\textbf p}^*_a).
\end{equation}
Putting all the pieces together, we can rewrite Eq.~(\ref{eq:Fscdef}) as
\begin{equation}
F_{a \ell m,a' \ell' m'}(P,L)= \delta_{aa'}\frac{iq^*_a}{8\pi E^*}\xi_a \left[\delta_{\ell\ell'}\delta_{m m'} +i\sum_{\ell'' m''}\frac{(4\pi)^{3/2}}{q_a^{*({\ell''}+1)}}c_{a \ell'' m''}^{{\pmb \Delta}}(q_a^{*2};{L})  \int d\Omega~Y^*_{ \ell m}(\hat {\textbf k}^*_a) Y^*_{\ell'' m''}(\hat {\textbf k}^*_a) Y_{\ell' m'}(\hat {\textbf k}^*_a) \right],
\label{eq:Ffunct_KSS}
\end{equation}
where $c_{a \ell m}^{{\pmb \Delta}}(q_a^{*2};{L})$ is defined as
\begin{equation}
c_{a \ell m}^{{\pmb \Delta}}(q^{*2}_a;{L})
=
\left[\frac{1}{L^3}\sum_{\mathbf{k}}\hspace{-.5cm}\int~\right]
\frac{\omega_{a1}^*k_a^{*\ell}}{\omega_{a1}}
\mathcal P\frac{ \sqrt{4 \pi}  Y_{\ell m}(\hat {\textbf k}^*_a)  }{  k_a^{*2}-q_a^{*2}}.
\end{equation}
Alternatively, this function can be written in terms of the generalized Zeta functions~\cite{Kim:2005gf}
\begin{eqnarray}
\label{eq:clm}
c^{\pmb \Delta}_{a\ell m}(q_a^{*2}; {L})
=\frac{\sqrt{4\pi}}{\gamma L^3}\left(\frac{2\pi}{L}\right)^{\ell-2}\mathcal{Z}^{\pmb \Delta}_{a\ell m}[1;(q_a^* {L}/2\pi)^2],
\hspace{1cm}
\mathcal{Z}^{\pmb \Delta}_{a \ell m}[s;x^2]
= \sum_{\mathbf r \in \mathcal{P}_{{\pmb \Delta}}}\frac{{r}^\ell Y_{\ell m}(\hat {\mathbf{r}})}{(r^2-x^2)^s} \label{eq:clm} \,,
\end{eqnarray} 
where 
\begin{align}
P_{\pmb \Delta} & \equiv  \bigg \{  \, \textbf r \ \bigg \vert \ \textbf r =   \pmb \gamma^{-1}\left( \textbf n - \frac12 \pmb \Delta \right) \,, \ \textbf n \in \mathbb Z^3 \bigg \} \,, \\
\pmb \Delta & \equiv \frac{\textbf P L}{2 \pi} \left (1 + \frac{m_{a1}^2 - m^{a2}_2}{E^{*2}} \right) \,, \\
\pmb \gamma^{-1} \textbf p & \equiv \gamma^{-1} \textbf p_{\parallel} + \textbf p_{\bot} = (E/E^*)^{-1} \textbf p_{\parallel} + \textbf p_{\bot}  \,,
\end{align}
and where $\textbf p_{\parallel}$ and $\textbf p_{\bot}$ are the parallel and perpendicular components of $\textbf p$ with respect to the fixed total momentum $\textbf P$. We close by giving a particularly efficient form for evaluating these quantities~\cite{Leskovec:2012gb},
\begin{multline}
Z_{a\ell m}^{\pmb \Delta}(1,x^2) = \sum_{\textbf r \in P_{\pmb \Delta}} \frac{ r^\ell  Y_{\ell m}(\hat {\textbf r})}{ r^2 - x^2} e^{-( r^2 - x^2)} + \gamma \frac{\pi}{2} \delta_{\ell 0} \delta_{m0} G(x) \\
+ \gamma \pi^{3/2} \int_0^1 dt \frac{e^{tx^2}}{t^{3/2}} \left( \frac{\pi i}{t} \right)^{\ell} \sum_{\textbf n \neq 0} e^{-i \pi \textbf n \cdot \pmb \Delta} \vert \pmb \gamma \textbf n \vert ^{\ell}  Y_{\ell m}( \hat {\pmb \gamma \textbf n}) e^{-(\pi \pmb \gamma \textbf n)^2/t} \,,
\end{multline}
where $\pmb \gamma \textbf p  \equiv \gamma \textbf p_{\parallel} + \textbf p_{\bot}$ and
\begin{equation}
G(x)  \equiv \int_0^1 dt \frac{e^{t x^2} - 1}{t^{3/2}}   - 2  \,.
\end{equation}

 \section{Reducing $G$}

\label{app:Gnum}

In this appendix we describe how one to reduce the complicated function $G$,
\begin{multline}
G_{a \ell_f m_{f},a' \ell_f' m_{f}'; b' \ell_i' m_{i}', b \ell_i m_{i}}(P_f,P_i,L)  \equiv \\
\hspace{-4cm}
\delta_{aa'} \delta_{bb'}
\left[\frac{1}{L^3}\sum_{\mathbf{k}}\hspace{-.5cm}\int~\right]
~\frac{1}{2\omega_{a1}}
\frac{ 4 \pi  Y_{\ell_fm_{ f}}(\hat {\textbf k}^*_{af})
Y^*_{\ell_f'm_{f}'}(\hat {\textbf k}^*_{af})
 }{2 \omega_{{a2f}}(E_f -  \omega_{a1} - \omega_{a2f} + i \epsilon )} 
  \bigg (\frac{k^{*}_{af}}{q^*_{af}} \bigg)^{\ell_f+\ell_f'}
\frac{ 4 \pi  Y_{\ell_i' m_{i}'}(\hat {\textbf k}^*_{bi})
Y^*_{\ell_i m_{i}}(\hat {\textbf k}^*_{bi})
 }{2 \omega_{b2i}(E_i -  \omega_{b1} - \omega_{b2i} + i \epsilon )} 
 \bigg (\frac{k^{*}_{bi}}{q^*_{bi}} \bigg)^{\ell_i+\ell_i'}
  \,.
\end{multline}

\subsection{Single degenerate channel, $P_i = P_f = (iE, 0)$, $s$-wave}

To get started, we consider the simplest possible scenario, a single channel of degenerate scalar particles with total momenta $P_i = P_f = (iE, 0)$. As mentioned in the main text, $\mathcal W$ diverges for these kinematics. Nonetheless, $\Wdf$ is finite and constraining its value here could help to determine the full $\textbf 2 + \mathcal J \rightarrow \textbf 2$ transition amplitude away from this singular point. In this subsection we further assume that scattering is dominated by the $s$-wave so that all higher partial waves can be neglected. Then $G$ reduces to a single function of $E$ and $L$ given by
\begin{equation}
G(E,L)  \equiv  G_S(E,L) - G_I(E) \,,
\end{equation}
where
\begin{align}
G_S(E,L) & =  \frac{1}{L^3}\sum_{\mathbf{k}} \frac{1}{8 \omega^3 }
\frac{ 1
 }{  (E - 2 \omega )^2} \,, \\
 G_I(E) & = \int \! \! \frac{d \textbf k}{(2 \pi)^3} \ 
\frac{1}{8 \omega^3 }
\frac{ 1
 }{  (E - 2 \omega + i \epsilon )^2} \,,
\end{align}
and where we have introduced the shorthand $\omega = \sqrt{\textbf k^2 + m^2}$.

Our main task here is to reduce the integral. We begin by rewriting the integral in terms of the magnitude and direction. The latter is trivial and so we reach
\begin{equation}
 G_I(E)  = \frac{4 \pi}{(2 \pi)^3} \int_0^\infty dk k^2 
\frac{1}{8 \omega^3}
\frac{ 1
 }{  (E - 2 \omega+ i \epsilon )^2} \bigg \vert_{\omega = \sqrt{k^2 + m^2}} \,.
\end{equation}
We now change variables, first by substituting $dk = (\omega/k) d \omega$ and then by shifting via $x = \omega-E/2$
\begin{equation}
\label{eq:xint}
G_I(E) =  \int_{-a}^\infty dx  f(x) \frac{1}{(x - i \epsilon)^2} \,,
\end{equation}
where $a \equiv E/2-m$ and
\begin{equation}
\label{eq:fxdef}
f(x) \equiv  \frac{1}{2 \pi^2} \frac{\sqrt{\omega^2 - m^2} }{32 \omega^2} \bigg \vert_{\omega = x + E/2 } \,.
\end{equation}
To reduce further we substitute $f(x) = f(x- i \epsilon) - f(0) + f(0) = g(x) (x - i \epsilon) + f(0) $ where $g(x) \equiv [f(x- i \epsilon) - f(0)]/(x - i \epsilon)$ has the same analytic properties as $f(x)$
\begin{equation}
G_I(E) \equiv \int_{-a}^\infty dx  g(x) \frac{1}{(x - i \epsilon)} + f(0) \int_{-a}^\infty dx   \frac{1}{(x - i \epsilon)^2} = \mathcal P \int_{-a}^\infty dx  g(x) \frac{1}{x} + i \pi g(0) - \frac{f(0)}{a} \,.
\end{equation}
Substituting for $g$ we conclude
\begin{equation}
\label{eq:simpresult}
G_I(E) \equiv  \mathcal P \int_{-a}^\infty dx  \frac{f(x) - f(0)}{x^2} + i \pi f'(0) - \frac{f(0)}{a} \,.
\end{equation}

Finally we substitute the definition of $f(x)$, Eq.~(\ref{eq:fxdef}), and combine results to conclude
\begin{multline}
G(E,L) \equiv  \frac{1}{L^3}\sum_{\mathbf{k}} \frac{1}{8 \omega^3 }
\frac{ 1
 }{  (E - 2 \omega )^2} - \frac{1}{2 \pi^2} \mathcal P \int_{m}^\infty d \omega \left [  \frac{\sqrt{\omega^2 - m^2}}{8 \omega^2}  - \frac{\sqrt{E^2/4-m^2}}{2 E^2}\right ]\frac{1}{(E - 2 \omega )^2} \\ +  \frac{1}{2 \pi^2}\frac{\sqrt{E^2/4-m^2}}{8 E^2 (E/2 - m)} -  i \frac{1}{2 \pi \sqrt{E^2/4-m^2}}  \left[ \frac{E^2/4 + m^2}{4 E^3 }  \right ] \,.
\end{multline}
This is our final form for the simplest version of $G(E,L)$.  

\subsection{Single degenerate channel, general $P_i =  P_f$, general angular momentum}

We now turn to the general case in which the poles coincide for all $\textbf k$. This occurs whenever the particles of channel one have the same masses as those of channel two and also $P_i =  P_f = P$. The two particles within each channel may, however, still be nondegenerate. As already mentioned in the previous subsection, $\mathcal W$ diverges for these kinematics but it may nonetheless be useful to constrain $\Wdf$. Unlike the previous subsection, here we also accommodate general angular momentum. Again we focus on reducing the integral part of $G$
\begin{multline}
G_{I; \ell_f m_{f}, \ell_f' m_{f}';  \ell_i' m_{i}',  \ell_i m_{i}}(P)  \equiv \\
\hspace{-4cm}
\int \! \! \frac{d \textbf k}{(2 \pi)^3}
~\frac{1}{8 \omega_{1} \omega_{2}^2} 4 \pi  Y_{\ell_fm_{ f}}(\hat {\textbf k}^* )
Y^*_{\ell_f'm_{f}'}(\hat {\textbf k}^* ) 4 \pi  Y_{\ell_i' m_{i}'}(\hat {\textbf k}^* )
Y^*_{\ell_i m_{i}}(\hat {\textbf k}^* )   \bigg (\frac{k^{*} }{q^* } \bigg)^{\ell_f+\ell_f'+\ell_i+\ell_i'} \left[ \frac{1}{(E -  \omega_{1} - \omega_{2} + i \epsilon )} \right ]^2
  \,.
\end{multline}

We begin by rewriting the integral as 
\begin{equation}
\label{eq:GwithF1}
G_{I}(P)  = 
\int \frac{d \textbf k^*}{(2 \pi)^3} \frac{1}{2 \omega_{1}^*}
\mathcal F(\textbf k^*) (E^* - \omega_1^* + \omega_2^*)^2 \left[  \frac{1}{(E  -  \omega_{1})^2 - \omega_{2}^2 +    i\epsilon}\right ]^2
  \,,
\end{equation}
where we have left the harmonic indices on $G_I$ implicit and where
\begin{equation}
 \mathcal F(\textbf k^*) \equiv  \frac{(E -  \omega_{1} + \omega_{2}  )^2}{4 \omega_{2}^2 (E^* - \omega_1^* + \omega_2^*)^2} 4 \pi  Y_{\ell_fm_{ f}}(\hat {\textbf k}^* )
Y^*_{\ell_f'm_{f}'}(\hat {\textbf k}^* ) 4 \pi  Y_{\ell_i' m_{i}'}(\hat {\textbf k}^* )
Y^*_{\ell_i m_{i}}(\hat {\textbf k}^* )   \bigg (\frac{k^{*} }{q^* } \bigg)^{\ell_f+\ell_f'+\ell_i+\ell_i'}\,.
\end{equation}
Here we have also used the fact that ${d \textbf k}/{\omega_{1}}={d \textbf k^*}/{\omega_{1}^*}$. The next step is to rewrite the double pole in CM frame variables
\begin{equation}
\label{eq:ipolesub}
(E -  \omega_{1})^2 - \omega_{2}^2 = [(E, \textbf P) - (\omega_{1}, \textbf k)]^2 - m_{2}^2 = [(E^*, 0) - (\omega_{1}^*, \textbf k^* )]^2 - m_{2}^2 = (E^* - \omega_{1}^*)^2 - \omega_{2}^{*2} \,.
\end{equation}
Substituting this into Eq.~(\ref{eq:GwithF1}) and also substituting
\begin{equation}
\mathcal F(k^*) \equiv \int \! d \Omega\ \mathcal F(\textbf k^*) \,,
\end{equation}
then gives
\begin{equation}
G_{I}(P)  = 
\int_{0}^\infty \frac{d k^* k^{*2}}{(2 \pi)^3} \frac{1}{2 \omega_{1}^*}
 \frac{\mathcal F( k^*)}{(E^*  -  \omega_{1}^* - \omega_{2}^* +    i\epsilon)^2}
  \,.
\end{equation}

The final step is to observe
\begin{equation}
\frac{1}{E^*  -  \omega_{1}^* - \omega_{2}^* +    i\epsilon} = \frac{H(k^*)}{q^* - k^* + i \epsilon} \,,
\end{equation} 
where $E^{*}=\sqrt{m_1^2+q^{*2}}+\sqrt{m_2^2+q^{*2}}$ and 
\begin{equation}
H(k^*) = \frac{(E^*  +  \omega_{1}^* + \omega_{2}^*)(E^{*2} -  \omega_{1}^{*2} - \omega_{2}^{*2} + 2\omega_{1}^*  \omega_{2}^*)}{ 4 E^{*2}(q^*+k^*)} \,.
\end{equation}
This equality follows from
\begin{align}
(E^{*2} -  \omega_{1}^{*2} - \omega_{2}^{*2} + 2\omega_{1}^*  \omega_{2}^*) (E^{*2} -  \omega_{1}^{*2} - \omega_{2}^{*2} - 2\omega_{1}^*  \omega_{2}^*)
\nn\\
&\hspace{-6cm} =E^{*4}  + (2 k^{*2} + m_1^2 + m_2^2)^2  - 2 E^{*2} (2 k^{*2} + m_1^2 + m_2^2) - 4 ( k^{*2} + m_1^2 ) ( k^{*2} + m_2^2 )  \nn\\
&\hspace{-6cm} =E^{*4} - 2 E^{*2} (m_1^2 + m_2^2) + m^4_1 +m^4_2 -2m^2_1m^2_2-4E^{*2}k^{*2}\nn\\
&\hspace{-6cm} =E^{*2}\left(E^{*2} - 2  (m_1^2 + m_2^2) + \frac{\left(m^2_1 -m^2_2\right)^{2}}{E^{*2}}\right)-4E^{*2}k^{*2}
\nn\\
&\hspace{-6cm} = 4 E^{*2}(q^{*2}-k^{*2}).
\end{align}
Finally, we can rewrite the integral as
\begin{equation}
G_{I}(P)  = 
\int_{-q^*}^\infty d x \left [ \frac{ (x+q^*)^{2}}{(2 \pi)^3} \frac{1}{2 \sqrt{(x+q^*)^2 + m_1^2}} \mathcal F( x+q^*) H(x+q^*)^2 \right ]
 \frac{1}{(x -    i\epsilon)^2}
  \,.
\end{equation}
At this stage the integral be may reduced following the method outlined after Eq.~(\ref{eq:xint}) above.

\subsection{General $P_i \neq P_f$}

 In this section we analyze $G$ for all scenarios in which the poles do not coincide. More precisely for all cases where the set of $\textbf k$ for which both poles diverge is a one-(or-less)-dimensional subspace of the three-dimensional $\textbf k$ space. This is the case whenever $ P_i \neq  P_f$ or whenever the current changes the incoming particle to a new species with a different mass. As above our goal is to simplify
\begin{multline}
G_{I;a \ell_f m_{f},a' \ell_f' m_{f}'; b' \ell_i' m_{i}', b \ell_i m_{i}}(P_f,P_i)  \equiv \\
\hspace{-4cm}
\delta_{aa'} \delta_{bb'}
\int \! \! \frac{d \textbf k}{(2 \pi)^3}
~\frac{1}{2\omega_{a1}}
\frac{ 4 \pi  Y_{\ell_fm_{ f}}(\hat {\textbf k}^*_{af})
Y^*_{\ell_f'm_{f}'}(\hat {\textbf k}^*_{af})
 }{2 \omega_{{a2f}}(E_f -  \omega_{a1} - \omega_{a2f} + i \epsilon )} 
  \bigg (\frac{k^{*}_{af}}{q^*_{af}} \bigg)^{\ell_f+\ell_f'}
\frac{ 4 \pi  Y_{\ell_i' m_{i}'}(\hat {\textbf k}^*_{bi})
Y^*_{\ell_i m_{i}}(\hat {\textbf k}^*_{bi})
 }{2 \omega_{b2i}(E_i -  \omega_{b1} - \omega_{b2i} + i \epsilon )} 
 \bigg (\frac{k^{*}_{bi}}{q^*_{bi}} \bigg)^{\ell_i+\ell_i'}
  \,.
\end{multline}
As we will see in the course of this analysis, it turns out that one is justified to treat the two poles as independent single poles. In other words, the fact that the two poles can diverge simultaneously {\em does not} complicate the integral because the region where they coincide is at most a one-dimensional subspace of $\textbf k$ space. 

To see this in detail, first observe that the set of $\textbf k$ for which both poles diverge forms a one-(or less)-dimensional subspace of the three-dimensional $\textbf k$ space. One can visualize this by first recalling that, in the incoming CM frame, the momentum for which $E_i - \omega_{b1} - \omega_{b2i}$ vanishes is a sphere with radius $k_{bi}^* = q_{bi}^*$. Boosting this to the finite-volume frame gives an ellipsoid in that frame. Further, one can use the same analysis to define a second ellipsoid, for the set of momentum for which $E_f - \omega_{a1} - \omega_{a2f}$ vanishes. Finally, for $P_i \neq P_f$, the intersection of these two ellipsoids is a one-dimensional ellipse (or else a point or an empty set).

The fact that the double-pole space has a lower dimension than the single-pole space implies that it is not necessary to specially treat the case of both poles simultaneously diverging. In short, no special treatment is needed because the difference between the double and single pole expressions is measure zero and does not contribute to the integral. This can be seen directly by evaluating the integral. As a simple example consider
\begin{equation}
\int_{-1}^1 dx \int_{-1}^1 dy \int_{-1}^1 dz  \frac{1}{x + z - i \epsilon} \frac{1}{x + y + z - i \epsilon} =  \int_{-1}^1 dx \int_{-1}^1 dz \frac{1}{x + z - i \epsilon} \log \left [ \frac{x + z + 1 - i \epsilon }{x + z - 1 - i \epsilon} \right ] \,.
\end{equation}
Here the $y$ integral reduces us the integrand to a single pole and the fact that both poles diverge for $y = 0$ and $x+z = 0$ does not require any special attention. We will see that $G_{I}(P_f,P_i)$ is very similar.

We begin by rewriting the integral as 
\begin{equation}
\label{eq:Grw1}
G_{I}(P_f,P_i)  = 
\int \frac{d \textbf k}{(2 \pi)^3} \frac{1}{2 \omega_{a1}}
\mathcal F_i(\textbf k_{bi}^*) \left[  \frac{1}{(E_f -  \omega_{a1})^2 - \omega_{a2f}^2 +    i\epsilon}\right ]
  \left[  \frac{1 }{(E_i -  \omega_{b1})^2 - \omega_{b2i}^2 +    i\epsilon}\right ]
  \,,
\end{equation}
where
\begin{multline}
 \mathcal F_{i}(\textbf k_{bi}^*) \equiv \delta_{aa'} \delta_{bb'} (E_f -  \omega_{a1} + \omega_{a2f}  )(E_i -  \omega_{b1} + \omega_{b2i}  ) \\ \times
\frac{ 4 \pi  Y_{\ell_fm_{ f}}(\hat {\textbf k}^*_{af})
Y^*_{\ell_f'm_{f}'}(\hat {\textbf k}^*_{af})
 }{2 \omega_{{a2f}}} 
  \bigg (\frac{k^{*}_{af}}{q^*_{af}} \bigg)^{\ell_f+\ell_f'}
\frac{ 4 \pi  Y_{\ell_i' m_{i}'}(\hat {\textbf k}^*_{bi})
Y^*_{\ell_i m_{i}}(\hat {\textbf k}^*_{bi})
 }{2 \omega_{b2i}} 
 \bigg (\frac{k^{*}_{bi}}{q^*_{bi}} \bigg)^{\ell_i+\ell_i'} \,.
\end{multline}
Note that one can express this as a function of only $\textbf k_{bi}^*$, together with implicit $P_i$ and $P_f$. The next step is to rewrite all variables in the incoming CM frame. This is most straightforward for the incoming pole
\begin{equation}
\label{eq:ipolesub}
(E_i -  \omega_{b1})^2 - \omega_{b2i}^2 = [(E_i, \textbf P_i) - (\omega_{b1}, \textbf k)]^2 - m_{b2}^2 = [(E_i^*, 0) - (\omega_{b1i}^*, \textbf k_{bi}^* )]^2 - m_{b2}^2 = (E_i^* - \omega_{b1i}^*)^2 - \omega_{b2i}^{*2} \,.
\end{equation}
For the outgoing pole we must introduce new notation. We define $(E_f^{(i*)} , \textbf P_f^{(i*)})$ by boosting $(E_f, \textbf P_f)$ to the {\em incoming} two-particle CM frame. This allows us to write
\begin{align}
(E_f -  \omega_{a1})^2 - \omega_{a2f}^2 & = [(E_f, \textbf P_f) - (\omega_{a1}, \textbf k)]^2 - m_{a2}^2 = [(E_f^{(i*)}, \textbf P_f^{(i*)}) - (\omega_{b1i}^{*}, \textbf k_{bi}^{*} )]^2 - m_{a2}^2 \\
& = E_f^{*2}   + m_{a1}^2 -  m_{a2}^2  + 2 E_f^{(i*)}  \omega_{b1i}^{*}     -2 \textbf P_f^{(i*)} \cdot \textbf k_{bi}^{*}      \,.
\label{eq:fpolesub}
\end{align}
Substituting Eqs.~(\ref{eq:ipolesub}) and (\ref{eq:fpolesub}) into Eq.~(\ref{eq:Grw1}) we reach
\begin{align}
\begin{split}
G_{I}(P_f,P_i ) & = 
\int \frac{d \textbf k_{bi}^*}{(2 \pi)^3} \frac{1}{2 \omega_{b1i}^*}
\mathcal F_i(\textbf k_{bi}^*)     \bigg[  \frac{1}{  E_f^{*2}   + m_{a1}^2 -  m_{a2}^2  + 2 E_f^{(i*)}  \omega_{b1i}^{*}     -2 \textbf P_f^{(i*)} \cdot \textbf k_{bi}^{*} + i \epsilon}\bigg ]\\
& \hspace{270pt} \times \bigg[  \frac{1}{ (E_i^* - \omega_{b1i}^*)^2 - \omega_{b2i}^{*2} + i \epsilon } \bigg ]
  \,,\end{split} \\
  \begin{split}
  & = \int \frac{d  k_{bi}^*  k_{bi}^{*2} d \phi d z}{(2 \pi)^3} \frac{1}{2 \omega_{b1i}^*}
\mathcal F_i( k_{bi}^*, z, \phi)    \bigg[  \frac{1}{  E_f^{*2}   + m_{a1}^2 -  m_{a2}^2  + 2 E_f^{(i*)}  \omega_{b1i}^{*}     -2  P_f^{(i*)}  k_{bi}^{*} z+ i \epsilon}\bigg ] \\
& \hspace{270pt} \times \bigg[  \frac{1}{ (E_i^* - \omega_{b1i}^*)^2 - \omega_{b2i}^{*2}+ i \epsilon }\bigg ]
  \end{split}\,,
\end{align}
where $z=\cos\theta$. Note that with this boost we are treating the problem asymmetrically, arbitrarily focusing on the incoming frame. We could just as well work with the outgoing frame. Either way, we have found that a CM frame must be chosen to reduce the problem.

Next we  split the second, $z$-independent pole into principal value and delta function and also substitute 
\begin{equation}
\mathcal F(k_{bi}^*, z) = \int_0^{2 \pi} d \phi \mathcal F(k_{bi}^*, z, \phi) \,,
\end{equation}
to reach
\begin{multline}
 G_I(P_f,P_i)  = \int \frac{d  k_{bi}^*  k_{bi}^{*2}  d z}{(2 \pi)^3} \frac{1}{2 \omega_{b1i}^*}  \frac{\mathcal F_i( k_{bi}^*, z) }{  E_f^{*2}   + m_{a1}^2 -  m_{a2}^2  + 2 E_f^{(i*)}  \omega_{b1i}^{*}     -2  P_f^{(i*)}  k_{bi}^{*} z + i \epsilon}  \mathcal P \frac{1}{ (E_i^* - \omega_{b1i}^*)^2 - \omega_{b2i}^{*2} }  \\
 - i \frac{q^*_{bi}}{32 \pi^2 E_i^*} \int_{-1}^1 dz  \frac{\mathcal F_i( q_{bi}^*, z)}{  E_f^{*2}   + m_{a1}^2 -  m_{a2}^2  + 2 E_f^{(i*)}  \omega_{b1qi}^{*}     -2  P_f^{(i*)}  q_{bi}^{*} z + i \epsilon} \,.
\end{multline}
This separation is valid regardless over the entire range of both the $k_{bi}^*$ and $z$ integrals. This observation gives precise meaning to the statement made at the beginning of this subsection, that we can treat the two poles as independent.
  
Finally we break the remaining pole into principal value and delta function to conclude
\begin{multline}
\label{eq:IfinPiPfdiff}
 G_I(P_f,P_i)  = \int \frac{d  k_{bi}^*  k_{bi}^{*2}  d z}{(2 \pi)^3} \frac{\mathcal F_i( k_{bi}^*, z) }{2 \omega_{b1i}^*} \mathcal P \frac{1}{  E_f^{*2}   + m_{a1}^2 -  m_{a2}^2  + 2 E_f^{(i*)}  \omega_{b1i}^{*}     -2  P_f^{(i*)}  k_{bi}^{*} z }  \mathcal P \frac{1}{ (E_i^* - \omega_{b1i}^*)^2 - \omega_{b2i}^{*2} }  \\
 - i \frac{q^*_{bi}}{32 \pi^2 E_i^*} \int_{-1}^1 dz \ \mathcal P  \frac{\mathcal F_i( q_{bi}^*, z)}{  E_f^{*2}   + m_{a1}^2 -  m_{a2}^2  + 2 E_f^{(i*)}  \omega_{b1qi}^{*}     -2  P_f^{(i*)}  q_{bi}^{*} z}  \\ - i \frac{q^*_{af}}{32 \pi^2 E_f^*} \int_{-1}^1 dz \ \mathcal P \frac{\mathcal F_f( q_{af}^*, z)}{  E_i^{*2}   + m_{b1}^2 -  m_{b2}^2  + 2 E_i^{(f*)}  \omega_{a1qf}^{*}     -2  P_i^{(f*)}  q_{af}^{*} z }   -  \frac{q^*_{bi}}{32 \pi E_i^*} \frac{ \mathcal F_i( q_{bi}^*, z_i) }{2  P_f^{(i*)}  q_{bi}^{*}}   \,,
\end{multline}
where $z_i \equiv ( E_f^{*2}   + m_{a1}^2 -  m_{a2}^2  + 2 E_f^{(i*)}  \omega_{b1qi}^{*}  )/(2  P_f^{(i*)}  q_{bi}^{*} )$.

To reach Eq.~(\ref{eq:IfinPiPfdiff}) we have rewritten the first term appearing on the last line. This is the term that comes from the delta-function part of the $z$ pole and the principal-value part of the $k^*_i$ pole, in other words the ``principal-value initial state and delta-function final state'' term. To rewrite this term we have used the fact that it is given by swapping all $i$ and $f$ labels on the term of the second line. This is the ``delta-function initial state and principal-value final state'' term, so it must be related to the first term on the last line by swapping labels as indicated. We are restoring some of the symmetry that we lost when we chose to work in the incoming CM frame. 

We next comment that the last term of Eq.~(\ref{eq:IfinPiPfdiff}) is given by taking the delta function terms from both poles
\begin{multline}
-  \frac{q^*_{bi}}{32 \pi E_i^*} \frac{ \mathcal F_i( q_{bi}^*, z_i) }{2  P_f^{(i*)}  q_{bi}^{*}}  = - \pi^2  
\delta_{aa'} \delta_{bb'}
\int \! \! \frac{d \textbf k}{(2 \pi)^3}
~\frac{1}{2\omega_{a1}}
\frac{ 4 \pi  Y_{\ell_fm_{ f}}(\hat {\textbf k}^*_{af})
Y^*_{\ell_f'm_{f}'}(\hat {\textbf k}^*_{af})
 }{2 \omega_{{a2f}}} 
  \bigg (\frac{k^{*}_{af}}{q^*_{af}} \bigg)^{\ell_f+\ell_f'} \\ \times
\frac{ 4 \pi  Y_{\ell_i' m_{i}'}(\hat {\textbf k}^*_{bi})
Y^*_{\ell_i m_{i}}(\hat {\textbf k}^*_{bi})
 }{2 \omega_{b2i}} 
 \bigg (\frac{k^{*}_{bi}}{q^*_{bi}} \bigg)^{\ell_i+\ell_i'}\delta(E_f -  \omega_{a1} - \omega_{a2f}) \delta(E_i -  \omega_{b1} - \omega_{b2i}) \,.
\end{multline}
It is very important to remember that this term is only present if there exists some $\textbf k$ for which $(E_f -  \omega_{a1} - \omega_{a2f}) = (E_i -  \omega_{b1} - \omega_{b2i}) = 0$. That is, the two ellipsoids in $\textbf k$-space, defined by the two pole conditions, must have some non-zero intersection for this term to appear. Note also that this term is unchanged if we swap all $i$ and $f$ indices. Even though this ``double delta function'' term is perfectly symmetric with respect to $i$ and $f$, we can only solve the integral by choosing a specific frame. 

Finally we comment that the first term in Eq.~(\ref{eq:IfinPiPfdiff}) is equivalent to
\begin{multline}
\delta_{aa'} \delta_{bb'}
\int \! \! \frac{d \textbf k}{(2 \pi)^3}
~\frac{1}{2\omega_{a1}} \mathcal P \left[
\frac{ 4 \pi  Y_{\ell_fm_{ f}}(\hat {\textbf k}^*_{af})
Y^*_{\ell_f'm_{f}'}(\hat {\textbf k}^*_{af})
 }{2 \omega_{{a2f}}(E_f -  \omega_{a1} - \omega_{a2f} + i \epsilon )}  \right ]  \bigg (\frac{k^{*}_{af}}{q^*_{af}} \bigg)^{\ell_f+\ell_f'}\\ \times \mathcal P \left [
\frac{ 4 \pi  Y_{\ell_i' m_{i}'}(\hat {\textbf k}^*_{bi})
Y^*_{\ell_i m_{i}}(\hat {\textbf k}^*_{bi})
 }{2 \omega_{b2i}(E_i -  \omega_{b1} - \omega_{b2i} + i \epsilon )}  \right ]
 \bigg (\frac{k^{*}_{bi}}{q^*_{bi}} \bigg)^{\ell_i+\ell_i'} \,.
\end{multline}
That is, it is just given by replacing the original two $i \epsilon$ poles with principal value poles. This is the only term in Eq.~(\ref{eq:IfinPiPfdiff}) which still contains a divergent integral. In a numerical evaluation this term will be combined with the sum to reach a numerically tractable sum-integral-difference. The UV divergence of course cancels between the sum and integral in this difference.

\bibliographystyle{apsrev} 
\bibliography{bibi} 

\begin{thebibliography}{79}
\expandafter\ifx\csname natexlab\endcsname\relax\def\natexlab#1{#1}\fi
\expandafter\ifx\csname bibnamefont\endcsname\relax
  \def\bibnamefont#1{#1}\fi
\expandafter\ifx\csname bibfnamefont\endcsname\relax
  \def\bibfnamefont#1{#1}\fi
\expandafter\ifx\csname citenamefont\endcsname\relax
  \def\citenamefont#1{#1}\fi
\expandafter\ifx\csname url\endcsname\relax
  \def\url#1{\texttt{#1}}\fi
\expandafter\ifx\csname urlprefix\endcsname\relax\def\urlprefix{URL }\fi
\providecommand{\bibinfo}[2]{#2}
\providecommand{\eprint}[2][]{\url{#2}}

\bibitem[{\citenamefont{Shultz et~al.}(2015)\citenamefont{Shultz, Dudek, and
  Edwards}}]{Shultz:2015pfa}
\bibinfo{author}{\bibfnamefont{C.~J.} \bibnamefont{Shultz}},
  \bibinfo{author}{\bibfnamefont{J.~J.} \bibnamefont{Dudek}}, \bibnamefont{and}
  \bibinfo{author}{\bibfnamefont{R.~G.} \bibnamefont{Edwards}},
  \bibinfo{journal}{Phys.Rev.} \textbf{\bibinfo{volume}{D91}},
  \bibinfo{pages}{114501} (\bibinfo{year}{2015}), \eprint{1501.07457}.

\bibitem[{\citenamefont{Alexandrou}(2012)}]{Alexandrou:2011ga}
\bibinfo{author}{\bibfnamefont{C.}~\bibnamefont{Alexandrou}},
  \bibinfo{journal}{AIP Conf.Proc.} \textbf{\bibinfo{volume}{1432}},
  \bibinfo{pages}{62} (\bibinfo{year}{2012}), \eprint{1108.4112}.

\bibitem[{\citenamefont{Alexandrou et~al.}(2011)\citenamefont{Alexandrou,
  Gregory, Korzec, Koutsou, Negele et~al.}}]{Alexandrou:2011py}
\bibinfo{author}{\bibfnamefont{C.}~\bibnamefont{Alexandrou}},
  \bibinfo{author}{\bibfnamefont{E.~B.} \bibnamefont{Gregory}},
  \bibinfo{author}{\bibfnamefont{T.}~\bibnamefont{Korzec}},
  \bibinfo{author}{\bibfnamefont{G.}~\bibnamefont{Koutsou}},
  \bibinfo{author}{\bibfnamefont{J.~W.} \bibnamefont{Negele}},
  \bibnamefont{et~al.}, \bibinfo{journal}{Phys.Rev.Lett.}
  \textbf{\bibinfo{volume}{107}}, \bibinfo{pages}{141601}
  (\bibinfo{year}{2011}), \eprint{1106.6000}.

\bibitem[{\citenamefont{Green et~al.}(2014)\citenamefont{Green, Negele,
  Pochinsky, Syritsyn, Engelhardt, and Krieg}}]{Green:2014xba}
\bibinfo{author}{\bibfnamefont{J.~R.} \bibnamefont{Green}},
  \bibinfo{author}{\bibfnamefont{J.~W.} \bibnamefont{Negele}},
  \bibinfo{author}{\bibfnamefont{A.~V.} \bibnamefont{Pochinsky}},
  \bibinfo{author}{\bibfnamefont{S.~N.} \bibnamefont{Syritsyn}},
  \bibinfo{author}{\bibfnamefont{M.}~\bibnamefont{Engelhardt}},
  \bibnamefont{and} \bibinfo{author}{\bibfnamefont{S.}~\bibnamefont{Krieg}},
  \bibinfo{journal}{Phys. Rev.} \textbf{\bibinfo{volume}{D90}},
  \bibinfo{pages}{074507} (\bibinfo{year}{2014}), \eprint{1404.4029}.

\bibitem[{\citenamefont{Feng et~al.}(2015)\citenamefont{Feng, Aoki, Hashimoto,
  and Kaneko}}]{Feng:2014gba}
\bibinfo{author}{\bibfnamefont{X.}~\bibnamefont{Feng}},
  \bibinfo{author}{\bibfnamefont{S.}~\bibnamefont{Aoki}},
  \bibinfo{author}{\bibfnamefont{S.}~\bibnamefont{Hashimoto}},
  \bibnamefont{and} \bibinfo{author}{\bibfnamefont{T.}~\bibnamefont{Kaneko}},
  \bibinfo{journal}{Phys.Rev.} \textbf{\bibinfo{volume}{D91}},
  \bibinfo{pages}{054504} (\bibinfo{year}{2015}), \eprint{1412.6319}.

\bibitem[{\citenamefont{Brice\~no et~al.}(2015)\citenamefont{Brice\~no, Dudek,
  Edwards, Shultz, Thomas, and Wilson}}]{Briceno:2015dca}
\bibinfo{author}{\bibfnamefont{R.~A.} \bibnamefont{Brice\~no}},
  \bibinfo{author}{\bibfnamefont{J.~J.} \bibnamefont{Dudek}},
  \bibinfo{author}{\bibfnamefont{R.~G.} \bibnamefont{Edwards}},
  \bibinfo{author}{\bibfnamefont{C.~J.} \bibnamefont{Shultz}},
  \bibinfo{author}{\bibfnamefont{C.~E.} \bibnamefont{Thomas}},
  \bibnamefont{and} \bibinfo{author}{\bibfnamefont{D.~J.} \bibnamefont{Wilson}}
  (\bibinfo{year}{2015}), \eprint{1507.06622}.

\bibitem[{\citenamefont{Bertero et~al.}(1982)\citenamefont{Bertero, Boccacci,
  and Pike}}]{Bertero}
\bibinfo{author}{\bibfnamefont{M.}~\bibnamefont{Bertero}},
  \bibinfo{author}{\bibfnamefont{P.}~\bibnamefont{Boccacci}}, \bibnamefont{and}
  \bibinfo{author}{\bibfnamefont{E.~R.} \bibnamefont{Pike}},
  \bibinfo{journal}{Proc. R. Soc. Lond.} \textbf{\bibinfo{volume}{A383}}
  (\bibinfo{year}{1982}).

\bibitem[{\citenamefont{Luscher}(1986)}]{Luscher:1986pf}
\bibinfo{author}{\bibfnamefont{M.}~\bibnamefont{Luscher}},
  \bibinfo{journal}{Commun.Math.Phys.} \textbf{\bibinfo{volume}{105}},
  \bibinfo{pages}{153} (\bibinfo{year}{1986}).

\bibitem[{\citenamefont{Luscher}(1991)}]{Luscher:1990ux}
\bibinfo{author}{\bibfnamefont{M.}~\bibnamefont{Luscher}},
  \bibinfo{journal}{Nucl.Phys.} \textbf{\bibinfo{volume}{B354}},
  \bibinfo{pages}{531} (\bibinfo{year}{1991}).

\bibitem[{\citenamefont{Rummukainen and Gottlieb}(1995)}]{Rummukainen:1995vs}
\bibinfo{author}{\bibfnamefont{K.}~\bibnamefont{Rummukainen}} \bibnamefont{and}
  \bibinfo{author}{\bibfnamefont{S.~A.} \bibnamefont{Gottlieb}},
  \bibinfo{journal}{Nucl. Phys.} \textbf{\bibinfo{volume}{B450}},
  \bibinfo{pages}{397} (\bibinfo{year}{1995}), \eprint{hep-lat/9503028}.

\bibitem[{\citenamefont{He et~al.}(2005)\citenamefont{He, Feng, and
  Liu}}]{He:2005ey}
\bibinfo{author}{\bibfnamefont{S.}~\bibnamefont{He}},
  \bibinfo{author}{\bibfnamefont{X.}~\bibnamefont{Feng}}, \bibnamefont{and}
  \bibinfo{author}{\bibfnamefont{C.}~\bibnamefont{Liu}},
  \bibinfo{journal}{JHEP} \textbf{\bibinfo{volume}{07}}, \bibinfo{pages}{011}
  (\bibinfo{year}{2005}), \eprint{hep-lat/0504019}.

\bibitem[{\citenamefont{Kim et~al.}(2005)\citenamefont{Kim, Sachrajda, and
  Sharpe}}]{Kim:2005gf}
\bibinfo{author}{\bibfnamefont{C.}~\bibnamefont{Kim}},
  \bibinfo{author}{\bibfnamefont{C.}~\bibnamefont{Sachrajda}},
  \bibnamefont{and} \bibinfo{author}{\bibfnamefont{S.~R.}
  \bibnamefont{Sharpe}}, \bibinfo{journal}{Nucl.Phys.}
  \textbf{\bibinfo{volume}{B727}}, \bibinfo{pages}{218} (\bibinfo{year}{2005}),
  \eprint{hep-lat/0507006}.

\bibitem[{\citenamefont{Christ et~al.}(2005)\citenamefont{Christ, Kim, and
  Yamazaki}}]{Christ:2005gi}
\bibinfo{author}{\bibfnamefont{N.~H.} \bibnamefont{Christ}},
  \bibinfo{author}{\bibfnamefont{C.}~\bibnamefont{Kim}}, \bibnamefont{and}
  \bibinfo{author}{\bibfnamefont{T.}~\bibnamefont{Yamazaki}},
  \bibinfo{journal}{Phys.Rev.} \textbf{\bibinfo{volume}{D72}},
  \bibinfo{pages}{114506} (\bibinfo{year}{2005}), \eprint{hep-lat/0507009}.

\bibitem[{\citenamefont{Leskovec and Prelovsek}(2012)}]{Leskovec:2012gb}
\bibinfo{author}{\bibfnamefont{L.}~\bibnamefont{Leskovec}} \bibnamefont{and}
  \bibinfo{author}{\bibfnamefont{S.}~\bibnamefont{Prelovsek}},
  \bibinfo{journal}{Phys.Rev.} \textbf{\bibinfo{volume}{D85}},
  \bibinfo{pages}{114507} (\bibinfo{year}{2012}), \eprint{1202.2145}.

\bibitem[{\citenamefont{Brice\~no and
  Davoudi}(2013{\natexlab{a}})}]{Briceno:2012yi}
\bibinfo{author}{\bibfnamefont{R.~A.} \bibnamefont{Brice\~no}}
  \bibnamefont{and} \bibinfo{author}{\bibfnamefont{Z.}~\bibnamefont{Davoudi}},
  \bibinfo{journal}{Phys. Rev. D. 88,} \textbf{\bibinfo{volume}{094507}},
  \bibinfo{pages}{094507} (\bibinfo{year}{2013}{\natexlab{a}}),
  \eprint{1204.1110}.

\bibitem[{\citenamefont{Hansen and Sharpe}(2012)}]{Hansen:2012tf}
\bibinfo{author}{\bibfnamefont{M.~T.} \bibnamefont{Hansen}} \bibnamefont{and}
  \bibinfo{author}{\bibfnamefont{S.~R.} \bibnamefont{Sharpe}},
  \bibinfo{journal}{Phys.Rev.} \textbf{\bibinfo{volume}{D86}},
  \bibinfo{pages}{016007} (\bibinfo{year}{2012}), \eprint{1204.0826}.

\bibitem[{\citenamefont{Brice\~no}(2014)}]{Briceno:2014oea}
\bibinfo{author}{\bibfnamefont{R.~A.} \bibnamefont{Brice\~no}},
  \bibinfo{journal}{Phys.Rev.} \textbf{\bibinfo{volume}{D89}},
  \bibinfo{pages}{074507} (\bibinfo{year}{2014}), \eprint{1401.3312}.

\bibitem[{\citenamefont{Hansen and Sharpe}(2014)}]{Hansen:2014eka}
\bibinfo{author}{\bibfnamefont{M.~T.} \bibnamefont{Hansen}} \bibnamefont{and}
  \bibinfo{author}{\bibfnamefont{S.~R.} \bibnamefont{Sharpe}},
  \bibinfo{journal}{Phys.Rev.} \textbf{\bibinfo{volume}{D90}},
  \bibinfo{pages}{116003} (\bibinfo{year}{2014}), \eprint{1408.5933}.

\bibitem[{\citenamefont{Hansen and Sharpe}(2015)}]{Hansen:2015zga}
\bibinfo{author}{\bibfnamefont{M.~T.} \bibnamefont{Hansen}} \bibnamefont{and}
  \bibinfo{author}{\bibfnamefont{S.~R.} \bibnamefont{Sharpe}}
  (\bibinfo{year}{2015}), \eprint{1504.04248}.

\bibitem[{\citenamefont{Brice\~no and
  Davoudi}(2013{\natexlab{b}})}]{Briceno:2012rv}
\bibinfo{author}{\bibfnamefont{R.~A.} \bibnamefont{Brice\~no}}
  \bibnamefont{and} \bibinfo{author}{\bibfnamefont{Z.}~\bibnamefont{Davoudi}},
  \bibinfo{journal}{Phys.Rev.} \textbf{\bibinfo{volume}{D87}},
  \bibinfo{pages}{094507} (\bibinfo{year}{2013}{\natexlab{b}}),
  \eprint{1212.3398}.

\bibitem[{\citenamefont{Polejaeva and Rusetsky}(2012)}]{Polejaeva:2012ut}
\bibinfo{author}{\bibfnamefont{K.}~\bibnamefont{Polejaeva}} \bibnamefont{and}
  \bibinfo{author}{\bibfnamefont{A.}~\bibnamefont{Rusetsky}},
  \bibinfo{journal}{Eur.Phys.J.} \textbf{\bibinfo{volume}{A48}},
  \bibinfo{pages}{67} (\bibinfo{year}{2012}), \eprint{1203.1241}.

\bibitem[{\citenamefont{Wilson et~al.}(2015)\citenamefont{Wilson, Brice\~no,
  Dudek, Edwards, and Thomas}}]{Wilson:2015dqa}
\bibinfo{author}{\bibfnamefont{D.~J.} \bibnamefont{Wilson}},
  \bibinfo{author}{\bibfnamefont{R.~A.} \bibnamefont{Brice\~no}},
  \bibinfo{author}{\bibfnamefont{J.~J.} \bibnamefont{Dudek}},
  \bibinfo{author}{\bibfnamefont{R.~G.} \bibnamefont{Edwards}},
  \bibnamefont{and} \bibinfo{author}{\bibfnamefont{C.~E.} \bibnamefont{Thomas}}
  (\bibinfo{year}{2015}), \eprint{1507.02599}.

\bibitem[{\citenamefont{Wilson et~al.}(2014)\citenamefont{Wilson, Dudek,
  Edwards, and Thomas}}]{Wilson:2014cna}
\bibinfo{author}{\bibfnamefont{D.~J.} \bibnamefont{Wilson}},
  \bibinfo{author}{\bibfnamefont{J.~J.} \bibnamefont{Dudek}},
  \bibinfo{author}{\bibfnamefont{R.~G.} \bibnamefont{Edwards}},
  \bibnamefont{and} \bibinfo{author}{\bibfnamefont{C.~E.} \bibnamefont{Thomas}}
  (\bibinfo{year}{2014}), \eprint{1411.2004}.

\bibitem[{\citenamefont{Dudek et~al.}(2014)\citenamefont{Dudek, Edwards,
  Thomas, and Wilson}}]{Dudek:2014qha}
\bibinfo{author}{\bibfnamefont{J.~J.} \bibnamefont{Dudek}},
  \bibinfo{author}{\bibfnamefont{R.~G.} \bibnamefont{Edwards}},
  \bibinfo{author}{\bibfnamefont{C.~E.} \bibnamefont{Thomas}},
  \bibnamefont{and} \bibinfo{author}{\bibfnamefont{D.~J.} \bibnamefont{Wilson}}
  (\bibinfo{year}{2014}), \eprint{1406.4158}.

\bibitem[{\citenamefont{Dudek et~al.}(2013)\citenamefont{Dudek, Edwards, and
  Thomas}}]{Dudek:2012xn}
\bibinfo{author}{\bibfnamefont{J.~J.} \bibnamefont{Dudek}},
  \bibinfo{author}{\bibfnamefont{R.~G.} \bibnamefont{Edwards}},
  \bibnamefont{and} \bibinfo{author}{\bibfnamefont{C.~E.}
  \bibnamefont{Thomas}}, \bibinfo{journal}{Phys.Rev.}
  \textbf{\bibinfo{volume}{D87}}, \bibinfo{pages}{034505}
  (\bibinfo{year}{2013}), \eprint{1212.0830}.

\bibitem[{\citenamefont{Lang et~al.}(2011)\citenamefont{Lang, Mohler,
  Prelovsek, and Vidmar}}]{Lang:2011mn}
\bibinfo{author}{\bibfnamefont{C.~B.} \bibnamefont{Lang}},
  \bibinfo{author}{\bibfnamefont{D.}~\bibnamefont{Mohler}},
  \bibinfo{author}{\bibfnamefont{S.}~\bibnamefont{Prelovsek}},
  \bibnamefont{and} \bibinfo{author}{\bibfnamefont{M.}~\bibnamefont{Vidmar}},
  \bibinfo{journal}{Phys. Rev.} \textbf{\bibinfo{volume}{D84}},
  \bibinfo{pages}{054503} (\bibinfo{year}{2011}), \bibinfo{note}{[Erratum:
  Phys. Rev.D89,no.5,059903(2014)]}, \eprint{1105.5636}.

\bibitem[{\citenamefont{Lang et~al.}(2015)\citenamefont{Lang, Mohler,
  Prelovsek, and Woloshyn}}]{Lang:2015hza}
\bibinfo{author}{\bibfnamefont{C.~B.} \bibnamefont{Lang}},
  \bibinfo{author}{\bibfnamefont{D.}~\bibnamefont{Mohler}},
  \bibinfo{author}{\bibfnamefont{S.}~\bibnamefont{Prelovsek}},
  \bibnamefont{and} \bibinfo{author}{\bibfnamefont{R.~M.}
  \bibnamefont{Woloshyn}}, \bibinfo{journal}{Phys. Lett.}
  \textbf{\bibinfo{volume}{B750}}, \bibinfo{pages}{17} (\bibinfo{year}{2015}),
  \eprint{1501.01646}.

\bibitem[{\citenamefont{Lang et~al.}(2014)\citenamefont{Lang, Leskovec, Mohler,
  Prelovsek, and Woloshyn}}]{Lang:2014yfa}
\bibinfo{author}{\bibfnamefont{C.~B.} \bibnamefont{Lang}},
  \bibinfo{author}{\bibfnamefont{L.}~\bibnamefont{Leskovec}},
  \bibinfo{author}{\bibfnamefont{D.}~\bibnamefont{Mohler}},
  \bibinfo{author}{\bibfnamefont{S.}~\bibnamefont{Prelovsek}},
  \bibnamefont{and} \bibinfo{author}{\bibfnamefont{R.~M.}
  \bibnamefont{Woloshyn}}, \bibinfo{journal}{Phys. Rev.}
  \textbf{\bibinfo{volume}{D90}}, \bibinfo{pages}{034510}
  (\bibinfo{year}{2014}), \eprint{1403.8103}.

\bibitem[{\citenamefont{Martínez~Torres
  et~al.}(2015)\citenamefont{Martínez~Torres, Oset, Prelovsek, and
  Ramos}}]{Torres:2014vna}
\bibinfo{author}{\bibfnamefont{A.}~\bibnamefont{Martínez~Torres}},
  \bibinfo{author}{\bibfnamefont{E.}~\bibnamefont{Oset}},
  \bibinfo{author}{\bibfnamefont{S.}~\bibnamefont{Prelovsek}},
  \bibnamefont{and} \bibinfo{author}{\bibfnamefont{A.}~\bibnamefont{Ramos}},
  \bibinfo{journal}{JHEP} \textbf{\bibinfo{volume}{05}}, \bibinfo{pages}{153}
  (\bibinfo{year}{2015}), \eprint{1412.1706}.

\bibitem[{\citenamefont{Feng et~al.}(2011)\citenamefont{Feng, Jansen, and
  Renner}}]{Feng:2010es}
\bibinfo{author}{\bibfnamefont{X.}~\bibnamefont{Feng}},
  \bibinfo{author}{\bibfnamefont{K.}~\bibnamefont{Jansen}}, \bibnamefont{and}
  \bibinfo{author}{\bibfnamefont{D.~B.} \bibnamefont{Renner}},
  \bibinfo{journal}{Phys. Rev.} \textbf{\bibinfo{volume}{D83}},
  \bibinfo{pages}{094505} (\bibinfo{year}{2011}), \eprint{1011.5288}.

\bibitem[{\citenamefont{Pelissier and Alexandru}(2013)}]{Pelissier:2012pi}
\bibinfo{author}{\bibfnamefont{C.}~\bibnamefont{Pelissier}} \bibnamefont{and}
  \bibinfo{author}{\bibfnamefont{A.}~\bibnamefont{Alexandru}},
  \bibinfo{journal}{Phys. Rev.} \textbf{\bibinfo{volume}{D87}},
  \bibinfo{pages}{014503} (\bibinfo{year}{2013}), \eprint{1211.0092}.

\bibitem[{\citenamefont{Prelovsek et~al.}(2013)\citenamefont{Prelovsek,
  Leskovec, Lang, and Mohler}}]{Prelovsek:2013ela}
\bibinfo{author}{\bibfnamefont{S.}~\bibnamefont{Prelovsek}},
  \bibinfo{author}{\bibfnamefont{L.}~\bibnamefont{Leskovec}},
  \bibinfo{author}{\bibfnamefont{C.~B.} \bibnamefont{Lang}}, \bibnamefont{and}
  \bibinfo{author}{\bibfnamefont{D.}~\bibnamefont{Mohler}},
  \bibinfo{journal}{Phys. Rev.} \textbf{\bibinfo{volume}{D88}},
  \bibinfo{pages}{054508} (\bibinfo{year}{2013}), \eprint{1307.0736}.

\bibitem[{\citenamefont{Aoki et~al.}(2011)}]{Aoki:2011yj}
\bibinfo{author}{\bibfnamefont{S.}~\bibnamefont{Aoki}} \bibnamefont{et~al.}
  (\bibinfo{collaboration}{CS}), \bibinfo{journal}{Phys. Rev.}
  \textbf{\bibinfo{volume}{D84}}, \bibinfo{pages}{094505}
  (\bibinfo{year}{2011}), \eprint{1106.5365}.

\bibitem[{\citenamefont{Aoki et~al.}(2007)}]{Aoki:2007rd}
\bibinfo{author}{\bibfnamefont{S.}~\bibnamefont{Aoki}} \bibnamefont{et~al.}
  (\bibinfo{collaboration}{CP-PACS}), \bibinfo{journal}{Phys. Rev.}
  \textbf{\bibinfo{volume}{D76}}, \bibinfo{pages}{094506}
  (\bibinfo{year}{2007}), \eprint{0708.3705}.

\bibitem[{\citenamefont{Bolton et~al.}(2015)\citenamefont{Bolton, Brice\~no,
  and Wilson}}]{Bolton:2015psa}
\bibinfo{author}{\bibfnamefont{D.~R.} \bibnamefont{Bolton}},
  \bibinfo{author}{\bibfnamefont{R.~A.} \bibnamefont{Brice\~no}},
  \bibnamefont{and} \bibinfo{author}{\bibfnamefont{D.~J.} \bibnamefont{Wilson}}
  (\bibinfo{year}{2015}), \eprint{1507.07928}.

\bibitem[{\citenamefont{Lellouch and Luscher}(2001)}]{Lellouch:2000pv}
\bibinfo{author}{\bibfnamefont{L.}~\bibnamefont{Lellouch}} \bibnamefont{and}
  \bibinfo{author}{\bibfnamefont{M.}~\bibnamefont{Luscher}},
  \bibinfo{journal}{Commun.Math.Phys.} \textbf{\bibinfo{volume}{219}},
  \bibinfo{pages}{31} (\bibinfo{year}{2001}), \eprint{hep-lat/0003023}.

\bibitem[{\citenamefont{Lin et~al.}(2001)\citenamefont{Lin, Martinelli,
  Sachrajda, and Testa}}]{Lin:2001ek}
\bibinfo{author}{\bibfnamefont{C.~D.} \bibnamefont{Lin}},
  \bibinfo{author}{\bibfnamefont{G.}~\bibnamefont{Martinelli}},
  \bibinfo{author}{\bibfnamefont{C.~T.} \bibnamefont{Sachrajda}},
  \bibnamefont{and} \bibinfo{author}{\bibfnamefont{M.}~\bibnamefont{Testa}},
  \bibinfo{journal}{Nucl.Phys.} \textbf{\bibinfo{volume}{B619}},
  \bibinfo{pages}{467} (\bibinfo{year}{2001}), \eprint{hep-lat/0104006}.

\bibitem[{\citenamefont{Agadjanov et~al.}(2014)\citenamefont{Agadjanov,
  Bernard, Mei$\ss$ner, and Rusetsky}}]{Agadjanov:2014kha}
\bibinfo{author}{\bibfnamefont{A.}~\bibnamefont{Agadjanov}},
  \bibinfo{author}{\bibfnamefont{V.}~\bibnamefont{Bernard}},
  \bibinfo{author}{\bibfnamefont{U.-G.} \bibnamefont{Mei$\ss$ner}},
  \bibnamefont{and} \bibinfo{author}{\bibfnamefont{A.}~\bibnamefont{Rusetsky}},
  \bibinfo{journal}{Nucl.Phys.} \textbf{\bibinfo{volume}{B886}},
  \bibinfo{pages}{1199} (\bibinfo{year}{2014}), \eprint{1405.3476}.

\bibitem[{\citenamefont{Meyer}(2012)}]{Meyer:2012wk}
\bibinfo{author}{\bibfnamefont{H.~B.} \bibnamefont{Meyer}}
  (\bibinfo{year}{2012}), \eprint{1202.6675}.

\bibitem[{\citenamefont{Bernard et~al.}(2012)\citenamefont{Bernard, Hoja,
  Mei$\ss$ner, and Rusetsky}}]{Bernard:2012bi}
\bibinfo{author}{\bibfnamefont{V.}~\bibnamefont{Bernard}},
  \bibinfo{author}{\bibfnamefont{D.}~\bibnamefont{Hoja}},
  \bibinfo{author}{\bibfnamefont{U.}~\bibnamefont{Mei$\ss$ner}},
  \bibnamefont{and} \bibinfo{author}{\bibfnamefont{A.}~\bibnamefont{Rusetsky}},
  \bibinfo{journal}{JHEP} \textbf{\bibinfo{volume}{1209}}, \bibinfo{pages}{023}
  (\bibinfo{year}{2012}), \eprint{1205.4642}.

\bibitem[{\citenamefont{Brice\~no and Hansen}(2015)}]{Briceno:2015csa}
\bibinfo{author}{\bibfnamefont{R.~A.} \bibnamefont{Brice\~no}}
  \bibnamefont{and} \bibinfo{author}{\bibfnamefont{M.~T.} \bibnamefont{Hansen}}
  (\bibinfo{year}{2015}), \eprint{1502.04314}.

\bibitem[{\citenamefont{Brice\~no et~al.}(2014)\citenamefont{Brice\~no, Hansen,
  and Walker-Loud}}]{Briceno:2014uqa}
\bibinfo{author}{\bibfnamefont{R.~A.} \bibnamefont{Brice\~no}},
  \bibinfo{author}{\bibfnamefont{M.~T.} \bibnamefont{Hansen}},
  \bibnamefont{and}
  \bibinfo{author}{\bibfnamefont{A.}~\bibnamefont{Walker-Loud}}
  (\bibinfo{year}{2014}), \eprint{1406.5965}.

\bibitem[{\citenamefont{Detmold and Flynn}(2015)}]{Detmold:2014fpa}
\bibinfo{author}{\bibfnamefont{W.}~\bibnamefont{Detmold}} \bibnamefont{and}
  \bibinfo{author}{\bibfnamefont{M.}~\bibnamefont{Flynn}},
  \bibinfo{journal}{Phys.Rev.} \textbf{\bibinfo{volume}{D91}},
  \bibinfo{pages}{074509} (\bibinfo{year}{2015}), \eprint{1412.3895}.

\bibitem[{\citenamefont{Lee and Yang}(1956)}]{Lee:1956zz}
\bibinfo{author}{\bibfnamefont{T.}~\bibnamefont{Lee}} \bibnamefont{and}
  \bibinfo{author}{\bibfnamefont{C.}~\bibnamefont{Yang}},
  \bibinfo{journal}{Phys.Rev.} \textbf{\bibinfo{volume}{104}},
  \bibinfo{pages}{822} (\bibinfo{year}{1956}).

\bibitem[{\citenamefont{Wu et~al.}(1957)\citenamefont{Wu, Ambler, Hayward,
  Hoppes, and Hudson}}]{Wu:1957my}
\bibinfo{author}{\bibfnamefont{C.}~\bibnamefont{Wu}},
  \bibinfo{author}{\bibfnamefont{E.}~\bibnamefont{Ambler}},
  \bibinfo{author}{\bibfnamefont{R.}~\bibnamefont{Hayward}},
  \bibinfo{author}{\bibfnamefont{D.}~\bibnamefont{Hoppes}}, \bibnamefont{and}
  \bibinfo{author}{\bibfnamefont{R.}~\bibnamefont{Hudson}},
  \bibinfo{journal}{Phys.Rev.} \textbf{\bibinfo{volume}{105}},
  \bibinfo{pages}{1413} (\bibinfo{year}{1957}).

\bibitem[{\citenamefont{Tanner}(1957)}]{tanner}
\bibinfo{author}{\bibfnamefont{N.}~\bibnamefont{Tanner}},
  \bibinfo{journal}{Phys.Rev.} \textbf{\bibinfo{volume}{107}},
  \bibinfo{pages}{1233} (\bibinfo{year}{1957}).

\bibitem[{\citenamefont{Krane et~al.}(1971{\natexlab{a}})\citenamefont{Krane,
  Olsen, Sites, and Steyert}}]{Krane:1971zza}
\bibinfo{author}{\bibfnamefont{K.}~\bibnamefont{Krane}},
  \bibinfo{author}{\bibfnamefont{C.}~\bibnamefont{Olsen}},
  \bibinfo{author}{\bibfnamefont{J.~R.} \bibnamefont{Sites}}, \bibnamefont{and}
  \bibinfo{author}{\bibfnamefont{W.}~\bibnamefont{Steyert}},
  \bibinfo{journal}{Phys.Rev.} \textbf{\bibinfo{volume}{C4}},
  \bibinfo{pages}{1906} (\bibinfo{year}{1971}{\natexlab{a}}).

\bibitem[{\citenamefont{Krane et~al.}(1971{\natexlab{b}})\citenamefont{Krane,
  Olsen, Sites, and Steyert}}]{Krane:1971zz}
\bibinfo{author}{\bibfnamefont{K.}~\bibnamefont{Krane}},
  \bibinfo{author}{\bibfnamefont{C.}~\bibnamefont{Olsen}},
  \bibinfo{author}{\bibfnamefont{J.~R.} \bibnamefont{Sites}}, \bibnamefont{and}
  \bibinfo{author}{\bibfnamefont{W.}~\bibnamefont{Steyert}},
  \bibinfo{journal}{Phys.Rev.Lett.} \textbf{\bibinfo{volume}{26}},
  \bibinfo{pages}{1579} (\bibinfo{year}{1971}{\natexlab{b}}).

\bibitem[{\citenamefont{Yuan et~al.}(1991)\citenamefont{Yuan, Bowman, Bowman,
  Bush, Delheij et~al.}}]{Yuan:1991zz}
\bibinfo{author}{\bibfnamefont{V.}~\bibnamefont{Yuan}},
  \bibinfo{author}{\bibfnamefont{C.}~\bibnamefont{Bowman}},
  \bibinfo{author}{\bibfnamefont{J.}~\bibnamefont{Bowman}},
  \bibinfo{author}{\bibfnamefont{J.~E.} \bibnamefont{Bush}},
  \bibinfo{author}{\bibfnamefont{P.}~\bibnamefont{Delheij}},
  \bibnamefont{et~al.}, \bibinfo{journal}{Phys.Rev.}
  \textbf{\bibinfo{volume}{C44}}, \bibinfo{pages}{2187} (\bibinfo{year}{1991}).

\bibitem[{\citenamefont{Eversheim et~al.}(1991)\citenamefont{Eversheim,
  Schmitt, Kuhn, Hinterberger, von Rossen et~al.}}]{Eversheim:1991tg}
\bibinfo{author}{\bibfnamefont{P.}~\bibnamefont{Eversheim}},
  \bibinfo{author}{\bibfnamefont{W.}~\bibnamefont{Schmitt}},
  \bibinfo{author}{\bibfnamefont{S.}~\bibnamefont{Kuhn}},
  \bibinfo{author}{\bibfnamefont{F.}~\bibnamefont{Hinterberger}},
  \bibinfo{author}{\bibfnamefont{P.}~\bibnamefont{von Rossen}},
  \bibnamefont{et~al.}, \bibinfo{journal}{Phys.Lett.}
  \textbf{\bibinfo{volume}{B256}}, \bibinfo{pages}{11} (\bibinfo{year}{1991}).

\bibitem[{\citenamefont{Balzer et~al.}(1980)\citenamefont{Balzer, Henneck,
  Jacquemart, Lang, Simonius et~al.}}]{Balzer:1980dn}
\bibinfo{author}{\bibfnamefont{R.}~\bibnamefont{Balzer}},
  \bibinfo{author}{\bibfnamefont{R.}~\bibnamefont{Henneck}},
  \bibinfo{author}{\bibfnamefont{C.}~\bibnamefont{Jacquemart}},
  \bibinfo{author}{\bibfnamefont{J.}~\bibnamefont{Lang}},
  \bibinfo{author}{\bibfnamefont{M.}~\bibnamefont{Simonius}},
  \bibnamefont{et~al.}, \bibinfo{journal}{Phys.Rev.Lett.}
  \textbf{\bibinfo{volume}{44}}, \bibinfo{pages}{699} (\bibinfo{year}{1980}).

\bibitem[{\citenamefont{Balzer et~al.}(1984)\citenamefont{Balzer, Henneck,
  Jacquemart, Lang, Nessi-Tedaldi et~al.}}]{Balzer:1985au}
\bibinfo{author}{\bibfnamefont{R.}~\bibnamefont{Balzer}},
  \bibinfo{author}{\bibfnamefont{R.}~\bibnamefont{Henneck}},
  \bibinfo{author}{\bibfnamefont{C.}~\bibnamefont{Jacquemart}},
  \bibinfo{author}{\bibfnamefont{J.}~\bibnamefont{Lang}},
  \bibinfo{author}{\bibfnamefont{F.}~\bibnamefont{Nessi-Tedaldi}},
  \bibnamefont{et~al.}, \bibinfo{journal}{Phys.Rev.}
  \textbf{\bibinfo{volume}{C30}}, \bibinfo{pages}{1409} (\bibinfo{year}{1984}).

\bibitem[{\citenamefont{Berdoz et~al.}(2001)}]{Berdoz:2001nu}
\bibinfo{author}{\bibfnamefont{A.}~\bibnamefont{Berdoz}} \bibnamefont{et~al.}
  (\bibinfo{collaboration}{TRIUMF E497 Collaboration}),
  \bibinfo{journal}{Phys.Rev.Lett.} \textbf{\bibinfo{volume}{87}},
  \bibinfo{pages}{272301} (\bibinfo{year}{2001}), \eprint{nucl-ex/0107014}.

\bibitem[{\citenamefont{Kistryn et~al.}(1987)\citenamefont{Kistryn, Lang,
  Liechti, Maier, Muller et~al.}}]{Kistryn:1987tq}
\bibinfo{author}{\bibfnamefont{S.}~\bibnamefont{Kistryn}},
  \bibinfo{author}{\bibfnamefont{J.}~\bibnamefont{Lang}},
  \bibinfo{author}{\bibfnamefont{J.}~\bibnamefont{Liechti}},
  \bibinfo{author}{\bibfnamefont{T.}~\bibnamefont{Maier}},
  \bibinfo{author}{\bibfnamefont{R.}~\bibnamefont{Muller}},
  \bibnamefont{et~al.}, \bibinfo{journal}{Phys.Rev.Lett.}
  \textbf{\bibinfo{volume}{58}}, \bibinfo{pages}{1616} (\bibinfo{year}{1987}).

\bibitem[{\citenamefont{Berdoz et~al.}(2003)}]{Berdoz:2002sn}
\bibinfo{author}{\bibfnamefont{A.}~\bibnamefont{Berdoz}} \bibnamefont{et~al.}
  (\bibinfo{collaboration}{TRIUMF E497 Collaboration}),
  \bibinfo{journal}{Phys.Rev.} \textbf{\bibinfo{volume}{C68}},
  \bibinfo{pages}{034004} (\bibinfo{year}{2003}), \eprint{nucl-ex/0211020}.

\bibitem[{\citenamefont{Cavaignac et~al.}(1977)\citenamefont{Cavaignac, Vignon,
  and Wilson}}]{Cavaignac:1977uk}
\bibinfo{author}{\bibfnamefont{J.}~\bibnamefont{Cavaignac}},
  \bibinfo{author}{\bibfnamefont{B.}~\bibnamefont{Vignon}}, \bibnamefont{and}
  \bibinfo{author}{\bibfnamefont{R.}~\bibnamefont{Wilson}},
  \bibinfo{journal}{Phys.Lett.} \textbf{\bibinfo{volume}{B67}},
  \bibinfo{pages}{148} (\bibinfo{year}{1977}).

\bibitem[{\citenamefont{Gericke et~al.}(2011)\citenamefont{Gericke, Alarcon,
  Balascuta, Barron-Palos, Blessinger et~al.}}]{Gericke:2011zz}
\bibinfo{author}{\bibfnamefont{M.}~\bibnamefont{Gericke}},
  \bibinfo{author}{\bibfnamefont{R.}~\bibnamefont{Alarcon}},
  \bibinfo{author}{\bibfnamefont{S.}~\bibnamefont{Balascuta}},
  \bibinfo{author}{\bibfnamefont{L.}~\bibnamefont{Barron-Palos}},
  \bibinfo{author}{\bibfnamefont{C.}~\bibnamefont{Blessinger}},
  \bibnamefont{et~al.}, \bibinfo{journal}{Phys.Rev.}
  \textbf{\bibinfo{volume}{C83}}, \bibinfo{pages}{015505}
  (\bibinfo{year}{2011}).

\bibitem[{\citenamefont{Knyazkov et~al.}(1984)\citenamefont{Knyazkov,
  Kolomenskii, Lobashev, Nazarenko, Pirozhkov et~al.}}]{Knyazkov:1984zz}
\bibinfo{author}{\bibfnamefont{V.}~\bibnamefont{Knyazkov}},
  \bibinfo{author}{\bibfnamefont{E.}~\bibnamefont{Kolomenskii}},
  \bibinfo{author}{\bibfnamefont{V.}~\bibnamefont{Lobashev}},
  \bibinfo{author}{\bibfnamefont{V.}~\bibnamefont{Nazarenko}},
  \bibinfo{author}{\bibfnamefont{A.}~\bibnamefont{Pirozhkov}},
  \bibnamefont{et~al.}, \bibinfo{journal}{Nucl.Phys.}
  \textbf{\bibinfo{volume}{A417}}, \bibinfo{pages}{209} (\bibinfo{year}{1984}).

\bibitem[{\citenamefont{Snow et~al.}(2011)\citenamefont{Snow, Bass, Bass,
  Crawford, Gan et~al.}}]{Snow:2011zza}
\bibinfo{author}{\bibfnamefont{W.}~\bibnamefont{Snow}},
  \bibinfo{author}{\bibfnamefont{C.}~\bibnamefont{Bass}},
  \bibinfo{author}{\bibfnamefont{T.}~\bibnamefont{Bass}},
  \bibinfo{author}{\bibfnamefont{B.}~\bibnamefont{Crawford}},
  \bibinfo{author}{\bibfnamefont{K.}~\bibnamefont{Gan}}, \bibnamefont{et~al.},
  \bibinfo{journal}{Phys.Rev.} \textbf{\bibinfo{volume}{C83}},
  \bibinfo{pages}{022501} (\bibinfo{year}{2011}).

\bibitem[{\citenamefont{Phillips et~al.}(2009)\citenamefont{Phillips,
  Schindler, and Springer}}]{Phillips:2008hn}
\bibinfo{author}{\bibfnamefont{D.~R.} \bibnamefont{Phillips}},
  \bibinfo{author}{\bibfnamefont{M.~R.} \bibnamefont{Schindler}},
  \bibnamefont{and} \bibinfo{author}{\bibfnamefont{R.~P.}
  \bibnamefont{Springer}}, \bibinfo{journal}{Nucl.Phys.}
  \textbf{\bibinfo{volume}{A822}}, \bibinfo{pages}{1} (\bibinfo{year}{2009}),
  \eprint{0812.2073}.

\bibitem[{\citenamefont{Griesshammer et~al.}(2012)\citenamefont{Griesshammer,
  Schindler, and Springer}}]{Griesshammer:2011md}
\bibinfo{author}{\bibfnamefont{H.~W.} \bibnamefont{Griesshammer}},
  \bibinfo{author}{\bibfnamefont{M.~R.} \bibnamefont{Schindler}},
  \bibnamefont{and} \bibinfo{author}{\bibfnamefont{R.~P.}
  \bibnamefont{Springer}}, \bibinfo{journal}{Eur.Phys.J.}
  \textbf{\bibinfo{volume}{A48}}, \bibinfo{pages}{7} (\bibinfo{year}{2012}),
  \eprint{1109.5667}.

\bibitem[{\citenamefont{Schindler and Springer}(2010)}]{Schindler:2009wd}
\bibinfo{author}{\bibfnamefont{M.~R.} \bibnamefont{Schindler}}
  \bibnamefont{and} \bibinfo{author}{\bibfnamefont{R.~P.}
  \bibnamefont{Springer}}, \bibinfo{journal}{Nucl.Phys.}
  \textbf{\bibinfo{volume}{A846}}, \bibinfo{pages}{51} (\bibinfo{year}{2010}),
  \eprint{0907.5358}.

\bibitem[{\citenamefont{Shin et~al.}(2010)\citenamefont{Shin, Ando, and
  Hyun}}]{Shin:2009hi}
\bibinfo{author}{\bibfnamefont{J.}~\bibnamefont{Shin}},
  \bibinfo{author}{\bibfnamefont{S.}~\bibnamefont{Ando}}, \bibnamefont{and}
  \bibinfo{author}{\bibfnamefont{C.}~\bibnamefont{Hyun}},
  \bibinfo{journal}{Phys.Rev.} \textbf{\bibinfo{volume}{C81}},
  \bibinfo{pages}{055501} (\bibinfo{year}{2010}), \eprint{0907.3995}.

\bibitem[{\citenamefont{Savage and Springer}(2001)}]{Savage:1999cm}
\bibinfo{author}{\bibfnamefont{M.~J.} \bibnamefont{Savage}} \bibnamefont{and}
  \bibinfo{author}{\bibfnamefont{R.~P.} \bibnamefont{Springer}},
  \bibinfo{journal}{Nucl.Phys.} \textbf{\bibinfo{volume}{A686}},
  \bibinfo{pages}{413} (\bibinfo{year}{2001}), \eprint{nucl-th/9907069}.

\bibitem[{\citenamefont{Haxton and Holstein}(2013)}]{Haxton:2013aca}
\bibinfo{author}{\bibfnamefont{W.}~\bibnamefont{Haxton}} \bibnamefont{and}
  \bibinfo{author}{\bibfnamefont{B.}~\bibnamefont{Holstein}},
  \bibinfo{journal}{Prog.Part.Nucl.Phys.} \textbf{\bibinfo{volume}{71}},
  \bibinfo{pages}{185} (\bibinfo{year}{2013}), \eprint{1303.4132}.

\bibitem[{\citenamefont{Schindler and Springer}(2013)}]{Schindler:2013yua}
\bibinfo{author}{\bibfnamefont{M.}~\bibnamefont{Schindler}} \bibnamefont{and}
  \bibinfo{author}{\bibfnamefont{R.}~\bibnamefont{Springer}},
  \bibinfo{journal}{Prog.Part.Nucl.Phys.} \textbf{\bibinfo{volume}{72}},
  \bibinfo{pages}{1} (\bibinfo{year}{2013}), \eprint{1305.4190}.

\bibitem[{\citenamefont{Wasem}(2012)}]{Wasem:2011zz}
\bibinfo{author}{\bibfnamefont{J.}~\bibnamefont{Wasem}},
  \bibinfo{journal}{Phys.Rev.} \textbf{\bibinfo{volume}{C85}},
  \bibinfo{pages}{022501} (\bibinfo{year}{2012}), \eprint{1108.1151}.

\bibitem[{\citenamefont{Berkowitz et~al.}(2015)\citenamefont{Berkowitz, Kurth,
  Nicholson, Joo, Rinaldi, Strother, Vranas, and
  Walker-Loud}}]{Berkowitz:2015eaa}
\bibinfo{author}{\bibfnamefont{E.}~\bibnamefont{Berkowitz}},
  \bibinfo{author}{\bibfnamefont{T.}~\bibnamefont{Kurth}},
  \bibinfo{author}{\bibfnamefont{A.}~\bibnamefont{Nicholson}},
  \bibinfo{author}{\bibfnamefont{B.}~\bibnamefont{Joo}},
  \bibinfo{author}{\bibfnamefont{E.}~\bibnamefont{Rinaldi}},
  \bibinfo{author}{\bibfnamefont{M.}~\bibnamefont{Strother}},
  \bibinfo{author}{\bibfnamefont{P.~M.} \bibnamefont{Vranas}},
  \bibnamefont{and}
  \bibinfo{author}{\bibfnamefont{A.}~\bibnamefont{Walker-Loud}}
  (\bibinfo{year}{2015}), \eprint{1508.00886}.

\bibitem[{\citenamefont{Brice\~no
  et~al.}(2013{\natexlab{a}})\citenamefont{Brice\~no, Davoudi, and
  Luu}}]{Briceno:2013lba}
\bibinfo{author}{\bibfnamefont{R.~A.} \bibnamefont{Brice\~no}},
  \bibinfo{author}{\bibfnamefont{Z.}~\bibnamefont{Davoudi}}, \bibnamefont{and}
  \bibinfo{author}{\bibfnamefont{T.~C.} \bibnamefont{Luu}},
  \bibinfo{journal}{Phys. Rev.} \textbf{\bibinfo{volume}{D88}},
  \bibinfo{pages}{034502} (\bibinfo{year}{2013}{\natexlab{a}}),
  \eprint{1305.4903}.

\bibitem[{\citenamefont{Brice\~no
  et~al.}(2013{\natexlab{b}})\citenamefont{Brice\~no, Davoudi, Luu, and
  Savage}}]{Briceno:2013bda}
\bibinfo{author}{\bibfnamefont{R.~A.} \bibnamefont{Brice\~no}},
  \bibinfo{author}{\bibfnamefont{Z.}~\bibnamefont{Davoudi}},
  \bibinfo{author}{\bibfnamefont{T.}~\bibnamefont{Luu}}, \bibnamefont{and}
  \bibinfo{author}{\bibfnamefont{M.~J.} \bibnamefont{Savage}},
  \bibinfo{journal}{Phys.Rev.} \textbf{\bibinfo{volume}{D88}},
  \bibinfo{pages}{114507} (\bibinfo{year}{2013}{\natexlab{b}}),
  \eprint{1309.3556}.

\bibitem[{\citenamefont{Orginos et~al.}(2015)\citenamefont{Orginos, Parreno,
  Savage, Beane, Chang, and Detmold}}]{Orginos:2015aya}
\bibinfo{author}{\bibfnamefont{K.}~\bibnamefont{Orginos}},
  \bibinfo{author}{\bibfnamefont{A.}~\bibnamefont{Parreno}},
  \bibinfo{author}{\bibfnamefont{M.~J.} \bibnamefont{Savage}},
  \bibinfo{author}{\bibfnamefont{S.~R.} \bibnamefont{Beane}},
  \bibinfo{author}{\bibfnamefont{E.}~\bibnamefont{Chang}}, \bibnamefont{and}
  \bibinfo{author}{\bibfnamefont{W.}~\bibnamefont{Detmold}}
  (\bibinfo{year}{2015}), \eprint{1508.07583}.

\bibitem[{\citenamefont{Beane et~al.}(2006)\citenamefont{Beane, Bedaque,
  Orginos, and Savage}}]{Beane:2006mx}
\bibinfo{author}{\bibfnamefont{S.~R.} \bibnamefont{Beane}},
  \bibinfo{author}{\bibfnamefont{P.~F.} \bibnamefont{Bedaque}},
  \bibinfo{author}{\bibfnamefont{K.}~\bibnamefont{Orginos}}, \bibnamefont{and}
  \bibinfo{author}{\bibfnamefont{M.~J.} \bibnamefont{Savage}},
  \bibinfo{journal}{Phys. Rev. Lett.} \textbf{\bibinfo{volume}{97}},
  \bibinfo{pages}{012001} (\bibinfo{year}{2006}), \eprint{hep-lat/0602010}.

\bibitem[{\citenamefont{Beane et~al.}(2013)}]{Beane:2013br}
\bibinfo{author}{\bibfnamefont{S.~R.} \bibnamefont{Beane}} \bibnamefont{et~al.}
  (\bibinfo{collaboration}{NPLQCD}), \bibinfo{journal}{Phys. Rev.}
  \textbf{\bibinfo{volume}{C88}}, \bibinfo{pages}{024003}
  (\bibinfo{year}{2013}), \eprint{1301.5790}.

\bibitem[{\citenamefont{Beane et~al.}(2014)\citenamefont{Beane, Chang, Cohen,
  Detmold, Lin et~al.}}]{Beane:2014ora}
\bibinfo{author}{\bibfnamefont{S.}~\bibnamefont{Beane}},
  \bibinfo{author}{\bibfnamefont{E.}~\bibnamefont{Chang}},
  \bibinfo{author}{\bibfnamefont{S.}~\bibnamefont{Cohen}},
  \bibinfo{author}{\bibfnamefont{W.}~\bibnamefont{Detmold}},
  \bibinfo{author}{\bibfnamefont{H.}~\bibnamefont{Lin}}, \bibnamefont{et~al.},
  \bibinfo{journal}{Phys.Rev.Lett.} \textbf{\bibinfo{volume}{113}},
  \bibinfo{pages}{252001} (\bibinfo{year}{2014}), \eprint{1409.3556}.

\bibitem[{\citenamefont{Beane et~al.}(2015)\citenamefont{Beane, Chang, Detmold,
  Orginos, Parreño et~al.}}]{Beane:2015yha}
\bibinfo{author}{\bibfnamefont{S.~R.} \bibnamefont{Beane}},
  \bibinfo{author}{\bibfnamefont{E.}~\bibnamefont{Chang}},
  \bibinfo{author}{\bibfnamefont{W.}~\bibnamefont{Detmold}},
  \bibinfo{author}{\bibfnamefont{K.}~\bibnamefont{Orginos}},
  \bibinfo{author}{\bibfnamefont{A.}~\bibnamefont{Parreño}},
  \bibnamefont{et~al.} (\bibinfo{year}{2015}), \eprint{1505.02422}.

\bibitem[{\citenamefont{Detmold and Savage}(2004)}]{Detmold:2004qn}
\bibinfo{author}{\bibfnamefont{W.}~\bibnamefont{Detmold}} \bibnamefont{and}
  \bibinfo{author}{\bibfnamefont{M.~J.} \bibnamefont{Savage}},
  \bibinfo{journal}{Nucl. Phys.} \textbf{\bibinfo{volume}{A743}},
  \bibinfo{pages}{170} (\bibinfo{year}{2004}), \eprint{hep-lat/0403005}.

\bibitem[{\citenamefont{Dudek et~al.}(2010)\citenamefont{Dudek, Edwards,
  Peardon, Richards, and Thomas}}]{Dudek:2010wm}
\bibinfo{author}{\bibfnamefont{J.~J.} \bibnamefont{Dudek}},
  \bibinfo{author}{\bibfnamefont{R.~G.} \bibnamefont{Edwards}},
  \bibinfo{author}{\bibfnamefont{M.~J.} \bibnamefont{Peardon}},
  \bibinfo{author}{\bibfnamefont{D.~G.} \bibnamefont{Richards}},
  \bibnamefont{and} \bibinfo{author}{\bibfnamefont{C.~E.}
  \bibnamefont{Thomas}}, \bibinfo{journal}{Phys. Rev.}
  \textbf{\bibinfo{volume}{D82}}, \bibinfo{pages}{034508}
  (\bibinfo{year}{2010}), \eprint{1004.4930}.

\bibitem[{\citenamefont{Thomas et~al.}(2012)\citenamefont{Thomas, Edwards, and
  Dudek}}]{Thomas:2011rh}
\bibinfo{author}{\bibfnamefont{C.~E.} \bibnamefont{Thomas}},
  \bibinfo{author}{\bibfnamefont{R.~G.} \bibnamefont{Edwards}},
  \bibnamefont{and} \bibinfo{author}{\bibfnamefont{J.~J.} \bibnamefont{Dudek}},
  \bibinfo{journal}{Phys. Rev.} \textbf{\bibinfo{volume}{D85}},
  \bibinfo{pages}{014507} (\bibinfo{year}{2012}), \eprint{1107.1930}.

\bibitem[{\citenamefont{Dudek et~al.}(2012)\citenamefont{Dudek, Edwards, and
  Thomas}}]{Dudek:2012gj}
\bibinfo{author}{\bibfnamefont{J.~J.} \bibnamefont{Dudek}},
  \bibinfo{author}{\bibfnamefont{R.~G.} \bibnamefont{Edwards}},
  \bibnamefont{and} \bibinfo{author}{\bibfnamefont{C.~E.}
  \bibnamefont{Thomas}}, \bibinfo{journal}{Phys. Rev.}
  \textbf{\bibinfo{volume}{D86}}, \bibinfo{pages}{034031}
  (\bibinfo{year}{2012}), \eprint{1203.6041}.

\end{thebibliography}

\end{document}